\documentclass{article}

\usepackage{microtype}
\usepackage{graphicx}
\usepackage{subfigure}
\usepackage{booktabs} % for professional tables

\usepackage{hyperref}

\usepackage[accepted]{mlsys2025}

\usepackage[utf8]{inputenc} % allow utf-8 input
\usepackage[T1]{fontenc}    % use 8-bit T1 fonts
\usepackage{url}            % simple URL typesetting
\usepackage{amsfonts}       % blackboard math symbols
\usepackage{nicefrac}       % compact symbols for 1/2, etc.
\usepackage{caption}
\usepackage{cleveref}
\usepackage{appendix}
\usepackage{listingsutf8}
\usepackage[inline]{enumitem}
\usepackage{inconsolata}
\usepackage{xspace}
\usepackage{subcaption}
\usepackage[export]{adjustbox}
\usepackage{pifont}
\usepackage[dvipsnames]{xcolor}
\usepackage{makecell}
\usepackage{amssymb}
\usepackage{pdfrender}
\usepackage{xurl}
\usepackage{multirow}
\usepackage{array}

\newcommand{\blackxmark}{\ding{55}}%

\crefname{figure}{Fig.}{Figs.}
\crefname{section}{Sec.}{Sec.}
\crefname{subsection}{Sec.}{Sec.}
\crefname{table}{Table}{Tables}

\newcommand{\sys}[0] {Autocomp\xspace}

\newif\ifcomments
\commentstrue
\ifcomments
    \providecommand{\sahil}[1]{{\protect\color{red}{\bf [sahil: #1]}}}
    \providecommand{\alvin}[1]{{\protect\color{purple}{\bf [alvin: #1]}}}
    \providecommand{\charles}[1]{{\protect\color{teal}{\bf [charles: #1]}}}

\else
    \providecommand{\sahil}[1]{}
    \providecommand{\charles}[1]{}
    \providecommand{\alvin}[1]{}
    
\fi

\lstdefinelanguage{cpp}{
    morekeywords={void, for, int},
    escapeinside={(&}{&)},
    numbers=left, firstnumber=1, numberstyle=\tiny\color{gray},
    xleftmargin=0.5cm,
    keywordstyle=\color{blue},
    basicstyle=\footnotesize\ttfamily,
    showlines=true,
    columns=fullflexible    
}

\newcommand{\ourtitle}[0]{\sys{}: A Powerful and Portable Code Optimizer for Tensor Accelerators}

\mlsystitlerunning{\ourtitle{}}

\begin{document}

\twocolumn[
\mlsystitle{\ourtitle{}}

% unless you've provided the [accepted] option to the mlsys2025


% \mlsyssetsymbol{equal}{*}

\begin{mlsysauthorlist}
\mlsysauthor{Charles Hong}{berk}
\mlsysauthor{Sahil Bhatia}{berk}
\mlsysauthor{Alvin Cheung}{berk}
\mlsysauthor{Yakun Sophia Shao}{berk}
\end{mlsysauthorlist}

\mlsysaffiliation{berk}{UC Berkeley, Berkeley, CA, USA}

\mlsyscorrespondingauthor{Charles Hong}{charleshong@berkeley.edu}

\mlsyskeywords{Large language models, Hardware accelerators, Compilers, Code optimization}

\vskip 0.3in

\begin{abstract}
Hardware accelerators, especially those designed for tensor processing, have become ubiquitous in today's computing landscape.
However, even with significant efforts in building compilers, 
programming these tensor accelerators remains challenging, leaving much of their potential underutilized.
Recently, large language models (LLMs), trained on large amounts of code, have shown significant promise in code generation and optimization tasks, but generating low-resource languages, such as specialized tensor accelerator code still poses a significant challenge.
We tackle this challenge with \sys{}, an approach that empowers accelerator programmers to leverage domain knowledge and hardware feedback to optimize code via an automated LLM-driven search.
We accomplish this by:
1) formulating each optimization pass as a structured two-phase prompt, divided into planning and code generation phases, 
2) inserting domain knowledge during planning via a concise and adaptable optimization menu, and
3) integrating correctness and performance metrics from hardware as feedback at each search iteration.
Across three distinct hardware platforms, we demonstrate that \sys-optimized code runs 5.6$\times$ faster than the vendor-provided library (Gemmini), outperforms expert-level hand-tuned code by 1.9$\times$ (AWS Trainium), and achieves 3.8$\times$ higher performance than a machine learning-based cost model for GPUs (NVIDIA L40S).
% Across three categories of representative workloads and two different accelerators, we demonstrate that \sys-optimized code runs 5.6$\times$ (GEMM) and 2.7$\times$ (convolution) faster than the vendor-provided library, and outperforms expert-level hand-tuned code by 1.4$\times$ (GEMM), 1.1$\times$ (convolution), and 1.3$\times$ (fine-grained linear algebra).
Additionally, we demonstrate that optimization schedules generated from \sys{} can be reused across similar tensor operations, improving speedups by up to 24\% under a fixed sample budget. 
\end{abstract}
]



\printAffiliationsAndNotice{}  % leave blank if no need to mention equal contribution
% \printAffiliationsAndNotice{\mlsysEqualContribution} % otherwise use the standard text.

\section{Introduction}
\label{sec:intro}

Hardware accelerators~\cite{nickolls2008cuda, tpu-isca2016} have become a critical driving force for the recent breakthroughs~\cite{alexnet, resnet, transformer, chatgpt} in machine learning. They provide orders-of-magnitude improvements in performance and energy efficiency in running deep neural networks (DNNs), and this has led to an increasing number of accelerators for tensor processing in recent years~\cite{ane, nvdla, lauterbach2021cerebras, intelamx, armsme}. However, extracting that performance requires writing high-performance accelerator code, which is time-consuming and requires a deep understanding of the underlying hardware.

To address this challenge, various compilers and domain-specific languages (DSLs) have appeared.
For deep learning applications, compilers such as XLA, TVM, and Triton generate high-performance code, but they only support a few hardware backends, particularly CPUs and GPUs~\cite{xla, chen2018tvm}. Unfortunately, adapting compilers and DSLs to new hardware platforms with vendor-specific instruction set architectures (ISAs) and implementation-specific dataflow patterns requires significant engineering effort. In fact, software alone comprises 40-50\% of the development cost for new hardware~\cite{reuters2025openaichipcost}, even before considering the effort needed for end users to write and debug software for a newly developed chip. Prior work in DSLs like Halide and Exo~\cite{jrk2013halide,exo} targets accelerators by providing primitives that make it easier to express tensor computation, but the onus of optimizing code written in such DSLs still lies on the accelerator programmer.

Even once a compiler exists, generating performant code runs into the classical ``scheduling'' problem, i.e., deciding which optimizations to apply and in what order. For general-purpose backends (CPUs and GPUs), these passes have been iteratively developed and refined over many years through a combination of experts and auto-tuning frameworks. Recent work has gone further in exploring data-driven approaches such as supervised learning~\cite{zheng2021tenset}, reinforcement learning~\cite{cummins2022compilergym}, and even LLM training~\cite{cummins2023compiler, metallmcompiler} to tackle the combinatorial explosion of pass sequences. While these data-driven approaches have shown promise, they depend on vast amounts of performance data to train, which is painfully scarce for domain-specific hardware accelerators. 

In this paper, we present \sys{}, which solves the problems with prior approaches with an iterative LLM-driven search framework to optimize accelerator code. 
Unlike previous compilers, \sys{} can adapt to new hardware platforms and ISAs by simply changing prompts. Unlike existing tensor DSLs, \sys{} automatically generates optimized code without manual tuning.
And unlike data-driven approaches targeting CPUs and GPUs, \sys{} requires no model training, instead leveraging LLM in-context reasoning and pretrained knowledge of common optimizations. 

In each iteration, \sys{} first {\em plans} by choosing an optimization from a predefined menu, i.e., a list of common hardware accelerator optimizations like tiling and unrolling, then {\em applies} the optimization to generate optimized DSL code. The generated candidates are validated for correctness and benchmarked on the accelerator to collect performance metrics, providing feedback for the next iteration of search. By encoding DSL syntax, optimization rules, and performance feedback concisely in a prompt, \sys{} guides the LLM to generate optimized accelerator code.

% We use Gemmini~\cite{gemmini} to generate two different accelerators for evaluation. Gemmini is an accelerator generator that can generate systolic array and vector-style tensor accelerators with a wide range of data types and sizes. This is ideal for evaluating \sys{} as:
% \begin{enumerate*}[label=\arabic*)]
%     \item it delivers performance compared to commercial ones and is open-source enabling extraction of fine-grained performance feedback and easy instantiation of different accelerator instances,
%     % \item supports diverse software flows like vendor-provided library, DSLs and hand-optimized code providing a spectrum of baselines to compare against, and \alvin{is this point needed?}
%     \item low-resource nature eliminates data contamination and makes directly prompting LLMs challenging.
% \end{enumerate*}
% an accelerator that doesn't have much compiler support. 

A key advantage of this prompt-driven approach is its inherent \emph{portability}. Whereas traditional compilers require significant engineering effort to target a new hardware platform, \sys{} can be retargeted by modifying its structured prompt. This dramatically lowers the barrier to entry for optimizing code for new low-resource accelerators. We demonstrate this in our experiments by applying \sys{} to \emph{three} distinct hardware backends. 

In our evaluation, we find that \sys{} significantly outperforms prior approaches across hardware platforms. On \textbf{Gemmini}, \sys{} speeds up GEMMs by \textbf{5.6$\times$} compared to the vendor-provided library and \textbf{1.4$\times$} compared to expert-level hand-tuned code, surpassing the prior best known implementations with less human effort. On \textbf{AWS Trainium}, \sys{} reduces the runtime of a wide range of hand-optimized tensor operators by an average of \textbf{1.9$\times$}. On an \textbf{NVIDIA L40S GPU}, \sys{} generates KernelBench kernels that run \textbf{2$\times$} faster than PyTorch and \textbf{3.8$\times$} faster than TVM MetaSchedule.
Moreover, we show that \sys{}'s schedules can be reused as guidance when scheduling similar tensor operations, alleviating the cost of scheduling new code and delivering up to 24\% greater speedups under a fixed sample budget.

% \sahil{todo update this}
% In our evaluation, we apply \sys{} to two representative low-resource accelerators and generate code that runs 5.6$\times$ (GEMM) and 2.7$\times$ (convolution) faster than the vendor-provided library. Furthermore, it outperforms expert-level hand-tuned code by 1.4$\times$ (GEMM), 1.1$\times$ (convolution), and 1.3$\times$ (fine-grained linear algebra), surpassing the prior best known implementations with less human effort.
% Moreover, we show that \sys{}'s schedules can be reused as guidance when scheduling similar tensor operations, alleviating the cost of scheduling new code and delivering up to 24\% greater speedups under a fixed sample budget.
% By automating accelerator code generation, \sys{} eliminates the need for manual tuning and domain expertise, making high-performance accelerator code generation accessible to users.

In summary, we make the following contributions:
% \sahil{TODO: I will fix this}
\begin{enumerate}[nosep,leftmargin=1.5em,labelwidth=*,align=left]
    %\item To the best of our knowledge, \sys{} is the first approach to leverage LLMs for optimizing low-resource tensor accelerator code.
    \item We present \sys{}, the first LLM-driven code optimization approach for low-resource tensor accelerator code generation.
    \item \sys{} is a portable optimization framework that dramatically lowers the engineering effort to target new hardware. Our search incorporates domain knowledge, hardware feedback on correctness and performance, and novel strategies for response diversity to automatically generate performant code. 
    %\item We formulate accelerator code optimization as a novel LLM-driven search, divided into tractable steps. Our search incorporates domain knowledge, hardware feedback on correctness and performance, and novel strategies for response diversity to automatically generate performant code. 
    \item \sys{}-generated code significantly outperforms expert hand-optimized code across a wide range of workloads and across different tensor accelerators.
    \item We illustrate that schedules generated by \sys{} can be reused to optimize similar tensor operations, reducing search cost and demonstrating the \emph{a posteriori} usefulness of \sys{}-generated schedules beyond pure performance.
    \item Our implementation and prompts are fully open-source\footnote{\url{https://github.com/ucb-bar/autocomp}}.
\end{enumerate}

% Specifically:

% \begin{itemize}[leftmargin=*]
% \item We devise a two-stage prompting technique that enables us to 1) generate salient plans to optimize DSL code for our tensor accelerator, Gemmini, and 2) implement those plans as rewrites to the original code. 
% \alvin{need to say more about what's the secret sauce here}
% \alvin{I suggest moving gemmini to the last bullet as an experiment detail, i.e., to evaluate our technique we chose to generate code for gemmini, an accelerator that doesn't yet have much compiler support. don't think we need to say more about that here}

% \item We integrate our two-stage optimization method into an iterative search process that leverages LLMs' stochastic output generation to thoroughly explore the space of possible optimizations. \alvin{I don't think we can prove that we are doing an exhaustive aka thorough exploration?}

% \item We demonstrate that the \sys{} generates code 1.34$\times$ (matrix multiplication) and 1.03$\times$ (2D convolution) faster on average than expert-level hand-tuned code on a range of important tensor processing kernels, and even outperforms a hardened hardware implementation by 1.51$\times$ in robotics kernels, leading to both a reduction in engineering effort and an increase in hardware utilization.

% \sahil{we can make the introduction flow as : importance of hardware accelerators and quick iterations of compiler frameworks. what prior approaches do to tackle this. shortcomings of these approaches. short description of autocomp + how it solves this.}
% \end{itemize}
\section{Background}
% \alvin{this section is too long}

\begin{figure}
    \centering
    \includegraphics[trim={0.1cm 6.7cm 13.9cm 0.1cm},width=0.8\linewidth]{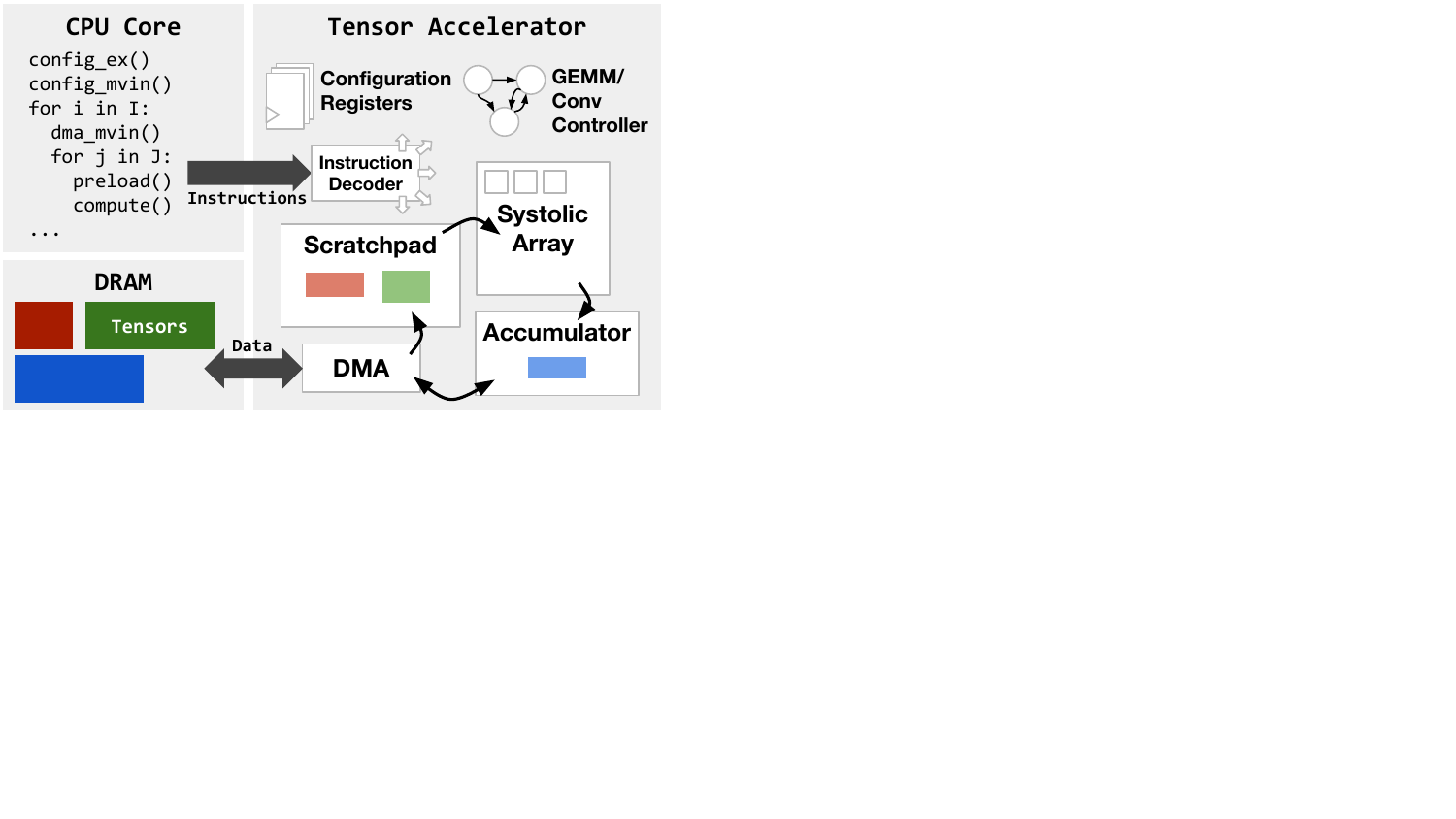}
    \vspace{-0.5em}
    \caption{Architecture and dataflow of a tensor accelerator system. Note that data movement is handled explicitly via accelerator direct memory access (DMA) instructions. 
    % GEMM/conv controllers refer to on-chip hardware implementations of matrix multiplications and convolutions, used as baselines in \cref{sec:eval}.
    }
    \label{fig:gemmini-arch}
    \vspace{-1em}
\end{figure}
\begin{figure}
\begin{lstlisting}[language=cpp,basicstyle=\scriptsize\ttfamily]
// CPU code
for (int i = 0; i < N; i++)
  for (int j = 0; j < N; j++) {
    C[i][j] = 0;
    for (int k = 0; k < N; k++)
      C[i][j] += A[i][k] * B[k][j];}

// Accelerator code
for (int ii = 0; ii < N; ii += T)
for (int jj = 0; jj < N; jj += T) {
  zero_accumulator(acc_addr+...);
  for (int kk = 0; kk < N; kk += T) {
    dma_mvin(A[ii*T][kk*T], A_spad_addr);
    dma_mvin(B[kk*T][jj*T], B_spad_addr);
    for (int i = 0; i < T; i+=16)
      for (int j = 0; j < T; j+=16) {
        for (int k = 0; k < T; k+=16)
          compute(A_spad_addr+..., B_spad_addr+...,
                  acc_addr+...);}}}
  dma_mvout(acc_addr, C[ii*T][jj*T]);}
\end{lstlisting}
\vspace{-0.5em}
\caption{Comparison of general-purpose CPU code and tensor accelerator code for matrix multiplication.}
\label{fig:accel-code-example}
\vspace{-1em}
\end{figure}

\begin{figure}[!t]
\begin{lstlisting}[language=cpp,basicstyle=\scriptsize\ttfamily]
// Unoptimized
for (int i = 0; i < 8; i++) {
    for (int j = 0; j < 32; j++) {
        for (int k = 0; k < 8; k++) {
            config_mvin(128); // A's stride is 128
            dma_mvin(A[i*16][k*16], spad_addr_1);
            config_mvin(256); // B's stride is 256
            dma_mvin(B[k*16][j*16], spad_addr_2);

// Optimized
config_mvin(128);
config_mvin_2(256);
for (int i = 0; i < 8; i++) {
    for (int j = 0; j < 32; j++) {
        for (int k = 0; k < 8; k++) {
            dma_mvin(A[i*16][k*16], spad_addr_1);
            dma_mvin_2(B[k*16][j*16], spad_addr_2);
\end{lstlisting}
\vspace{-0.5em}
\caption{Example of hoisting accelerator configuration instructions, which can block execution. In this case the accelerator supports multiple direct memory access (DMA) load instructions, each with its own configuration state.}
\label{fig:config-hoist-example}
\vspace{-0.7em}
\end{figure}
\begin{figure}
\begin{lstlisting}[language=cpp,basicstyle=\scriptsize\ttfamily]
// Unoptimized
for (int k = 0; k < 8; k++) {
    for (int i = 0; i < 32; i++) {
        dma_mvin(A[i*16][k*64], spad_addr);
        for (int k_i = 0; k_i < 4; k_i++) {
            compute(spad_addr + k_i * 16, ...);

// Optimized
for (int k = 0; k < 8; k++) {
    spad_addr = base_spad_addr;
    dma_mvin(A[0][k*64], spad_addr);
    for (int i = 0; i < 32; i++) {
        dma_mvin(A[(i+1)*16][k*64], spad_addr + 64);
        for (int k_i = 0; k_i < 4; k_i++) {
            compute(spad_addr + k_i * 16, ...);
        spad_addr += 64;
\end{lstlisting}
\vspace{-0.5em}
\caption{Example of software pipelining in tensor accelerators. The \texttt{A} matrix tile is spread throughout accelerator memory rather than repeatedly loaded to the same location, allowing data loading to run ahead and overlap with computation.}
\label{fig:pipelining-example}
\vspace{-0.7em}
\end{figure}
% \end{nolinenumbers}

\subsection{Code Optimization for Tensor Accelerators}
% As discussed in \cref{sec:intro}, tensor accelerators are essential to addressing today's exploding demand for machine learning and artificial intelligence.
% \alvin{don't think any of the text below is needed}
Programming tensor accelerators differs greatly from programming general-purpose CPUs. Tensor accelerators, depicted in \cref{fig:gemmini-arch}, generally focus on the efficient execution of fixed-size (e.g., 16$\times$16) matrix multiplication instructions, as shown in \cref{fig:accel-code-example}. Rather than trying to reduce the number or type of these instructions, which is often fixed, software optimization focuses on other aspects, such as:
\begin{itemize}[nosep,leftmargin=1.5em,labelwidth=*,align=left]
    \item Minimizing data movement between main memory and smaller accelerator-local memories (in \cref{fig:gemmini-arch}, the scratchpad and accumulator).
    \item Setting configuration state for computation and data movement.
    \item Scheduling or reordering operations to maximally overlap computation and data movement.
\end{itemize}

Code transformations that enable these optimizations range from low-level changes like arithmetic simplification or instruction selection, to higher-level changes like loop tiling, hoisting (\cref{fig:config-hoist-example}), or software pipelining (\cref{fig:pipelining-example}). These high-level changes, while improving performance, require loop nests, pointers, and indices to be modified in multiple locations, making them challenging to implement, especially in a low-resource DSL.

Prior work has explored some of this optimization space. For example, performance models like Timeloop~\cite{parashar2019timeloop} and MAESTRO~\cite{kwon2020maestro} use high-level hardware architectural models and software abstractions to represent tensor accelerators and their workloads. Much recent work has sought to automatically explore this space, using methods such as machine learning~\cite{dosa, vaesa, sakhuja2024polaris}, linear programming~\cite{huang2021cosa}, black-box optimization~\cite{zhang2022fast,sakhuja2023spotlight}, and reinforcement learning~\cite{xiao2021hasco}. While these abstractions capture some aspects of tensor accelerator code optimization, in particular the amount of data movement, they neglect other implementation-specific and instruction-level optimizations. In this work, LLM-based code generation allows us to directly rewrite accelerator code, expanding the search to include all potential axes of optimization.

% The introduction of LLM code generation to this search process allows us to expand the search space to capture all axes of tensor accelerator code optimization. 
% In this work, we expose optimizations such as instruction selection, instruction ordering, and fine-grained modifications to loops, indexing, addressing, or any other aspect of the program. 
% Furthermore, we can optimize compositions of multiple kernels, and run code on real hardware or in cycle-accurate simulation for accurate performance feedback.

\subsection{LLM-Based Code Optimization}
LLMs have been used in various code-related tasks~\cite{chen2021humaneval, nijkamp2022codegen, bhatia2024llmlift,hong2025hdl2vcodetranslationdataset}. 
For code optimization, researchers have successfully used LLMs in various paradigms, including evolutionary search~\cite{lehman2022evolutionlargemodels, romeraparedes2024funsearch}, retrieval-augmented generation~\cite{wang2025symrtlo, lin2025ecollmdrivenefficientcode}, iterative refinement using compiler feedback~\cite{peng2024perfcodegenimprovingperformancellm, damani2024warpdrive},
and model post-training~\cite{shypula2024learning, baronio2025kevinmultiturnrlgenerating}.

Some prior work, for example \citet{ouyang2025kernelbench} and \citet{loopvectorizer}, targets system-level performance programming, specifically CUDA and SIMD intrinsics. However, we are not aware of any works that address LLM code optimization for specialized hardware (i.e., not CPUs or GPUs), despite the growing need for effective code optimization methods.
% Low-level code for such accelerators is extremely low-resource, with few to no examples publicly available. 
\citet{hong2024llmaidedcompilation} show that zero-shot code generation for such languages is highly unreliable. Nonetheless, \sys{} successfully optimizes accelerator code via a combination of novel techniques.

\section{The \sys{} Approach}
\label{sec:approach}

% In this work, we optimize software code written in the ISA of the Gemmini accelerator~\cite{gemmini}. Gemmini is an open-source accelerator that supports a variety of tensor processing workloads, but is focused on deep learning. Gemmini's instructions are integrated into C code and are interleaved with instructions that run on a CPU core.

% terminology

\subsection{Rationale}
One naive way to generate optimized tensor accelerator code is to directly ask an LLM to rewrite the unoptimized code into its optimized counterpart. However, this approach fails for two reasons: 
\begin{enumerate}[nosep,leftmargin=1.5em,labelwidth=*,align=left]
    \item Tensor accelerator DSLs are low-resource languages (i.e., insufficiently represented in the LLM's training corpus), so the model produces neither semantically nor syntactically valid programs.
    \item Without guidance, the model has little notion of what optimizations to apply, or in which order.
\end{enumerate}

\begin{figure}[!t]
    \centering
    \includegraphics[trim={0 3.7cm 10.8cm 0}, clip, width=1\linewidth]{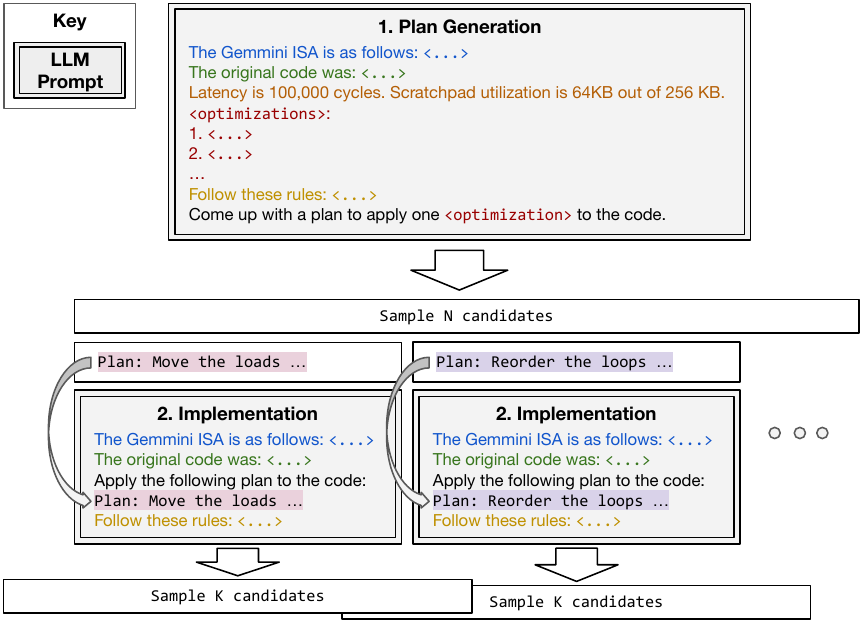}
    \caption{\sys{}'s two-phase optimization (\cref{sec:2-phase}), carried out at each iteration of beam search.}
    \label{fig:prompting}
    \vspace{-0.7em}
\end{figure}

% Prior work shows that decomposing tasks, including code generation tasks, into multiple steps can improve an LLM's ability to solve them~\cite{wei2022cot, gao2023pal, wang2023planandsolve, hong2024llmaidedcompilation, lee2024guess, wang2024planningnaturallanguageimproves}. 
Prior work shows that decomposing tasks, including code generation tasks, into multiple steps can improve an LLM's ability to solve them~\cite{gao2023pal, wang2024planningnaturallanguageimproves}. 
Therefore, as shown in \cref{fig:prompting}, we split our workflow into two phases: optimization plan generation and code implementation. We connect the two phases with an iterative beam search. Maintaining the best \textit{B} code candidates at each step helps us explore multiple optimization trajectories in parallel. We describe our two phases and search strategy in the next section. Detailed prompts are in Appendix~\ref{sec:prompts}.

In following sections, a \emph{plan} refers to a natural language description of a single step of optimization and how code is transformed in that step, whereas a \emph{schedule} refers to a sequence of plans that brings code from unoptimized to fully optimized.

\subsection{\textit{Plan-then-Implement}: Two-Phase Optimization}\label{sec:2-phase}
\textbf{Phase~1: Plan.}
We prompt an LLM to select \emph{one} optimization from a predefined menu of optimization options and to describe the concrete transformations required to apply it. \cref{fig:prompting} illustrates the prompt structure, which consists of the following parts:
\begin{enumerate}[nosep,leftmargin=1.5em,labelwidth=*,align=left]
    \item \textbf{Accelerator ISA.} A list of instructions in the accelerator's ISA. We describe the semantics of each instruction in natural language, provide a specification for the accelerator's internal memory addresses, and briefly describe the accelerator's structure.
    \item \textbf{Current Code.} In the initial iteration $t=0$, the original unoptimized code. When $t>0$, one of the \textit{B} candidates in our beam that has been selected for functional correctness and performance.
    \item \textbf{Feedback.} The latency of \textbf{Current Code} in cycles, as well as (for Gemmini) its scratchpad and accumulator utilization in kilobytes. Utilization is a common feedback metric across hardware platforms that reflects how effectively we are using the accelerator's hardware resources, and can help the model choose the next optimization to apply. For example, low scratchpad utilization can lead to the model suggesting a larger tile transformation. 
    \item \textbf{Optimization Menu.}
    A list of high-level optimizations, for instance, \emph{loop unrolling}, \emph{reordering}, \emph{fusion}, \emph{tiling}, and \emph{double buffering}. Note that only names or short descriptions of the optimization are included; we rely on the model to generate the implementation details for the selected optimization. Furthermore, the menu options range in specificity and include an option to select an optimization not listed, so the model's output is not overly constrained. The full list of optimizations for each hardware platform is in Appendix~\ref{sec:prompts}.
    % \alvin{should list the full set of items in a table somewhere and describe what each does. also explain where they come from. would be good if they are just common optimization tricks like those found in clang / gcc rather than highly specialized ones that only we know of}
    \item \textbf{Instruction.} A high-level description of the target accelerator, followed by a natural-language instruction to ``select exactly one optimization from the menu and output a corresponding transformation plan.''
    \item \textbf{Rules.} A set of constraints regarding the eventual code to be generated (see Appendix~\ref{sec:prompts}).
    % For example, we require all code to be contained within the body of a \texttt{test()} function like in the original code, restrict the usage of C preprocessor directives to streamline evaluation by evaluating multiple candidates in one simulation, and include a reminder to update all relevant indices and addresses when loop bounds are modified.
    \item \textbf{Beam Search Iteration.} The current iteration $t$ of $T$ beam search iterations. Helps guide optimization selection, as some optimizations are more relevant towards the beginning of search (e.g., loop splitting) or towards the end of search (e.g., loop unrolling).
\end{enumerate}

At each planning iteration, we sample $N$ independent plans, seeding the search with multiple diverse optimization trajectories that can be evaluated in parallel. 

\textbf{Phase~2: Implement.}
Once we generate the candidate plans, for each plan we prompt the LLM to apply the transformations in the plan to generate a new, semantically equivalent code. \cref{fig:prompting} shows the structure of our code generation prompt, which contains the following parts:
\begin{enumerate}[nosep,leftmargin=1.5em,labelwidth=*,align=left]
    \item \textbf{Accelerator ISA.} Same as in Phase 1.
    \item  \textbf{Current Code.} Same as in Phase 1.
    \item \textbf{Generated Plan.} The specific optimization plan generated for this code in Phase 1.
    \item \textbf{In-Context Learning (ICL) Example.} Depending on the target hardware platform, we include examples of certain difficult-to-implement optimizations.
    % We find that this does not improve end-to-end optimization performance, but does improve the number of correct code candidates and thereby sample efficiency.
    \item \textbf{Instruction.} A simple natural-language instruction to ``apply the above plan and output optimized accelerator code that is functionally equivalent to the current code.''
    \item \textbf{Rules.} Same as in Phase 1.
\end{enumerate}
We sample $K$ independent code candidates for each plan to improve the robustness of our search, since generating low-resource accelerator code is challenging and our task requires applying nontrivial transformations to the code. 
% Sampling multiple candidates helps improve the robustness of our search. 

\begin{figure}
    % \vspace{-1em}
    \centering
    \includegraphics[trim={0 2cm 11.8cm 0}, clip, width=0.93\linewidth]{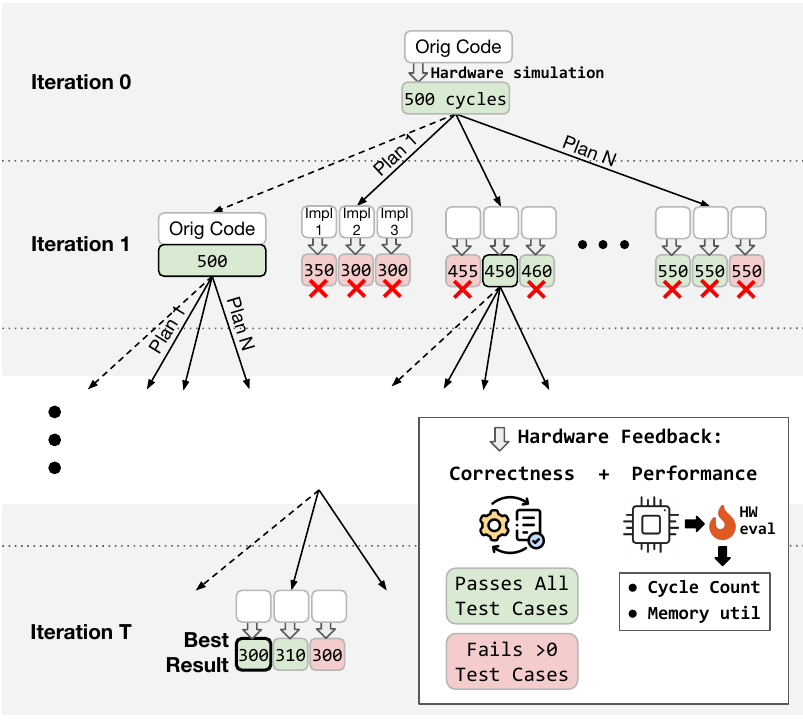}
    \caption{\sys{}'s beam search, described in \cref{sec:beam-search}.}
    \label{fig:search_tree}
    \vspace{-0.7em}
\end{figure}

\subsection{Beam Search}
\label{sec:beam-search}
We integrate our two-phase optimization inside an iterative \emph{beam search} of width $B$. Beam search allows us to efficiently explore several optimization trajectories in parallel. Since our code mostly consists of loop nests instead of sequential code, we find that merging candidates as in prior work~\cite{lehman2022evolutionlargemodels,ma2024eurekahumanlevelrewarddesign,novikov2025} 
% \sahil{not sure why alphaevlove is not rendering} 
is not suitable to tensor accelerator code. As illustrated in \cref{fig:search_tree}, candidates from the code generation step enter the beam only if they satisfy the criteria \emph{correctness} and \emph{performance}:
\begin{enumerate}[nosep,leftmargin=1.5em,labelwidth=*,align=left]
    \item \textbf{Correctness.} After each code generation step, every candidate is compiled and run against our functional test suite. Each input variable is initialized with random values and after running, the candidate's output is compared to that of a reference implementation. We first filter candidates by via a functional correctness check on the target hardware.
    \item \textbf{Performance.} We measure the latency of functionally correct candidates via cycle-accurate simulation or on-chip performance measurement (depending on the hardware platform). A candidate is retained only if it improves upon the parent from which it was derived.
\end{enumerate}

Of the functionally correct candidates, we keep the best (lowest latency) $B$ to seed the next iteration of beam search. Empirically, we find beam width $B=6$ to be a sweet spot that balances search quality and time trade-off.
% Choosing the beam width is a trade-off: a beam that is \textit{too small} may discard some promising candidates before they can be fully explored, whereas a beam that is \textit{too large} makes the candidate generation and evaluation prohibitively slow. \alvin{just say that this is a hyperparameter. so how is this chosen?}
We run this loop for a fixed budget of iterations.
% Combining candidates (as in evolutionary search) did not yield noticeable performance improvements in our experiments, likely because our code consists mainly of loop nests, rather than sequential code, and it is therefore difficult to combine multiple programs. 

\subsection{Increasing Plan and Code Diversity}
% \alvin{weird to have a section entirely consisting of bullet points}
We use the following two techniques to boost the diversity in plan (and in the case of LLM ensembling, code) generation and prevent the model from repeatedly selecting the same optimization:
\begin{itemize}[nosep,leftmargin=1.5em,labelwidth=*,align=left]
    \item \textbf{Optimization Menu Dropout.} Inspired by methods for preventing neural networks from overfitting~\cite{srivastava2014dropout}, we implement dropout in our optimization menu. Each time a plan is generated, each menu option in the list of optimizations has a chance to be removed.
    \item \textbf{LLM Ensembling.} Ensembling LLMs is known to improve diversity and quality of results~\cite{jiang2023llmblender}. To further increase the diversity of generated plans and code, whenever multiple candidates are sampled, we divide these requests between different LLMs.
\end{itemize}
We ablate these techniques, along with other components of \sys{}, in Appendix~\ref{sec:ablations}.

\subsection{Schedule Reuse}\label{sec:reuse-method}
Running \sys{}'s search starting with unoptimized code results in optimized code that outperforms all prior approaches, as we will discuss in~\cref{sec:eval}. However, using this approach with every new software workload can be costly, as \sys{} involves multiple LLM invocations and hardware simulations. A natural question, then, is whether the schedules discovered for one workload can be used to accelerate the optimization of others. We draw inspiration from traditional optimized libraries like BLAS~\cite{blackford2002blas}, where hand-tuned schedules are reused across GEMM shapes, and extend \sys{} with schedule reuse. 

To do so, we first record the best known schedule for a particular tensor operation. Then, during planning for new GEMMs with the same aspect ratios or with two shared dimensions, rather than exploring the full menu, we prompt the LLM to apply the menu options used in our recorded schedule, one at a time. As we are not exploring the full menu, we can use a smaller beam width and sample count, reducing both LLM calls and search time. After completing this lightweight search, we take the best-performing code so far and further refine it by invoking the full \sys{} search for a small number of iterations. This resembles the classic exploration-exploitation trade-off in optimization: by reusing a schedule we exploit a known high-quality schedule and avoid the initial exploration cost for a new workload.
\section{Evaluation}\label{sec:eval}

\subsection{Hardware Platforms}

\subsubsection{Gemmini}
Gemmini~\cite{gemmini} is an accelerator generator that can generate systolic array- and vector-style tensor accelerators with a wide range of data types and sizes. Gemmini is ideal for evaluating \sys{} as it:
\begin{enumerate*}[label=\arabic*)]
    \item generates accelerators that deliver performance comparable to commercial ones,
    \item is open-source, enabling instantiation of different accelerator instances, user modifications to the software toolchain, and extraction of fine-grained performance feedback, and
    \item supports fast and cycle-accurate hardware simulation via FireSim~\cite{firesim}.
\end{enumerate*}
The specialized and low-resource nature of generated Gemmini instances is ideal for our evaluation, as it  eliminates data contamination and makes directly prompting LLMs challenging. 
% We used AWS EC2 F1 instances and local AMD Alveo U250 FPGAs to run FireSim.

\subsubsection{AWS Trainium}
Trainium~\cite{trainium} is a family of state-of-the-art tensor accelerators built and deployed by Amazon Web Services (AWS). Trainium's software stack includes several different entry points for users, including PyTorch, JAX, and the Neuron Kernel Interface (NKI). In our evaluation, we optimize code written in NKI, which enables lower-level control of computation and data movement.
While Trainium accelerators are a real-world, high-performance industry backend, they are very low-resource, as Trainium was first deployed in 2022~\cite{trainium_released} with NKI only being released in 2024~\cite{trainium_nki_released}. This makes Trainium an ideal target for evaluating \sys{}. 
We optimize code for Trainium 1 (specifically, a \texttt{trn1.2xlarge} instance) as later generations are not yet widely available. 
This instance contains two NeuronCore-v2, each of which contains scalar, vector, and tensor (systolic array) engines, as well as on-chip scratchpad and accumulator memories (called SBUF and PSUM), which communicate with main memory, and supports a wide range of data types.

\subsubsection{NVIDIA L40S GPU}
As discussed in \cref{sec:intro}, ML-based compilers such as TVM Ansor~\cite{ansor} have been developed to generate high-performance deep learning code for CPUs and GPUs. However, these compilers do not support low-resource accelerators such as Gemmini and Trainium, and porting TVM to these platforms would require huge amounts of engineering effort. As a result, we additionally evaluate \sys{} on a GPU backend (specifically, NVIDIA L40S) to compare \sys{}'s performance against TVM's ML-based auto-scheduler and further demonstrate \sys{}'s portability and effectiveness. The L40S is a modern datacenter GPU that includes Tensor Cores specialized for tensor processing, and supports a wide range of data types and CUDA libraries.

\subsection{Discussion on \sys{}'s Portability}
\sys{}'s initial implementation targeted Gemmini. Once this initial implementation was complete, it was remarkably easy to port \sys{} to new hardware platforms. Qualitatively, \sys{} is easily portable thanks to the fact that for new hardware platforms, only prompts need to be changed---specifically the optimization menu and ISA description need to be updated---there is no complex compiler backend code to write. Quantitatively, we were able to get up and running with each of the Trainium and GPU backends in less than a day's effort by one graduate student.
% (with a couple more days needed to debug system infrastructure and tune the \textbf{Optimization Menu}). 
 When compared against the large industrial efforts behind multi-platform compilers such as XLA, this miniscule investment clearly illustrates \sys{}'s extreme portability and ease of use.

% \begin{table*}[]
% \centering
% \footnotesize
% \begin{tabular}{ l l l l l}
% \toprule
% \textbf{Approach} & \textbf{Performance} & \makecell[l]{\textbf{Portable across}\\\textbf{workloads}}& \makecell[l]{\textbf{Portable across}\\\textbf{accelerators}} & \makecell[l]{\textbf{No power/}\\\textbf{area cost}} \\ \midrule
% \makecell[l]{High-level library (e.g. PyTorch)} & Low & \cmark & \cmark & \cmark \\
% Hand tuning (e.g. Exo Opt) & Medium/High & \xmark & \cmark & \cmark \\
% ML cost model (e.g. AutoTVM) & Medium/High & \cmark & \xmark & \cmark \\
% Hardware FSM & High & \xmark & \xmark & \xmark \\
% \textbf{\sys} & \textbf{High} & \boldcheckmark & \boldcheckmark & \boldcheckmark \\
% \bottomrule
% \end{tabular}
% \caption{Qualitative comparison of \sys to baseline approaches.}
% \label{tab:baselines}
% % \vspace{-3.5em}
% \end{table*}

\begin{table*}[t]
\centering
\scriptsize
\begin{tabular}{ l | l l l l l}
\toprule
\textbf{Baseline} & \makecell[l]{\textbf{Gemmini (16x16 INT8)}}& \textbf{Gemmini (4x4 FP32)} & \makecell[l]{\textbf{AWS Trainium}} & \makecell[l]{\textbf{NVIDIA L40S}} \\ \midrule
\makecell[l]{High-level library} & Gemmini SW lib & \blackxmark & PyTorch NeuronX~\cite{trainium_xla} & PyTorch \\ \midrule
Unoptimized low-level & Exo unoptimized & \citet{dong2024designspaceexplorationembedded} & \makecell[l]{\texttt{nki-samples} Tutorial} & \blackxmark \\ \midrule
Optimized low-level & \makecell[l]{Exo optimized~\cite{exo}} & \blackxmark & \makecell[l]{\texttt{nki-samples} Tutorial/Advanced} & \blackxmark \\ \midrule
ML cost model & \blackxmark & \blackxmark & \blackxmark & \makecell[l]{TVM MetaSchedule\\\cite{shao2022metaschedule}} \\ \midrule
Hardware FSM & \makecell[l]{Gemmini GEMM/conv FSMs} & \makecell[l]{Expert hand-optimized (this work)} & \blackxmark & \blackxmark \\
\bottomrule
\end{tabular}
\caption{Summary of baselines used with each hardware platform.}
\label{tab:baselines}
\vspace{-1em}
\end{table*}

\subsection{Baselines}\label{sec:baselines}
We compare \sys{} with the following baselines. Note that not all baselines are available for each hardware platform.
% We compare \sys with the following baselines, summarized in~\cref{tab:baselines}. Note that not every baseline is available for each hardware platform.

\begin{figure*}[!htb]
  \vspace{-0.5em}
  \centering
  \begin{minipage}[t]{0.52\textwidth}
    \centering
    \includegraphics[trim={0.3cm 0cm 2.5cm 0cm},clip,width=0.97\linewidth]{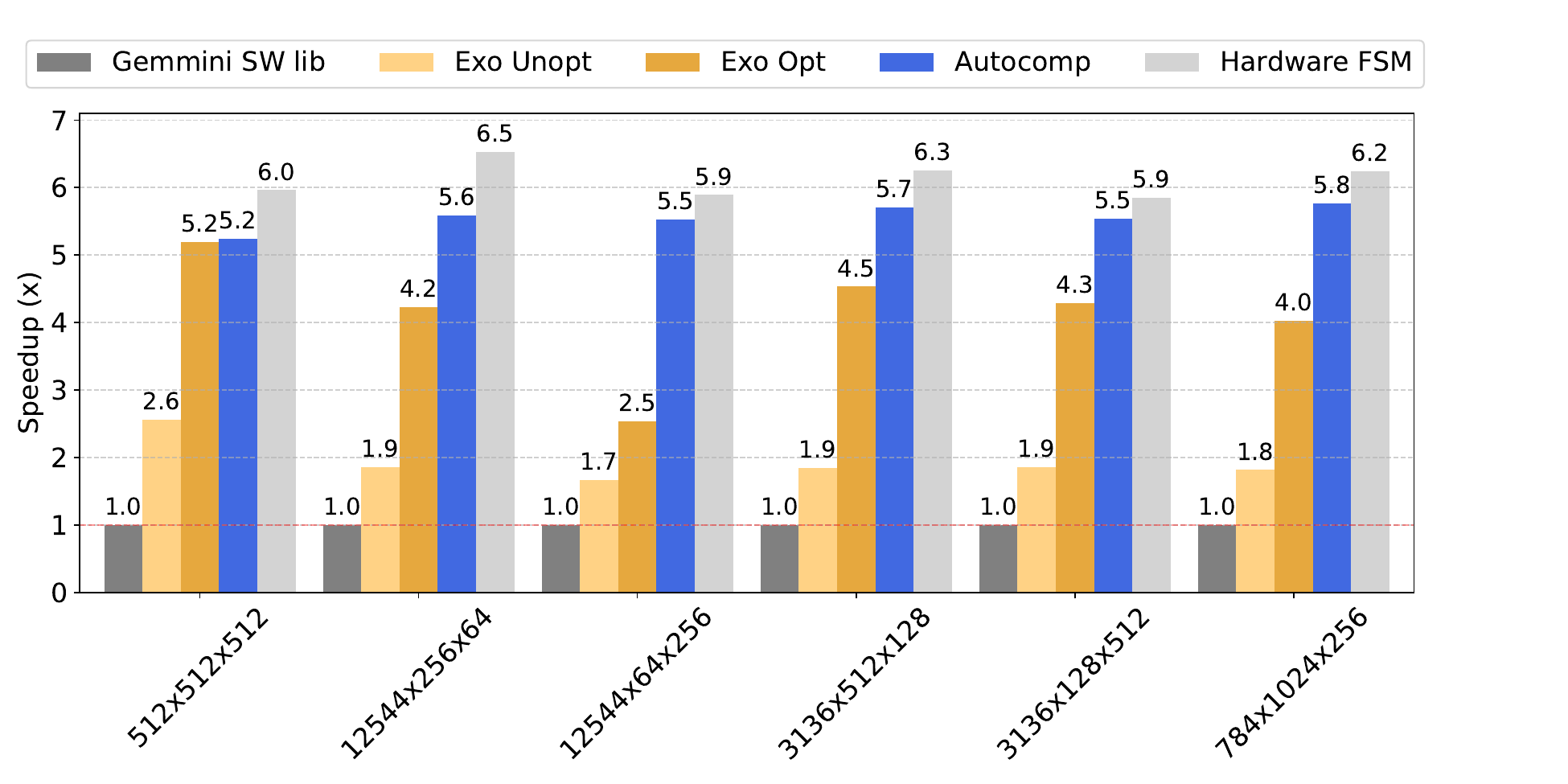}
    \vspace{-1em}
    \caption{Speedup for \textbf{Gemmini GEMM} benchmarks.}
    \label{fig:gemm}
  \end{minipage}%
  \begin{minipage}[t]{0.265\textwidth}
    \centering
    \includegraphics[trim={0.38cm 0cm 0.3cm 0cm}, clip, width=0.97\linewidth,valign=b]{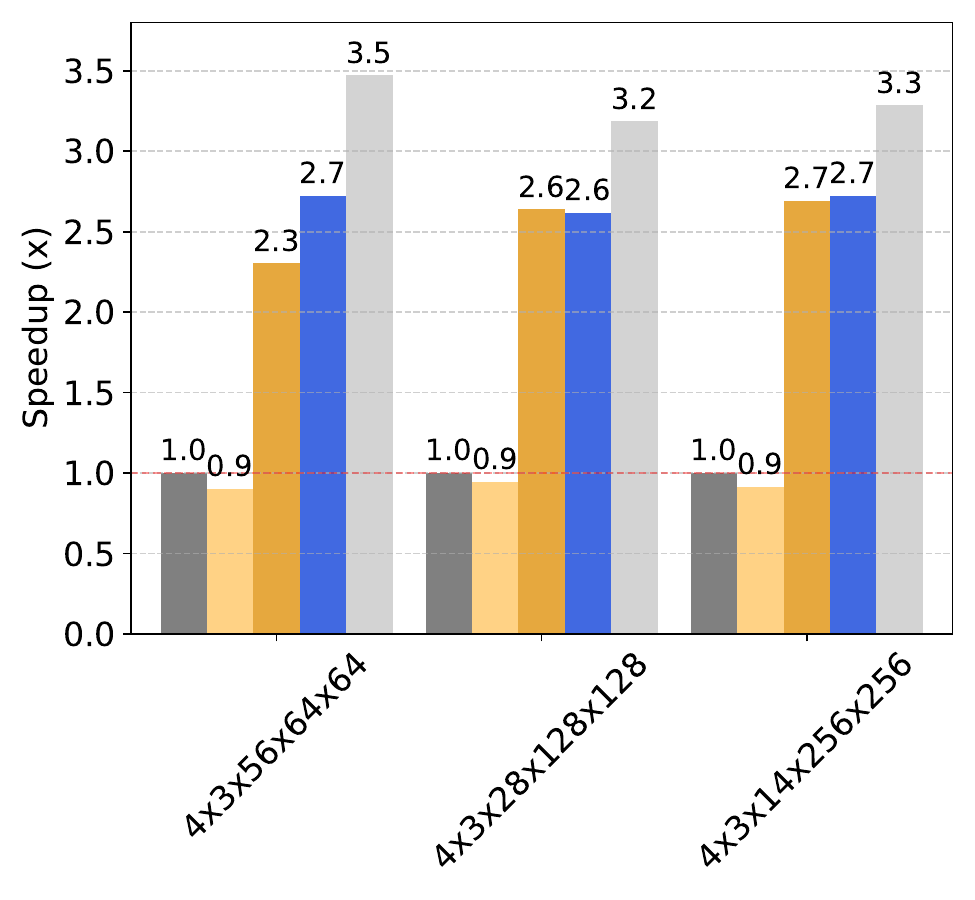}
    \vspace{-1.2em}
    \caption{Speedup for \textbf{Gemmini convolution} benchmarks.}
    \label{fig:conv}
  \end{minipage}
  \hfill
\begin{minipage}[t]{0.19\textwidth}
\centering
\raisebox{0.17cm}{\includegraphics[trim={0.3cm 0cm 0.2cm 1.5cm},clip, width=0.9\linewidth]{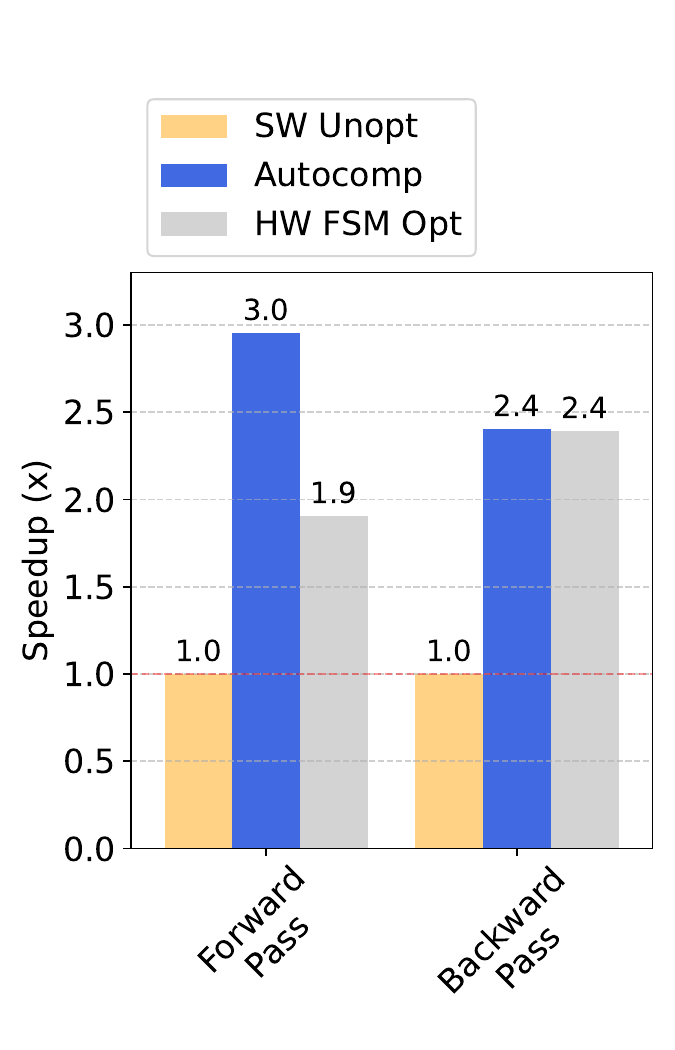}}
\vspace{-2em}
\captionof{figure}{Speedup for \textbf{Gemmini fine-grained linalg} benchmarks.}
\label{fig:admm}
\end{minipage}
\vspace{-1em}
\end{figure*}

\begin{enumerate}
[nosep,leftmargin=1.5em,labelwidth=*,align=left]
    \item \textbf{High-Level Software Library.} Gemmini ships with a software library that uses heuristics to tile and run GEMMs and convolutions on generated accelerators. As loop ordering is fixed and loop bounds, addresses, and indices must be computed at runtime, this implementation incurs significant software overhead and cannot fully utilize hardware resources.
    Trainium can be run using PyTorch as a frontend, but similarly, this prevents the user from implementing low-level optimizations. Instead, Trainium's NeuronX compiler, based on the XLA compiler~\cite{xla, trainium_xla}, can automatically optimize a PyTorch module by tracing it and taking advantage of fixed shapes and fixed control flows to produce a fused computation graph. Finally, NVIDIA GPUs are commonly used with PyTorch as a frontend for high-level programming.
    \item \textbf{Unoptimized Low-Level Code.}
    %% general pass, having a dsl to allow tuning,
    Exo~\cite{exo} is a DSL for tensor computation and scheduling. It comes with a basic compiler that emits GEMM or convolution code in Gemmini's ISA. However, without benchmark-specific optimization, performance is highly suboptimal, as hardware resources such as local memories tend to be underutilized.
    Trainium provides the \texttt{nki-samples} repository, which contains naive, unoptimized NKI implementations of several tensor operations in the directory \texttt{src/nki\_samples/tutorial}.
    \item \textbf{Hand-Optimized Low-Level Code.} In \citet{exo}, Exo's and Gemmini's developers spent significant effort manually writing and hand-tuning benchmark- and accelerator-specific schedules for each of the GEMM and convolution sizes in \cref{fig:gemm,fig:conv}. This is the previous best known software implementation for these benchmarks. Similarly, Trainium provides optimized versions of the naive tutorial implementations in \texttt{nki-samples}. In addition, \texttt{nki-samples} provides a set of advanced implementations that ``showcase cutting-edge optimizations and specialized implementations,'' in the directory \texttt{contributed/neuron-team-kernels}.
    \item \textbf{Machine Learning-Based Cost Model.} We use TVM MetaSchedule as our baseline for ML-based deep learning compilers for GPUs. MetaSchedule is TVM's latest iteration of auto-scheduler, succeeding TVM Ansor. This auto-scheduler runs a large number of samples, selected via evolutionary search, to train a cost model based on a large number of hand-crafted features~\cite{shao2022metaschedule}. No such ML-based compiler exists for the low-resource accelerators, Gemmini and Trainium.
    \item \textbf{Gemmini Hardware FSM.} Gemmini can generate accelerators with specialized hardware units for two coarse-grained operations: GEMM and convolution. These hardware units, implemented as finite state machines (FSMs), encode control sequences for each of these operations in hardware. If tuned correctly, the FSMs can exceed the theoretical maximum performance of any software-based implementation, as hardware is inherently parallel. 
    %, whereas Gemmini's software instructions are issued sequentially by a coupled CPU, incurring software overhead. 
    However, this is accomplished at the cost of scheduling flexibility, as well as increased area, power, and hardware complexity. We use the hardware FSM as a reference for the highest achievable performance, but do not expect to exceed its performance for these GEMM/convolution benchmarks, as its compute utilization exceeds 90\% for all but one benchmark.
    % Despite this, \sys-generated code approach hardware FSM performance for GEMM/convolution, and as seen in \cref{sec:admm}, even exceeds hardware FSM performance in end-to-end application performance thanks to \sys{}'s greater scheduling flexibility.
\end{enumerate}

% \vspace{-1em}

\subsection{Gemmini Evaluation}
On Gemmini-generated accelerators, we evaluate the effectiveness of \sys{} on three distinct types of workloads: 
\begin{enumerate*}[label=\arabic*)]
    \item matrix multiplication (GEMM) derived from ResNet-50,
    \item convolution derived from ResNet-50, and
    \item robotics code used for model-predictive control.
\end{enumerate*}
We ensemble OpenAI's o3-mini and gpt-4o (via the OpenAI API Platform) for both phases, with temperature 1.0. Menu options are dropped out with 70\% probability.
During the Implement phase of prompting, in cases where the optimization plan contains the string \texttt{``tiling''}, we provide an \textbf{ICL Example} example of code (from a different workload) before and after changing one tiling factor, as tiling is a key optimization that requires modifications across the program, making it challenging to implement.

For GEMM and convolution, we use unoptimized low-level code generated by Exo, which contains statically computed loops, addresses, and indices, as \sys{}'s starting point. This simplifies code generation and allows us to directly compare the effectiveness of \sys{} to hand-optimization. The target for these workloads is a Gemmini-generated accelerator with 16$\times$16 systolic array, 8-bit integer data type (accumulating in 32-bit), 256 KB scratchpad, and 64 KB accumulator, as used by \citet{exo}.

The third workload type, fine-grained linear algebra, contains sequences of element-wise operations and matrix-vector multiplications from robotics control. As this is not directly supported by Exo, we compare to an unoptimized software implementation ported to accelerator code by \citet{dong2024designspaceexplorationembedded} (also used as the starting point for \sys{}), and an extensively hand-optimized hardware FSM-based implementation written by an expert. These are run on a 32-bit floating point accelerator with a 4$\times$4 systolic array, and the same scratchpad and accumulator sizes as above, as used by \citet{dong2024designspaceexplorationembedded}.

  % \begin{figure}[t]
  %   \centering
  %   \includegraphics[width=\linewidth,trim={0.3cm 0cm 2.5cm 0cm},clip]{figs/gemm_speedup_chart.pdf}
  %   \vspace{-1em}
  %   \caption{Speedup for \textbf{Gemmini GEMM} benchmarks.}
  %   \label{fig:gemm}
  % \end{figure}%

\subsubsection{Matrix multiplication (GEMM)}
\label{sec:gemm}
We run \sys{} on a set of GEMM benchmarks selected by \citet{exo} from ResNet-50~\cite{resnet} for diversity in size and shape. We run search with beam size $B=6$, $N=6$ plans per element in the beam, $K=2$ code candidates per plan, and $T=15$ iterations. 
% This takes around 5 hours to run.

\Cref{fig:gemm} shows that \sys{} significantly outperforms even extensively hand-optimized code (Exo Opt) by a geomean of 1.4$\times$, Exo Unoptimized code (the starting point of \sys{}'s search) by 2.9$\times$, and Gemmini's software library by 5.6$\times$. \sys{} consistently achieves at least 85\% of the hardware FSM's utilization (91\% on average).

\sys{} especially outperforms prior implementations thanks to extensive exploration of software pipelining and double-buffering, which allows better overlapping of data movement and computation, for example by double-buffering both the scratchpad and accumulator. In many cases, \sys{}'s exploration also leads to different tiling and loop ordering choices than hand-optimized code, reducing data movement. We analyze \sys{}-generated GEMM code in further detail in Appendix~\ref{sec:code-examples}.

% \begin{figure}[!htb]
%   \vspace{-0.5em}
%   \centering
%   \begin{minipage}[t]{0.64\linewidth}
%     \centering
%     \includegraphics[trim={0.3cm 0cm 2.5cm 0cm},clip, height=6.2cm]{mlsys2025style/figs/conv_speedup_chart.pdf}
%     \vspace{-1em}
%     \caption{Speedup for \textbf{Gemmini convolution} benchmarks.}
%     \label{fig:gemm}
%   \end{minipage}%
%   \hfill
%   \begin{minipage}[t]{0.33\linewidth}
%     \centering
%     \includegraphics[trim={0.3cm 0cm 0.3cm 0cm}, clip, height=5.9cm,valign=b]{mlsys2025style/figs/admm_speedup_chart.pdf}
%     \vspace{-2em}
%     \caption{Speedup for \textbf{Gemmini fine-grained linear algebra} benchmarks.}
%     \label{fig:conv}
%   \end{minipage}
% \vspace{-0.9em}
% \end{figure}

\subsubsection{Convolution}\label{sec:conv}
We also optimize convolution benchmarks from ResNet-50 via the same process. Compared to the GEMM benchmarks, this code contains more loops and is more complex. In this case, we run beam search with beam size $B=6$, $N=12$ plans, and $K=4$ code candidates, for $T=10$ iterations. 
% which takes about 7 hours.

Compared to GEMM, convolution provides less room for improvement over both the Gemmini software library and \citet{exo}'s implementation, as even the hardware FSM only achieves a 3.3$\times$ geomean speedup over the software library, compared to 6.1$\times$ for GEMM. This is because on average the Gemmini software library achieves 28\% of the theoretical maximum compute utilization, compared to 16\% for GEMM. As discussed by \citet{gemmini}, at ResNet-50's tensor sizes, convolutions have greater arithmetic intensity than GEMMs, making them less memory-bound and causing the Gemmini software library’s suboptimal data orchestration to be less impactful.

Nonetheless, as shown in \cref{fig:conv}, \sys{} still exceeds the previous best known hand-optimized software ISA-based implementation (Exo Opt) by up to 1.2$\times$ and by a geomean of 1.1$\times$, via similar strategies as for GEMM. It also outperforms Exo Unoptimized code by 2.9$\times$, and Gemmini's software library by 2.6$\times$, and in all cases achieves at least 78\% of the hardware FSM's utilization.

% \begin{figure}[t]
% \centering
% \includegraphics[trim={0.3cm 0cm 0.2cm 0cm}, clip, height=6.2cm]{figs/admm_speedup_chart.pdf}
% \vspace{-1em}
% \captionof{figure}{Speedup for \textbf{Gemmini fine-grained linear algebra} benchmarks.}
% \label{fig:admm}
% \end{figure}

\subsubsection{Fine-Grained Linear Algebra}
\label{sec:admm}

Finally, we optimize fine-grained linear algebra benchmarks from the TinyMPC model-predictive control library~\cite{nguyen2024tinympc}, specifically the forward and backward passes of the primal update step. These benchmarks contain sequences of floating-point matrix-vector multiplications, interleaved with element-wise addition and subtraction. The inclusion of CPU-accelerator dependencies, low reuse, and a high ratio of data movement to computation leads to low accelerator utilization and makes this code challenging to optimize.

We compare \sys{}-generated code against \citet{dong2024designspaceexplorationembedded}'s unoptimized software-based implementation on a 4$\times$4 FP32 accelerator. For this work, we additionally had an expert hand-tune a hardware FSM-based implementation. The unoptimized software-based implementation is used as the starting point for search, and we use the same search parameters as for convolution, except with $T=15$ iterations.
% which takes about 12 hours. 
Some of the optimization menu options are different from those used for GEMM/convolution (see Appendix~\ref{sec:prompts}). As shown in \cref{fig:admm}, \sys{} outperforms even the expert-optimized hardware FSM implementation on the forward pass (by 1.6$\times$), and across benchmarks speeds up unoptimized code by a geomean of 2.7$\times$.

To outperform the hardware FSM implementation, \sys{} harnesses the flexibility of software-based implementation. It optimizes the code by hoisting data loads shared between kernels (reducing data movement beyond what is possible for the hardware FSM implementation), as well as utilizing fine-grained software pipelining and eliminating blocking operations where possible. This experiment highlights \sys{}'s adaptability: we optimize a new benchmark, running on an accelerator with a new size and data type, with highly different performance characteristics from previous experiments, by changing a few lines across the \textbf{Accelerator ISA} and \textbf{Optimization Menu} sections of the prompt.

\subsection{Trainium Evaluation}

As discussed above, we optimize two categories of Trainium benchmarks: \textit{Tutorial} (starting from naive code and comparing to optimized code) and \textit{Advanced} (starting from optimized code). Because Trainium has a larger ISA than Gemmini, for each benchmark, we manually specify relevant instructions, and include a description and examples (sourced from Trainium's documentation\footnote{\url{https://awsdocs-neuron.readthedocs-hosted.com/en/latest/nki/index.html}}) for only those instructions in the \textbf{Accelerator ISA} section of the prompt. Furthermore, as Trainium's compiler provides relatively good error messages for syntax errors, when a code implementation fails correctness checking, we prompt the LLM with the Accelerator ISA, the code, and the compiler-generated error message, and then evaluate the fixed code again.

For all experiments, we ensemble OpenAI's o4-mini and gpt-5 for both phases. We search with 70\% menu dropout, beam size $B=6$, $N=6$ plans, $K=2$ code candidates, and $T=10$ iterations.

\begin{figure}
    \centering
    \includegraphics[width=0.95\linewidth]{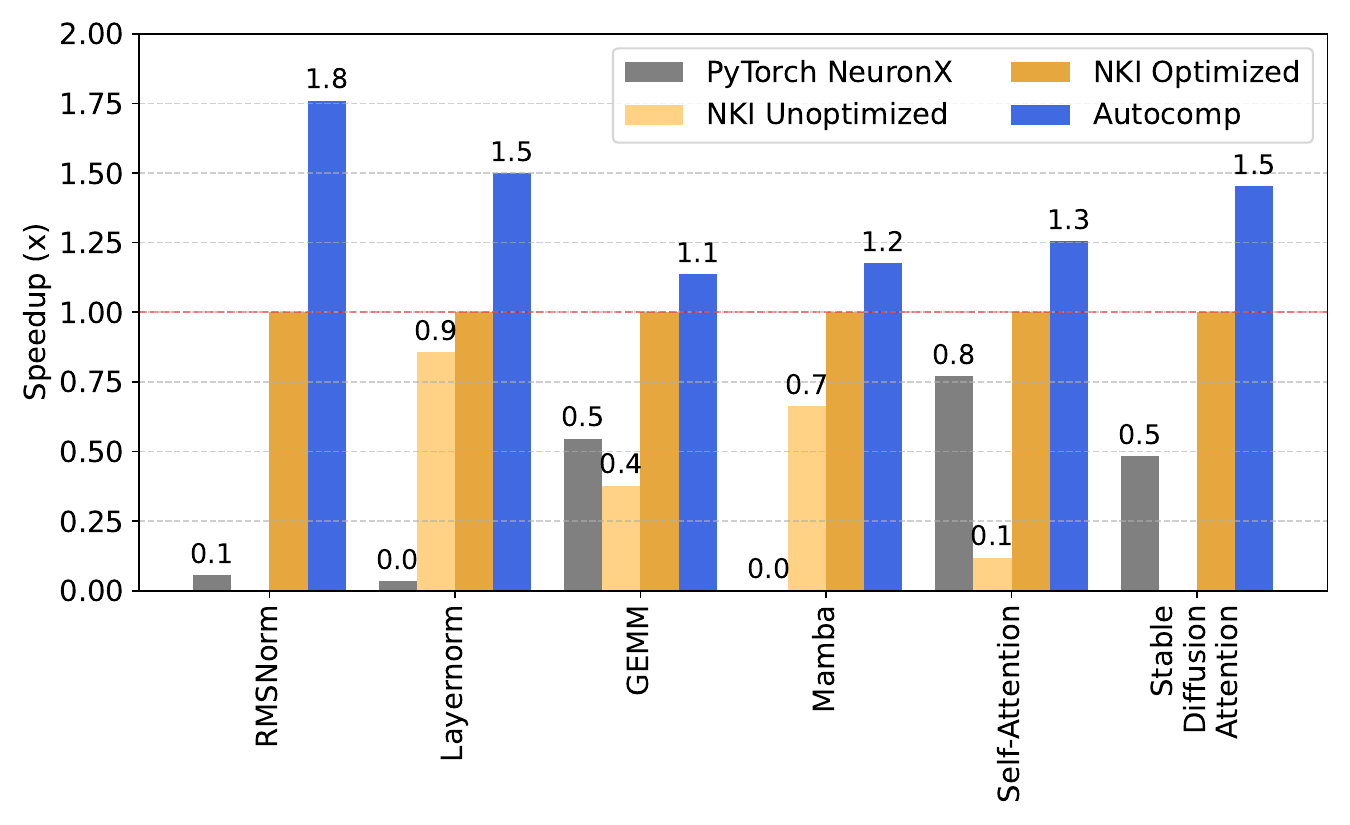}
    \vspace{-0.7em}
    \caption{Speedup for \textbf{Trainium tutorial} benchmarks, relative to NKI hand-optimized implementation.}
    \label{fig:trainium-tutorial}
    \vspace{-0.6em}
\end{figure}
\begin{figure}
    \centering
    \includegraphics[width=0.95\linewidth]{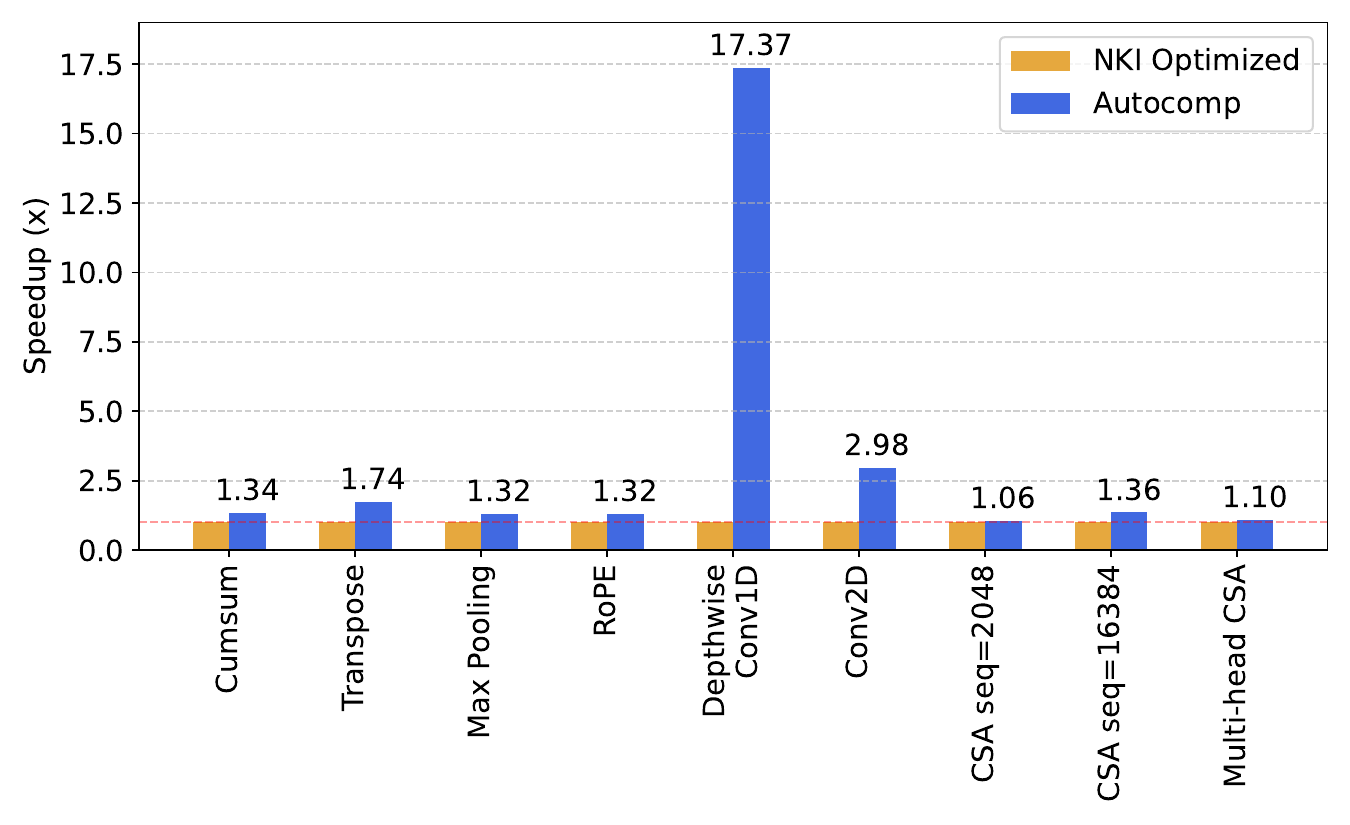}
    \vspace{-0.7em}
    \caption{Speedup for \textbf{Trainium advanced} benchmarks, relative to NKI hand-optimized implementation. CSA$=$causal self-attention.}
    \label{fig:trainium-advanced}
    \vspace{-0.7em}
\end{figure}

\subsubsection{Tutorial Workloads}
As discussed in \cref{sec:baselines}, AWS provides a set of tutorial NKI implementations that demonstrate varying levels of optimization. These workloads include key deep learning operators of varying scopes, detailed in Appendix~\ref{sec:experiments-appendix}. For these workloads, we start optimization from the unoptimized naive NKI implementation, if one is available (for RMSNorm and Stable Diffusion attention, we start from the optimized implementation).

Many of the optimizations in our \textbf{Optimization Menu} for Trainium are based on these tutorials, so we expect \sys{} to at least match the performance of the fully optimized code. \sys{} not only does so, but as shown in \cref{fig:trainium-tutorial}, outperforms hand-optimized code by a geomean of 1.36$\times$. In doing so, \sys{} speeds up the starting code used as input (either unoptimized or optimized NKI code) by a geomean of 2.51$\times$.

The \texttt{nki-samples} repository also contains PyTorch implementations of these operators, which we compile with Trainium's NeuronX compiler. \sys{}-generated code outperforms the code compiled from PyTorch by 13.52$\times$.

\subsubsection{Advanced Workloads}
AWS also provides a set of highly optimized NKI implementations written by expert kernel engineers. For these workloads, we start search from the already optimized code. Since these workloads are already optimized, any improvement is a highly positive result. As shown in \cref{fig:trainium-advanced}, we find that \sys{} is able to optimize these workloads by a geomean of 1.9$\times$, including speeding up 2D convolution by 2.98$\times$ and causal self-attention by up to 1.36$\times$. Unfortunately, no matching PyTorch implementation is provided for these workloads.

Notably, \sys{} speeds up 1D depthwise convolution by 17.37$\times$. It does so through a sequence of optimizations that takes advantage of both the specific target shape as well as the code's inherent inefficiencies: first, it decreases allocated tile sizes in the scratchpad to prevent spilling to main memory, which allows the next iteration to move the new smaller tile-sized accumulations into the accumulator. Then, it swaps loop ordering to increase filter reuse over batches, and finally, re-expands scratchpad tile size by adding a new level of tiling over the long output dimension, increasing data reuse in the scratchpad.

\subsection{NVIDIA L40S GPU Evaluation}
When optimizing GPU code, we remove the \textbf{Accelerator ISA} component of the prompt, as Python/CUDA are relatively high-resource languages compared to Gemmini and Trainium. The \textbf{Optimization Menu} is crafted from a variety of sources~\cite{mills_2024,lange2025robustagenticcudakernel,li2025cudal1improvingcudaoptimization,ouyang_liang_mirhoseini_2025}. For example, ``Use shared memory to reduce global memory bandwidth usage,'' or ``Minimize divergent branches within warps'' (the full list can be found in Appendix~\ref{sec:prompts}). We increase menu option dropout to 80\% due to the large resulting menu. We ensemble OpenAI's o4-mini and gpt-5 for both phases, with temperature 1.0. We use beam size $B=6$, $N=6$ plans, $K=2$ code candidates, and $T=10$ iterations.
During the Implement phase, when the string \texttt{``tensor core''} is detected in a plan, we include several \textbf{ICL Examples} of tensor core usage in CUDA, sourced from CUDA/cuBLAS/cuDNN documentation.

For our evaluation, we select a few kernels from KernelBench~\cite{ouyang2025kernelbench}. We focus on kernels from level 1 (individual operators) which do tensor-tensor computation, as we found that with more complex operations, porting PyTorch code to TVM was highly cumbersome and TVM's scheduler failed at a high rate or was extremely slow. We tune TVM MetaSchedule with 1000 trials.

KernelBench uses PyTorch code as a starting point for optimization. As a result, in the first two iterations of search, we focus on generating an initial inline CUDA-based implementation using a reduced menu (see Appendix~\ref{sec:prompts}).

\begin{figure}
    \centering
    \includegraphics[width=\linewidth]{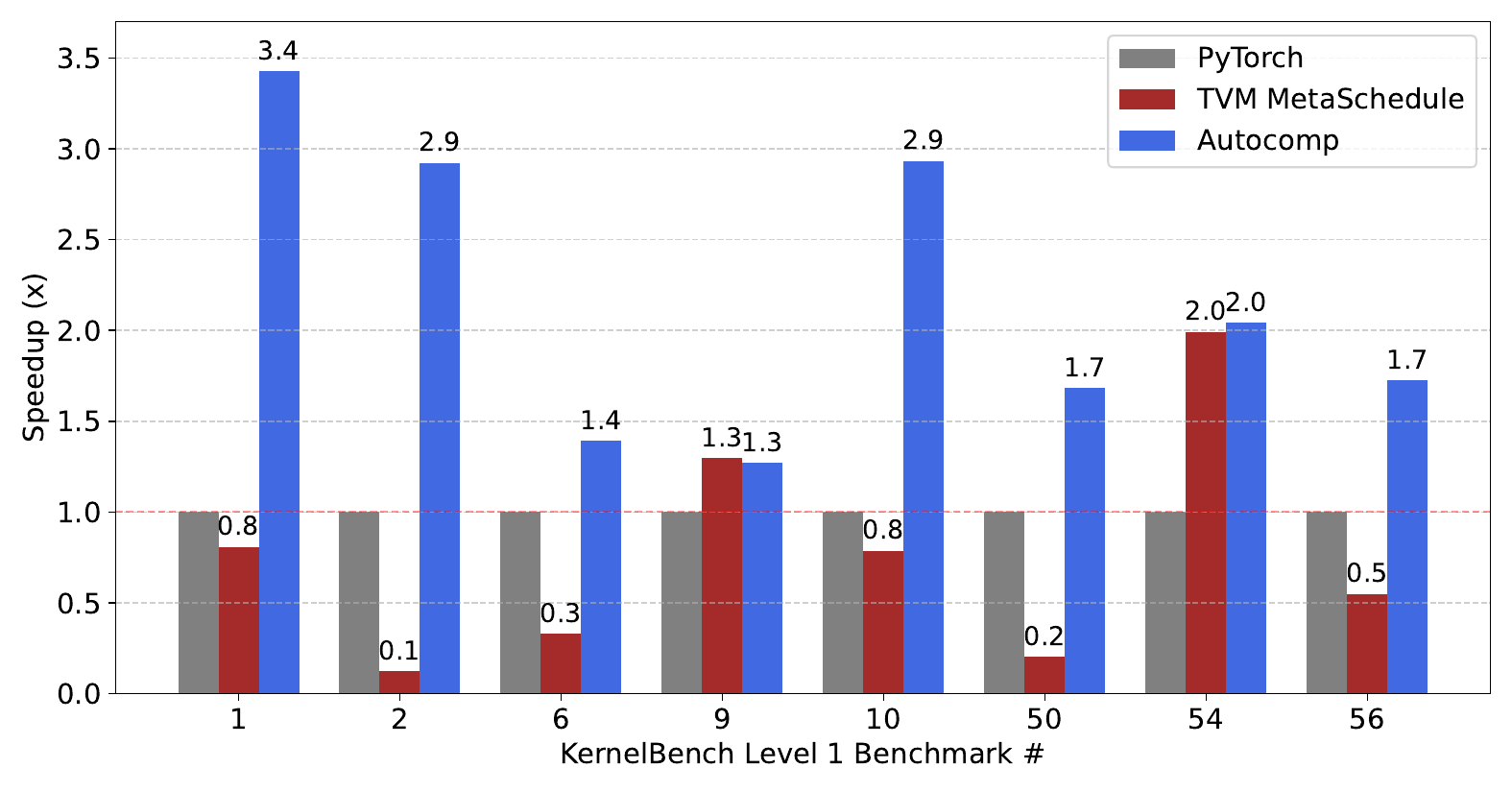}
    \vspace{-2em}
    \caption{Speedup for \textbf{NVIDIA L40S GPU} benchmarks, relative to PyTorch.}
    \label{fig:gpu-kernelbench}
    \vspace{-1.5em}
\end{figure}

As shown in~\cref{fig:gpu-kernelbench}, \sys{} generates code that is faster than PyTorch on every benchmark, speeding up the benchmarks by a geomean of 2.05$\times$, whereas TVM only outperforms PyTorch on two benchmarks, and on average is 0.54$\times$ as fast as PyTorch. TVM narrowly outperforms \sys{} on only benchmark 9, a GEMM with a small reduction dimension. One key to \sys{}'s performance on these benchmarks is its ability to explore different methods of enabling Tensor Core utilization, such as through PyTorch APIs and CUDA libraries such as cuBLASLt, and identify the best-performing implementation that passes KernelBench's accuracy check. On the other hand, since KernelBench focuses on 32-bit floating point operators, TVM MetaSchedule is unable to generate Tensor Core code without manual intervention.

These experiments show that not only is \sys{} more portable across hardware platforms and easier to use than TVM's auto-scheduler (since it can directly optimize PyTorch code), it is also significantly more performant for tensor operations from KernelBench's level 1.

\section{Improving Sample Efficiency With Schedule Reuse}
\label{sec:reuse-eval}
% \vspace{-0.5em}

In this experiment, we illustrate that optimization schedules generated by \sys{} for one tensor operation can be effectively reused to accelerate the optimization of similar tensor operations (as described in \cref{sec:reuse-method}). For this experiment, we generate additional benchmarks structurally similar (either same aspect ratio or two shared dimensions) to those in~\cref{sec:gemm}. For each new benchmark, in the plan reuse phase, we search with beam size $B=2$, $N=2$ plans, and $K=2$ code candidates. Once the reuse sequence is exhausted, we switch to a full \sys{} search with beam size $B=6$, $N=6$ plans, and $K=2$ code candidates.

We consider GEMMs from the following categories:
\begin{enumerate*}[label=\arabic*)]
    \item square,
    \item column-dominant and,
    \item row-dominant,
\end{enumerate*}
with dimensions specified in \cref{tab:gemm-reuse}. The ``base'' benchmark is optimized using full search, i.e., without any schedule reuse, and its \sys{} generated schedule is then reused to optimize the other two benchmarks in its category.

\begin{table}[t]
\centering
      \scriptsize
    \begin{tabular}{l l l}
      \toprule
      \textbf{Category} & \textbf{Base Benchmark} & \textbf{Reuse Targets} \\
      \midrule
      Square &
        \texttt{1024×1024×1024} &
        \texttt{512×512×512}, \texttt{256×256×256} \\
      Column-dominant &
        \texttt{12544×256x64} &
        \texttt{6272×256x64}, \texttt{12544×128x64} \\
      Row-dominant &
        \texttt{128×1024×1024} &
        \texttt{64×1024×1024}, \texttt{32×1024×1024} \\
      \bottomrule
    \end{tabular}
    \captionof{table}{GEMM shapes used for schedule reuse experiments.}
    \label{tab:gemm-reuse}
    \vspace{-2em}
\end{table}

\begin{figure}[t]
    \centering
    \includegraphics[trim={1cm 0 0 0},scale=0.29]{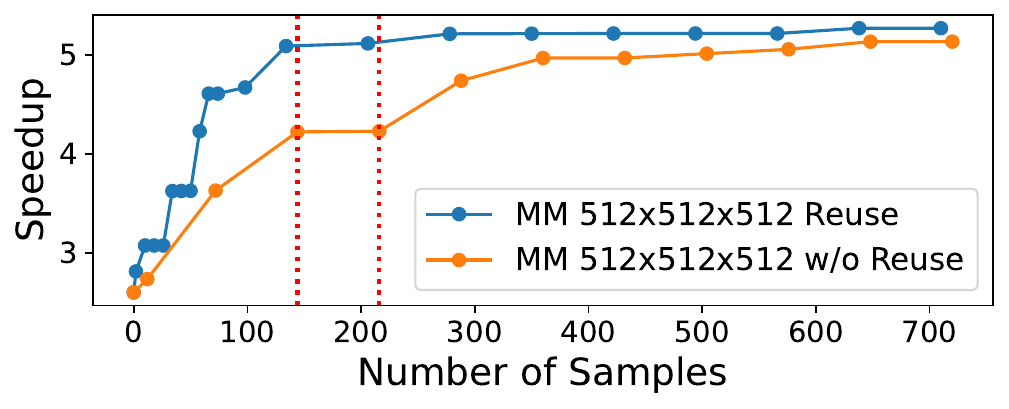}
    % \captionof{figure}{\texttt{512x512x512} schedule reuse.}
    % \label{fig:gemm-reuse-1}
    % \centering
    % \includegraphics[trim={1cm 0 0 0},scale=0.28]{figs/plan_reuse_gemm4.pdf}\vspace{-0.5em}
    % \captionof{figure}{\texttt{12544x128x64} schedule reuse.}
    % \label{fig:gemm-reuse-2}
    \vspace{-0.7em}
    \captionof{figure}{Example optimization trace for one of six reuse targets from \cref{tab:gemm-reuse}. With the same sample budget, \textcolor{RoyalBlue}{\sys{} with reuse} consistently delivers improved performance over \textcolor{orange}{\sys{} without reuse}.}\label{fig:gemm-reuse}
    \vspace{-0.7em}
\end{figure}

We include the full optimization traces 
% \alvin{you mean the schedules?} 
for \sys{} with and without reuse for one of these benchmarks: \texttt{512×512×512} (square GEMM), reused from a \texttt{1024x1024x1024} GEMM. As shown in \Cref{fig:gemm-reuse}, with reuse, \sys{} consistently achieves better speedups with the same number of samples.
% shows that reusing schedules can help reduce \sys{}'s sample complexity by achieving better speed-ups with fewer samples.
% \alvin{what do the vertical lines in figs 11 and 12 mean?} \sahil{iso sample performance}

For the remaining benchmarks, we perform an iso-samples comparison: we fix the total number of LLM calls and compare reuse to full search. In \cref{fig:gemm-reuse}, the \textcolor{red}{red vertical lines} compare the iso-sample performance for our two examples. Across all reuse targets in \cref{tab:gemm-reuse}, with a budget of 100 samples ($\approx$10\% of the calls used by full search), \sys{} achieves geomean speedups over Gemmini's software library of $4.6\times$ with reuse, compared to $3.7\times$ without reuse. With a 200 call budget, we observe $5.0\times$ speedup with reuse, compared to $4.2\times$ without reuse. This demonstrates \sys{} schedules are generalizable and can help reduce search cost when optimizing similar benchmarks.

\section{Conclusion}
\label{sec:conc}
% \vspace{-0.2em}
% \sahil{add limitation}

In this paper, we demonstrate how to construct an LLM-based flow to automatically optimize low-resource accelerator code at superhuman levels. 
Across three distinct hardware platforms, \sys{} generates code that significantly outperforms baselines on real-world benchmarks, while requiring far less manual effort than implementing a traditional compiler or hand-optimizing code. This demonstrates the effectiveness of \sys{}'s approach and its potential to serve as a key component in the accelerator design process.

\bibliography{mlsys2025style/refs}

@string {ASPLOS = "Proceedings of the International Conference on Architectural
		Support for Programming Languages and Operation Systems (ASPLOS)"}

@string {CGO = "International Symposium on Code Generation and Optimization (CGO)"}

@string {ISCA = "Proceedings of the International Symposium on Computer Architecture (ISCA)"}

@string {MICRO = "Proceedings of the International Symposium on Microarchitecture (MICRO)"}

@string {HPCA = "Proceedings of the International Symposium on High-Performance Computer Architecture (HPCA)"}

@string {OSDI = "USENIX Symposium on Operating Systems Design and Implementation (OSDI)"}

@string {ISPASS = "Proceedings of the International Symposium on Performance
  Analysis of Systems and Software (ISPASS)"}

@string {PLDI = "Proceedings of the Conference on Programming Language Design and Implementation (PLDI)"}

@string {DAC = "Design Automation Conference (DAC)"}

@string {NeurIPS = "Proceedings of the Conference on Neural Information Processing Systems (NeurIPS)"}

@string {CVPR = "Proceedings of the Conference on Computer Vision and Pattern Recognition (CVPR)"}

@string {ICLR = "Proceedings of the International Conference on Learning Representations (ICLR)"}

@string {ICML = "Proceedings of the International Conference on Machine Learning
  (ICML)"}

@misc{mills_2024, title={GPU MODE Lecture 8: CUDA Performance Checklist – Christian Mills}, url={https://christianjmills.com/posts/cuda-mode-notes/lecture-008/}, journal={Christian Mills}, author={Mills, Christian}, year={2024}, month={Sep} }

@misc{lange2025robustagenticcudakernel,
      title={Towards Robust Agentic CUDA Kernel Benchmarking, Verification, and Optimization}, 
      author={Robert Tjarko Lange and Qi Sun and Aaditya Prasad and Maxence Faldor and Yujin Tang and David Ha},
      year={2025},
      eprint={2509.14279},
      archivePrefix={arXiv},
      primaryClass={cs.SE},
      url={https://arxiv.org/abs/2509.14279}, 
}

@misc{li2025cudal1improvingcudaoptimization,
      title={CUDA-L1: Improving CUDA Optimization via Contrastive Reinforcement Learning}, 
      author={Xiaoya Li and Xiaofei Sun and Albert Wang and Jiwei Li and Chris Shum},
      year={2025},
      eprint={2507.14111},
      archivePrefix={arXiv},
      primaryClass={cs.AI},
      url={https://arxiv.org/abs/2507.14111}, 
}

@misc{lin2025ecollmdrivenefficientcode,
      title={ECO: An LLM-Driven Efficient Code Optimizer for Warehouse Scale Computers}, 
      author={Hannah Lin and Martin Maas and Maximilian Roquemore and Arman Hasanzadeh and Fred Lewis and Yusuf Simonson and Tzu-Wei Yang and Amir Yazdanbakhsh and Deniz Altinbüken and Florin Papa and Maggie Nolan Edmonds and Aditya Patil and Don Schwarz and Satish Chandra and Chris Kennelly and Milad Hashemi and Parthasarathy Ranganathan},
      year={2025},
      eprint={2503.15669},
      archivePrefix={arXiv},
      primaryClass={cs.SE},
      url={https://arxiv.org/abs/2503.15669}, 
}

@misc{baronio2025kevinmultiturnrlgenerating,
      title={Kevin: Multi-Turn RL for Generating CUDA Kernels}, 
      author={Carlo Baronio and Pietro Marsella and Ben Pan and Simon Guo and Silas Alberti},
      year={2025},
      eprint={2507.11948},
      archivePrefix={arXiv},
      primaryClass={cs.LG},
      url={https://arxiv.org/abs/2507.11948}, 
}

@article{novikov2025,
  title={AlphaEvolve: A coding agent for scientific and algorithmic discovery},
  author={Novikov, Alexander and V{\~u}, Ng{\^a}n and Eisenberger, Marvin and Dupont, Emilien and Huang, Po-Sen and Wagner, Adam Zsolt and Shirobokov, Sergey and Kozlovskii, Borislav and Ruiz, Francisco JR and Mehrabian, Abbas and others},
  journal={arXiv preprint arXiv:2506.13131},
  year={2025}
}

@misc{hong2025hdl2vcodetranslationdataset,
      title={hdl2v: A Code Translation Dataset for Enhanced LLM Verilog Generation}, 
      author={Charles Hong and Brendan Roberts and Huijae An and Alex Um and Advay Ratan and Yakun Sophia Shao},
      year={2025},
      eprint={2506.04544},
      archivePrefix={arXiv},
      primaryClass={cs.AR},
      url={https://arxiv.org/abs/2506.04544}, 
}

@misc{shao2022metaschedule,
      title={Tensor Program Optimization with Probabilistic Programs}, 
      author={Junru Shao and Xiyou Zhou and Siyuan Feng and Bohan Hou and Ruihang Lai and Hongyi Jin and Wuwei Lin and Masahiro Masuda and Cody Hao Yu and Tianqi Chen},
      year={2022},
      eprint={2205.13603},
      archivePrefix={arXiv},
      primaryClass={cs.LG},
      url={https://arxiv.org/abs/2205.13603}, 
}

@inproceedings{nickolls2008cuda,
author = {Nickolls, John and Buck, Ian and Garland, Michael and Skadron, Kevin},
title = {Scalable parallel programming with CUDA},
year = {2008},
isbn = {9781450378451},
publisher = {Association for Computing Machinery},
address = {New York, NY, USA},
url = {https://doi.org/10.1145/1401132.1401152},
doi = {10.1145/1401132.1401152},
booktitle = {ACM SIGGRAPH 2008 Classes},
articleno = {16},
numpages = {14},
location = {Los Angeles, California},
series = {SIGGRAPH '08}
}

@misc{ouyang_liang_mirhoseini_2025, 
title={Surprisingly Fast AI-Generated Kernels We Didn’t Mean to Publish (Yet)}, 
url={https://scalingintelligence.stanford.edu/blogs/fastkernels/}, 
journal={Scaling Intelligence Lab at Stanford University}, 
author={Ouyang, Anne and Liang, Percy and Mirhoseini, Azalia}, year={2025}, month={May} }

@misc{trainium_xla, title={PyTorch NeuronX Tracing API for Inference — AWS Neuron Documentation}, url={https://awsdocs-neuron.readthedocs-hosted.com/en/latest/frameworks/torch/torch-neuronx/api-reference-guide/inference/api-torch-neuronx-trace.html#torch-neuronx-trace-api}, journal={Readthedocs-hosted.com}, author={AWS}, year={2025} }

@misc{trainium, title={AWS Trainium}, url={https://aws.amazon.com/ai/machine-learning/trainium/}, journal={Amazon Web Services, Inc.}, author={AWS}, year={2025} }

@misc{trainium_released, title={Amazon EC2 Trn1 Instances for High-Performance Model Training are Now Available | Amazon Web Services}, url={https://aws.amazon.com/blogs/aws/amazon-ec2-trn1-instances-for-high-performance-model-training-are-now-available/}, journal={Amazon Web Services}, author={AWS}, year={2022}, month={Oct} }

@misc{trainium_nki_released, title={AWS Neuron introduces Neuron Kernel Interface (NKI), NxD Training, and JAX support for training - AWS}, url={https://aws.amazon.com/about-aws/whats-new/2024/09/aws-neuron-nki-nxd-training-jax/}, journal={Amazon Web Services, Inc.}, author={AWS}, year={2024} }

@inproceedings{tpu-isca2016,
	title = {{In-Datacenter Performance Analysis of a Tensor Processing Unit}},
	author  = {Norman P. Jouppi and Cliff Young and Nishant Patil and David Patterson
and Gaurav Agrawal and Raminder Bajwa and Sarah Bates and Suresh
Bhatia and Nan Boden and Al Borchers and Rick Boyle and Pierre-luc
Cantin and Clifford Chao and Chris Clark and Jeremy Coriell and Mike
Daley and Matt Dau and Jeffrey Dean and Ben Gelb and Tara Vazir
Ghaemmaghami and Rajendra Gottipati and William Gulland and Robert
Hagmann and C. Richard Ho and Doug Hogberg and John Hu and Robert
Hundt and Dan Hurt and Julian Ibarz and Aaron Jaffey and Alek Jaworski
and Alexander Kaplan and Harshit Khaitan and Daniel Killebrew and Andy
Koch and Naveen Kumar and Steve Lacy and James Laudon and James Law
and Diemthu Le and Chris Leary and Zhuyuan Liu and Kyle Lucke and Alan
Lundin and Gordon MacKean and Adriana Maggiore and Maire Mahony and
Kieran Miller and Rahul Nagarajan and Ravi Narayanaswami and Ray Ni
and Kathy Nix and Thomas Norrie and Mark Omernick and Narayana
Penukonda and Andy Phelps and Jonathan Ross and Matt Ross and Amir
Salek and Emad Samadiani and Chris Severn and Gregory Sizikov and
Matthew Snelham and Jed Souter and Dan Steinberg and Andy Swing and
Mercedes Tan and Gregory Thorson and Bo Tian and Horia Toma and Erick
Tuttle and Vijay Vasudevan and Richard Walter and Walter Wang and Eric
Wilcox and Doe Hyun Yoon},
	booktitle=ISCA,
	year={2017}
}

@misc{chatgpt,
  title = {Introducing ChatGPT},
  author = {{OpenAI}},
  url = {https://openai.com/index/chatgpt/},
  year = {2022},
}

@inproceedings{transformer,
 author = {Vaswani, Ashish and Shazeer, Noam and Parmar, Niki and Uszkoreit, Jakob and Jones, Llion and Gomez, Aidan N and Kaiser, \L ukasz and Polosukhin, Illia},
 booktitle = NIPS,
 title = {Attention is All you Need},
 year = {2017}
}

@inproceedings{alexnet,
	title={{Imagenet Classification with Deep Convolutional Neural Networks}},
	author={Krizhevsky, Alex and Sutskever, Ilya and Hinton, Geoffrey E.},
	booktitle=NIPS,
	year={2012}
}

@inproceedings{resnet,
	title={{Deep Residual Learning for Image Recognition}},
	author={He, Kaiming and Zhang, Xiangyu and Ren, Shaoqing and Sun, Jian},
	booktitle=CVPR,
	year={2016}
}

@misc{reuters2025openaichipcost, author={Anna Tong and Max A. Cherney and Krystal Hu}, title={Exclusive: OpenAI set to finalize first custom chip design this year}, url={https://www.reuters.com/technology/openai-set-finalize-first-custom-chip-design-this-year-2025-02-10/}, journal={Reuters}, publisher={Reuters}, year={2025}, month={Feb}}

@inproceedings{
  dosa,
  title={DOSA: Differentiable Model-Based One-Loop Search for DNN Accelerators},
  author={Charles Hong and Qijing Huang and Grace Dinh and Mahesh Subedar and Yakun Sophia Shao},
  booktitle={IEEE/ACM International Symposium on Microarchitecture (MICRO)},
  year={2023},
}

@inproceedings{gemmini,
author = {Genc, Hasan and Kim, Seah and Amid, Alon and Haj-Ali, Ameer and Iyer, Vighnesh and Prakash, Pranav and Zhao, Jerry and Grubb, Daniel and Liew, Harrison and Mao, Howard and Ou, Albert and Schmidt, Colin and Steffl, Samuel and Wright, John and Stoica, Ion and Ragan-Kelley, Jonathan and Asanovic, Krste and Nikolic, Borivoje and Shao, Yakun Sophia},
title = {Gemmini: Enabling Systematic Deep-Learning Architecture Evaluation via Full-Stack Integration},
year = {2021},
booktitle = {2021 58th ACM/IEEE Design Automation Conference (DAC)},
}

@article{lauterbach2021cerebras,
  title={The path to successful wafer-scale integration: the Cerebras story},
  author={Lauterbach, Gary},
  journal={IEEE Micro},
  volume={41},
  number={6},
  pages={52--57},
  year={2021},
  publisher={IEEE}
}

@misc{ane, 
title={Deploying Transformers on the Apple Neural Engine},
author={Atila Orhon and Aseem Wadhwa and Youchang Kim and Francesco Rossi and Vignesh Jagadeesh},
url={https://machinelearning.apple.com/research/neural-engine-transformers}, journal={Apple Machine Learning Research}, year={2022}, month={Jun}}

@misc{nvdla, author={NVIDIA}, url={https://nvdla.org/}, journal={NVIDIA Deep Learning Accelerator}, title={NVDLA}, year={2018}}

@ONLINE{intelamx,
  author={{Intel}},
  publisher={Intel Corporation},
  title = {Intel® Advanced Matrix Extensions Overview},
  url = {https://www.intel.com/content/www/us/en/products/docs/accelerator-engines/advanced-matrix-extensions/overview.html}
}

@ONLINE{armsme,
  author={Arm},
  publisher={Arm Limited},
  title = {The Scalable Matrix Extension (SME), for Armv9-A},
  url = {https://developer.arm.com/documentation/ddi0616/latest/},
  year = {2024},
}

@misc{xla,
  author={OpenXLA},
  publisher = {OpenXLA},
  title = {XLA Developer Guide},
  url = {https://openxla.org/xla},
  year = {2024},
}

@inproceedings{chen2018tvm,
  title="{TVM: An Automated End-to-end Optimizing Compiler for Deep Learning}",
  author={Chen, Tianqi and Moreau, Thierry and Jiang, Ziheng and Zheng, Lianmin and Yan, Eddie and Shen, Haichen and Cowan, Meghan and Wang, Leyuan and Hu, Yuwei and Ceze, Luis and  Guestrin, Carlos and Krishnamurthy, Arvind},
  booktitle=OSDI,
  year={2018}
}

@article{chen2021humaneval,
  title={Evaluating large language models trained on code},
  author={Chen, Mark and Tworek, Jerry and Jun, Heewoo and Yuan, Qiming and Pinto, Henrique Ponde de Oliveira and Kaplan, Jared and Edwards, Harri and Burda, Yuri and Joseph, Nicholas and Brockman, Greg and others},
  journal={arXiv preprint arXiv:2107.03374},
  year={2021}
}

@article{nijkamp2022codegen,
  title={Codegen: An open large language model for code with multi-turn program synthesis},
  author={Nijkamp, Erik and Pang, Bo and Hayashi, Hiroaki and Tu, Lifu and Wang, Huan and Zhou, Yingbo and Savarese, Silvio and Xiong, Caiming},
  journal={arXiv preprint arXiv:2203.13474},
  year={2022}
}

@inproceedings{nguyen2024tinympc,
      title={TinyMPC: Model-Predictive Control on Resource-Constrained Microcontrollers}, 
      author={Khai Nguyen and Sam Schoedel and Anoushka Alavilli and Brian Plancher and Zachary Manchester},
      booktitle={IEEE International Conference on Robotics and Automation (ICRA)},
      year = {2024}
}

@article{cummins2023compiler,
  title={Large language models for compiler optimization},
  author={Cummins, Chris and Seeker, Volker and Grubisic, Dejan and Elhoushi, Mostafa and Liang, Youwei and Roziere, Baptiste and Gehring, Jonas and Gloeckle, Fabian and Hazelwood, Kim and Synnaeve, Gabriel and others},
  journal={arXiv preprint arXiv:2309.07062},
  year={2023}
}

@inproceedings{vaesa,
  title={Learning A Continuous and Reconstructible Latent Space for Hardware Accelerator Design},
  author={Huang, Qijing and Hong, Charles and Wawrzynek, John and Subedar, Mahesh and Shao, Yakun Sophia},
  booktitle=ISPASS,
  year={2022},
}

@inproceedings{parashar2019timeloop,
  title={Timeloop: A Systematic Approach to DNN Accelerator Evaluation},
  author={Parashar, Angshuman and Raina, Priyanka and Shao, Yakun Sophia and Chen, Yu-Hsin and Ying, Victor A and Mukkara, Anurag and Venkatesan, Rangharajan and Khailany, Brucek and Keckler, Stephen W and Emer, Joel},
  booktitle=ISPASS,
  year={2019},
}

@inproceedings{jrk2013halide,
author = {Ragan-Kelley, Jonathan and Barnes, Connelly and Adams, Andrew and Paris, Sylvain and Durand, Fr\'{e}do and Amarasinghe, Saman},
title = {Halide: a language and compiler for optimizing parallelism, locality, and recomputation in image processing pipelines},
year = {2013},
isbn = {9781450320146},
publisher = {Association for Computing Machinery},
address = {New York, NY, USA},
url = {https://doi.org/10.1145/2491956.2462176},
doi = {10.1145/2491956.2462176},
booktitle = {Proceedings of the 34th ACM SIGPLAN Conference on Programming Language Design and Implementation},
pages = {519–530},
numpages = {12},
location = {Seattle, Washington, USA},
series = {PLDI '13}
}

@inproceedings{ansor,
  title={Ansor: Generating $\{$High-Performance$\}$ tensor programs for deep learning},
  author={Zheng, Lianmin and Jia, Chengfan and Sun, Minmin and Wu, Zhao and Yu, Cody Hao and Haj-Ali, Ameer and Wang, Yida and Yang, Jun and Zhuo, Danyang and Sen, Koushik and others},
  booktitle={14th USENIX symposium on operating systems design and implementation (OSDI 20)},
  pages={863--879},
  year={2020}
}

@inproceedings{exo,
author = {Ikarashi, Yuka and Bernstein, Gilbert Louis and Reinking, Alex and Genc, Hasan and Ragan-Kelley, Jonathan},
title = {Exocompilation for productive programming of hardware accelerators},
year = {2022},
isbn = {9781450392655},
publisher = {Association for Computing Machinery},
address = {New York, NY, USA},
url = {https://doi.org/10.1145/3519939.3523446},
doi = {10.1145/3519939.3523446},
booktitle = {Proceedings of the 43rd ACM SIGPLAN International Conference on Programming Language Design and Implementation},
pages = {703–718},
numpages = {16},
location = {San Diego, CA, USA},
series = {PLDI 2022}
}

@INPROCEEDINGS{damani2024warpdrive,
  author={Sana Damani and Siva Kumar Sastry Hari and Mark Stephenson and Christos Kozyrakis},
  booktitle={Machine Learning for Systems Workshop at NeurIPS 2024}, 
  title={WarpDrive: An Agentic Workflow for Ninja GPU Transformations}, 
  year={2024},
}

@INPROCEEDINGS{hong2024llmaidedcompilation,
  author={Hong, Charles and Bhatia, Sahil and Haan, Altan and Dong, Shengjun Kris and Nikiforov, Dima and Cheung, Alvin and Shao, Yakun Sophia},
  booktitle={2024 IEEE LLM Aided Design Workshop (LAD)}, 
  title={LLM-Aided Compilation for Tensor Accelerators}, 
  year={2024},
  volume={},
  number={},
  pages={1-14},
  doi={10.1109/LAD62341.2024.10691748}}

@misc{ouyang2025kernelbench,
      title={KernelBench: Can LLMs Write Efficient GPU Kernels?}, 
      author={Anne Ouyang and Simon Guo and Simran Arora and Alex L. Zhang and William Hu and Christopher Ré and Azalia Mirhoseini},
      year={2025},
      eprint={2502.10517},
      archivePrefix={arXiv},
      primaryClass={cs.LG},
      url={https://arxiv.org/abs/2502.10517}, 
}

@inproceedings{shypula2024learning,
  title={Learning Performance-Improving Code Edits},
  author={Shypula, Alexander and Madaan, Aman and Zeng, Yimeng and Alon, Uri and Gardner, Jacob R and Yang, Yiming and Hashemi, Milad and Neubig, Graham and Ranganathan, Parthasarathy and Bastani, Osbert and others},
  booktitle={ICLR},
  year={2024}
}

@inproceedings{bhatia2024llmlift,
 author = {Bhatia, Sahil and Qiu, Jie and Hasabnis, Niranjan and Seshia, Sanjit A. and Cheung, Alvin},
 booktitle = {Advances in Neural Information Processing Systems},
 editor = {A. Globerson and L. Mackey and D. Belgrave and A. Fan and U. Paquet and J. Tomczak and C. Zhang},
 pages = {41394--41424},
 publisher = {Curran Associates, Inc.},
 title = {Verified Code Transpilation with LLMs},
 url = {https://proceedings.neurips.cc/paper_files/paper/2024/file/48bb60a0c0aebb4142bf314bd1a5c6a0-Paper-Conference.pdf},
 volume = {37},
 year = {2024}
}

@article{romeraparedes2024funsearch,
  author = {Romera-Paredes, Bernardino and Barekatain, Mohammadamin and Novikov, Alexander and Balog, Matej and Kumar, M. Pawan and Dupont, Emilien and Ruiz, Francisco J. R. and Ellenberg, Jordan S. and Wang, Pengming and Fawzi, Omar and Kohli, Pushmeet and Fawzi, Alhussein},
  ee = {https://doi.org/10.1038/s41586-023-06924-6},
  journal = {Nat.},
  month = {January},
  number = 7995,
  pages = {468-475},
  title = {Mathematical discoveries from program search with large language models.},
  url = {http://dblp.uni-trier.de/db/journals/nature/nature625.html#RomeraParedesBNBKDREWFKF24},
  volume = 625,
  year = 2024
}

@misc{wang2025symrtlo,
      title={SymRTLO: Enhancing RTL Code Optimization with LLMs and Neuron-Inspired Symbolic Reasoning}, 
      author={Yiting Wang and Wanghao Ye and Ping Guo and Yexiao He and Ziyao Wang and Yexiao He and Bowei Tian and Shwai He and Guoheng Sun and Zheyu Shen and Sihan Chen and Ankur Srivastava and Qingfu Zhang and Gang Qu and Ang Li},
      year={2025},
      eprint={2504.10369},
      archivePrefix={arXiv},
      primaryClass={cs.AR},
      url={https://arxiv.org/abs/2504.10369}, 
}

@article{blackford2002blas,
  title={An updated set of basic linear algebra subprograms (BLAS)},
  author={Blackford, L Susan and Petitet, Antoine and Pozo, Roldan and Remington, Karin and Whaley, R Clint and Demmel, James and Dongarra, Jack and Duff, Iain and Hammarling, Sven and Henry, Greg and others},
  journal={ACM Transactions on Mathematical Software},
  volume={28},
  number={2},
  pages={135--151},
  year={2002}
}

@INPROCEEDINGS{firesim,
  author={Karandikar, Sagar and Mao, Howard and Kim, Donggyu and Biancolin, David and Amid, Alon and Lee, Dayeol and Pemberton, Nathan and Amaro, Emmanuel and Schmidt, Colin and Chopra, Aditya and Huang, Qijing and Kovacs, Kyle and Nikolic, Borivoje and Katz, Randy and Bachrach, Jonathan and Asanovic, Krste},
  booktitle={2018 ACM/IEEE 45th Annual International Symposium on Computer Architecture (ISCA)}, 
  title={FireSim: FPGA-Accelerated Cycle-Exact Scale-Out System Simulation in the Public Cloud}, 
  year={2018},
  volume={},
  number={},
  pages={29-42},
  doi={10.1109/ISCA.2018.00014}}

@misc{sakhuja2024polaris,
      title={Polaris: Multi-Fidelity Design Space Exploration of Deep Learning Accelerators}, 
      author={Chirag Sakhuja and Charles Hong and Calvin Lin},
      year={2024},
      eprint={2412.15548},
      archivePrefix={arXiv},
      primaryClass={cs.AR},
      url={https://arxiv.org/abs/2412.15548}, 
}

@inproceedings{zhang2022fast,
author = {Zhang, Dan and Huda, Safeen and Songhori, Ebrahim and Prabhu, Kartik and Le, Quoc and Goldie, Anna and Mirhoseini, Azalia},
title = {A full-stack search technique for domain optimized deep learning accelerators},
year = {2022},
isbn = {9781450392051},
publisher = {Association for Computing Machinery},
address = {New York, NY, USA},
url = {https://doi.org/10.1145/3503222.3507767},
doi = {10.1145/3503222.3507767},
booktitle = {Proceedings of the 27th ACM International Conference on Architectural Support for Programming Languages and Operating Systems},
pages = {27–42},
numpages = {16},
location = {Lausanne, Switzerland},
series = {ASPLOS '22}
}

@inproceedings{xiao2021hasco,
author = {Xiao, Qingcheng and Zheng, Size and Wu, Bingzhe and Xu, Pengcheng and Qian, Xuehai and Liang, Yun},
title = {HASCO: towards agile <u>ha</u>rdware and <u>s</u>oftware <u>co</u>-design for tensor computation},
year = {2021},
isbn = {9781450390866},
publisher = {IEEE Press},
url = {https://doi.org/10.1109/ISCA52012.2021.00086},
doi = {10.1109/ISCA52012.2021.00086},
booktitle = {Proceedings of the 48th Annual International Symposium on Computer Architecture},
pages = {1055–1068},
numpages = {14},
location = {Virtual Event, Spain},
series = {ISCA '21}
}

@inproceedings{sakhuja2023spotlight,
  title={Leveraging Domain Information for the Efficient
Automated Design of Deep Learning Accelerators},
  author={Sakhuja, Chirag and Shi, Zhan and Lin, Calvin},
  booktitle={International Symposium on High-Performance Computer Architectural (HPCA)},
  year={2023},
  organization={IEEE}
}

@INPROCEEDINGS{huang2021cosa,
  author={Huang, Qijing and Kang, Minwoo and Dinh, Grace  and Norell, Thomas and Kalaiah, Aravind and Demmel, James and Wawrzynek, John and Shao, Yakun Sophia},
  booktitle=ISCA, 
  title={{CoSA: Scheduling by Constrained Optimization for Spatial Accelerators}}, 
  year={2021},
  }

@article{srivastava2014dropout,
  author  = {Nitish Srivastava and Geoffrey Hinton and Alex Krizhevsky and Ilya Sutskever and Ruslan Salakhutdinov},
  title   = {Dropout: A Simple Way to Prevent Neural Networks from Overfitting},
  journal = {Journal of Machine Learning Research},
  year    = {2014},
  volume  = {15},
  number  = {56},
  pages   = {1929--1958},
  url     = {http://jmlr.org/papers/v15/srivastava14a.html}
}

@inproceedings{zheng2021tenset,
 author = {Zheng, Lianmin and Liu, Ruochen and Shao, Junru and Chen, Tianqi and Gonzalez, Joseph and Stoica, Ion and Haj-Ali, Ameer},
 booktitle = {Proceedings of the Neural Information Processing Systems Track on Datasets and Benchmarks},
 editor = {J. Vanschoren and S. Yeung},
 pages = {},
 title = {TenSet: A Large-scale Program Performance Dataset for Learned Tensor Compilers},
 url = {https://datasets-benchmarks-proceedings.neurips.cc/paper_files/paper/2021/file/a684eceee76fc522773286a895bc8436-Paper-round1.pdf},
 volume = {1},
 year = {2021}
}

@inproceedings{cummins2022compilergym,
author = {Cummins, Chris and Wasti, Bram and Guo, Jiadong and Cui, Brandon and Ansel, Jason and Gomez, Sahir and Jain, Somya and Liu, Jia and Teytaud, Olivier and Steiner, Benoit and Tian, Yuandong and Leather, Hugh},
title = {CompilerGym: robust, performant compiler optimization environments for AI research},
year = {2022},
isbn = {9781665405843},
publisher = {IEEE Press},
url = {https://doi.org/10.1109/CGO53902.2022.9741258},
doi = {10.1109/CGO53902.2022.9741258},
booktitle = {Proceedings of the 20th IEEE/ACM International Symposium on Code Generation and Optimization},
pages = {92–105},
numpages = {14},
location = {Virtual Event, Republic of Korea},
series = {CGO '22}
}

@ARTICLE{kwon2020maestro,

  author={Kwon, Hyoukjun and Chatarasi, Prasanth and Sarkar, Vivek and Krishna, Tushar and Pellauer, Michael and Parashar, Angshuman},

  journal={IEEE Micro}, 

  title={MAESTRO: A Data-Centric Approach to Understand Reuse, Performance, and Hardware Cost of DNN Mappings}, 

  year={2020},

  volume={40},

  number={3},

  pages={20-29},

  doi={10.1109/MM.2020.2985963}}

@misc{ma2024eurekahumanlevelrewarddesign,
      title={Eureka: Human-Level Reward Design via Coding Large Language Models}, 
      author={Yecheng Jason Ma and William Liang and Guanzhi Wang and De-An Huang and Osbert Bastani and Dinesh Jayaraman and Yuke Zhu and Linxi Fan and Anima Anandkumar},
      year={2024},
      eprint={2310.12931},
      archivePrefix={arXiv},
      primaryClass={cs.RO},
      url={https://arxiv.org/abs/2310.12931}, 
}

@misc{lehman2022evolutionlargemodels,
      title={Evolution through Large Models}, 
      author={Joel Lehman and Jonathan Gordon and Shawn Jain and Kamal Ndousse and Cathy Yeh and Kenneth O. Stanley},
      year={2022},
      eprint={2206.08896},
      archivePrefix={arXiv},
      primaryClass={cs.NE},
      url={https://arxiv.org/abs/2206.08896}, 
}

@inproceedings{gao2023pal,
author = {Gao, Luyu and Madaan, Aman and Zhou, Shuyan and Alon, Uri and Liu, Pengfei and Yang, Yiming and Callan, Jamie and Neubig, Graham},
title = {PAL: program-aided language models},
year = {2023},
publisher = {JMLR.org},
booktitle = {Proceedings of the 40th International Conference on Machine Learning},
articleno = {435},
numpages = {36},
location = {Honolulu, Hawaii, USA},
series = {ICML'23}
}

@inproceedings{loopvectorizer,
author = {Taneja, Jubi and Laird, Avery and Yan, Cong and Musuvathi, Madan and Lahiri, Shuvendu K.},
title = {LLM-Vectorizer: LLM-Based Verified Loop Vectorizer},
year = {2025},
isbn = {9798400712753},
publisher = {Association for Computing Machinery},
address = {New York, NY, USA},
url = {https://doi.org/10.1145/3696443.3708929},
doi = {10.1145/3696443.3708929},
booktitle = {Proceedings of the 23rd ACM/IEEE International Symposium on Code Generation and Optimization},
pages = {137–149},
numpages = {13},
location = {Las Vegas, NV, USA},
series = {CGO '25}
}

@misc{dong2024designspaceexplorationembedded,
      title={Design Space Exploration of Embedded SoC Architectures for Real-Time Optimal Control}, 
      author={Kris Shengjun Dong and Dima Nikiforov and Widyadewi Soedarmadji and Minh Nguyen and Christopher Fletcher and Yakun Sophia Shao},
      year={2024},
      url={https://github.com/ucb-bar/Accelerated-TinyMPC/blob/main/Design-Space-Exploration-of-Embedded-SoC-Architectures-for-Real-Time-Optimal-Control.pdf}, 
}

@misc{peng2024perfcodegenimprovingperformancellm,
      title={PerfCodeGen: Improving Performance of LLM Generated Code with Execution Feedback}, 
      author={Yun Peng and Akhilesh Deepak Gotmare and Michael Lyu and Caiming Xiong and Silvio Savarese and Doyen Sahoo},
      year={2024},
      eprint={2412.03578},
      archivePrefix={arXiv},
      primaryClass={cs.SE},
      url={https://arxiv.org/abs/2412.03578}, 
}

@inproceedings{jiang2023llmblender,
    title = "{LLM}-Blender: Ensembling Large Language Models with Pairwise Ranking and Generative Fusion",
    author = "Jiang, Dongfu  and
      Ren, Xiang  and
      Lin, Bill Yuchen",
    editor = "Rogers, Anna  and
      Boyd-Graber, Jordan  and
      Okazaki, Naoaki",
    booktitle = "Proceedings of the 61st Annual Meeting of the Association for Computational Linguistics (Volume 1: Long Papers)",
    month = jul,
    year = "2023",
    address = "Toronto, Canada",
    publisher = "Association for Computational Linguistics",
    url = "https://aclanthology.org/2023.acl-long.792/",
    doi = "10.18653/v1/2023.acl-long.792",
    pages = "14165--14178"
}

@misc{deepseekr1,
      title={DeepSeek-R1: Incentivizing Reasoning Capability in LLMs via Reinforcement Learning}, 
      author={DeepSeek-AI and Daya Guo and Dejian Yang and Haowei Zhang and Junxiao Song and Ruoyu Zhang and Runxin Xu and Qihao Zhu and Shirong Ma and Peiyi Wang and Xiao Bi and Xiaokang Zhang and Xingkai Yu and Yu Wu and Z. F. Wu and Zhibin Gou and Zhihong Shao and Zhuoshu Li and Ziyi Gao and Aixin Liu and Bing Xue and Bingxuan Wang and Bochao Wu and Bei Feng and Chengda Lu and Chenggang Zhao and Chengqi Deng and Chenyu Zhang and Chong Ruan and Damai Dai and Deli Chen and Dongjie Ji and Erhang Li and Fangyun Lin and Fucong Dai and Fuli Luo and Guangbo Hao and Guanting Chen and Guowei Li and H. Zhang and Han Bao and Hanwei Xu and Haocheng Wang and Honghui Ding and Huajian Xin and Huazuo Gao and Hui Qu and Hui Li and Jianzhong Guo and Jiashi Li and Jiawei Wang and Jingchang Chen and Jingyang Yuan and Junjie Qiu and Junlong Li and J. L. Cai and Jiaqi Ni and Jian Liang and Jin Chen and Kai Dong and Kai Hu and Kaige Gao and Kang Guan and Kexin Huang and Kuai Yu and Lean Wang and Lecong Zhang and Liang Zhao and Litong Wang and Liyue Zhang and Lei Xu and Leyi Xia and Mingchuan Zhang and Minghua Zhang and Minghui Tang and Meng Li and Miaojun Wang and Mingming Li and Ning Tian and Panpan Huang and Peng Zhang and Qiancheng Wang and Qinyu Chen and Qiushi Du and Ruiqi Ge and Ruisong Zhang and Ruizhe Pan and Runji Wang and R. J. Chen and R. L. Jin and Ruyi Chen and Shanghao Lu and Shangyan Zhou and Shanhuang Chen and Shengfeng Ye and Shiyu Wang and Shuiping Yu and Shunfeng Zhou and Shuting Pan and S. S. Li and Shuang Zhou and Shaoqing Wu and Shengfeng Ye and Tao Yun and Tian Pei and Tianyu Sun and T. Wang and Wangding Zeng and Wanjia Zhao and Wen Liu and Wenfeng Liang and Wenjun Gao and Wenqin Yu and Wentao Zhang and W. L. Xiao and Wei An and Xiaodong Liu and Xiaohan Wang and Xiaokang Chen and Xiaotao Nie and Xin Cheng and Xin Liu and Xin Xie and Xingchao Liu and Xinyu Yang and Xinyuan Li and Xuecheng Su and Xuheng Lin and X. Q. Li and Xiangyue Jin and Xiaojin Shen and Xiaosha Chen and Xiaowen Sun and Xiaoxiang Wang and Xinnan Song and Xinyi Zhou and Xianzu Wang and Xinxia Shan and Y. K. Li and Y. Q. Wang and Y. X. Wei and Yang Zhang and Yanhong Xu and Yao Li and Yao Zhao and Yaofeng Sun and Yaohui Wang and Yi Yu and Yichao Zhang and Yifan Shi and Yiliang Xiong and Ying He and Yishi Piao and Yisong Wang and Yixuan Tan and Yiyang Ma and Yiyuan Liu and Yongqiang Guo and Yuan Ou and Yuduan Wang and Yue Gong and Yuheng Zou and Yujia He and Yunfan Xiong and Yuxiang Luo and Yuxiang You and Yuxuan Liu and Yuyang Zhou and Y. X. Zhu and Yanhong Xu and Yanping Huang and Yaohui Li and Yi Zheng and Yuchen Zhu and Yunxian Ma and Ying Tang and Yukun Zha and Yuting Yan and Z. Z. Ren and Zehui Ren and Zhangli Sha and Zhe Fu and Zhean Xu and Zhenda Xie and Zhengyan Zhang and Zhewen Hao and Zhicheng Ma and Zhigang Yan and Zhiyu Wu and Zihui Gu and Zijia Zhu and Zijun Liu and Zilin Li and Ziwei Xie and Ziyang Song and Zizheng Pan and Zhen Huang and Zhipeng Xu and Zhongyu Zhang and Zhen Zhang},
      year={2025},
      eprint={2501.12948},
      archivePrefix={arXiv},
      primaryClass={cs.CL},
      url={https://arxiv.org/abs/2501.12948}, 
}

@misc{wang2024planningnaturallanguageimproves,
      title={Planning In Natural Language Improves LLM Search For Code Generation}, 
      author={Evan Wang and Federico Cassano and Catherine Wu and Yunfeng Bai and Will Song and Vaskar Nath and Ziwen Han and Sean Hendryx and Summer Yue and Hugh Zhang},
      year={2024},
      eprint={2409.03733},
      archivePrefix={arXiv},
      primaryClass={cs.LG},
      url={https://arxiv.org/abs/2409.03733}, 
}

@article{metallmcompiler,
  title={Meta large language model compiler: Foundation models of compiler optimization},
  author={Cummins, Chris and Seeker, Volker and Grubisic, Dejan and Roziere, Baptiste and Gehring, Jonas and Synnaeve, Gabriel and Leather, Hugh},
  journal={arXiv preprint arXiv:2407.02524},
  year={2024}
}
\bibliographystyle{mlsys2025}

%%%%%%%%%%%%%%%%%%%%%%%%%%%%%%%%%%%%%%%%%%%%%%%%%%%%%%%%%%%%%%%%%%%%%%%%%%%%%%%
%%%%%%%%%%%%%%%%%%%%%%%%%%%%%%%%%%%%%%%%%%%%%%%%%%%%%%%%%%%%%%%%%%%%%%%%%%%%%%%
% SUPPLEMENTAL CONTENT AS APPENDIX AFTER REFERENCES
%%%%%%%%%%%%%%%%%%%%%%%%%%%%%%%%%%%%%%%%%%%%%%%%%%%%%%%%%%%%%%%%%%%%%%%%%%%%%%%
%%%%%%%%%%%%%%%%%%%%%%%%%%%%%%%%%%%%%%%%%%%%%%%%%%%%%%%%%%%%%%%%%%%%%%%%%%%%%%%
\clearpage
\appendix
\section{Benchmark Details}
\label{sec:experiments-appendix}

\subsection{Gemmini Experiments}
We have uploaded the Gemmini benchmarks to Google Drive\footnote{\url{https://drive.google.com/drive/folders/1eceVTv7noM6Rp63ap9PTL5t6qMYOlfcr?}}. Gemmini's software library is from the \texttt{gemmini-rocc-tests} repository\footnote{\url{https://github.com/ucb-bar/gemmini-rocc-tests/tree/1a1a1c6bd60df6d7cae3d87aac96c8f406cae084}}. Exo optimized code was generated using using the published schedules in Exo's GitHub repository\footnote{\url{https://github.com/exo-lang/exo/tree/4c65f576b296025b8dcd86b2c4e41769507969cf/apps/gemmini/src/exo}}.

\subsection{Trainium Experiments}

\begin{table}[h!]
\centering
\small
\renewcommand{\arraystretch}{1.3} % add row spacing
\begin{tabular}{@{}%
    >{\raggedright\arraybackslash}p{3cm}%
    >{\raggedright\arraybackslash}p{4.8cm}@{}}
\toprule
\textbf{Operator} & \textbf{Configuration} \\ \midrule
RMSNorm & $4096 \times 512$ \\
LayerNorm & $4096 \times 8192$ \\
GEMM & $4096 \times 8192 \times 8192$ \\
Mamba & $batch=1$, $seq\_len=2048$, $channel=256$, $state\_size=16$ \\
Self-Attention & $d_{head}=128$, $seq\_len=4096$ \\
Stable Diffusion Attention & $d_{head}=64$, $seq\_len=4096$ \\
\bottomrule
\end{tabular}
\caption{Tutorial Operators and Configurations}
\label{tab:trainium-tutorial-details}
\end{table}

\begin{table}[h!]
\small
\centering
\renewcommand{\arraystretch}{1.3}
\begin{tabular}{@{}%
    >{\raggedright\arraybackslash}p{3cm}%
    >{\raggedright\arraybackslash}p{4.8cm}@{}}
\toprule
\textbf{Operator} & \textbf{Configuration} \\ \midrule
Cumsum & $4096 \times 4096$ \\
Transpose & $512 \times 512 \times 512, (012) \rightarrow (021)$ \\
Max Pooling & $448 \times 448$, pool $3 \times 3$ \\ 
RoPE & $128 \times 4096$ \\
Depthwise Conv1D & $8,512,1,2048$ (NCHW) \\
Conv2D & batch size $16$, input $128\times128$, kernel $3\times3$, input channels $128$, output channels $512$ \\
Causal Self-Attention seq=2048 & $d_{head}=128$, $seq\_len=2048$ \\
Causal Self-Attention seq=16384 & $d_{head}=128$, $seq_Q=128$, $seq_{KV}=16384$ \\
Multi-head Causal Self-Attention & $n_{head}=8$, $d_{head}=128$, $seq\_len=2048$ \\
\bottomrule
\end{tabular}
\caption{Advanced Operators and Configurations}
\label{tab:trainium-advanced-details}
\end{table}

\cref{tab:trainium-tutorial-details,tab:trainium-advanced-details} detail the shapes and configurations used for our Trainium evaluation.
Reference code was directly copied from the \texttt{nki-samples} repository\footnote{\url{https://github.com/aws-neuron/nki-samples/tree/5cfeabca92a64f642a154ef835cbcede609016a3}}, with some implementations requiring small modifications to run in isolation on a \texttt{trn1.2xlarge} instance.
We attempted to use the same shapes as in the original \texttt{nki-samples} code, whenever examples were provided and the shapes were large enough to show meaningful performance improvements. For causal self-attention with seq=16384, we measure the latency of calculating self attention between a query tile of length $seq_Q=128$ with key/value arrays of length $seq_{KV}=16384$.

\subsection{NVIDIA L40S GPU Experiments}

\begin{table}[h!]
\centering
\small
\renewcommand{\arraystretch}{1.3} % add row spacing
\begin{tabular}{@{}%
    >{\raggedright\arraybackslash}p{1.8cm}%
    >{\raggedright\arraybackslash}p{6cm}@{}}
\toprule
\textbf{Benchmark \#} & \textbf{Configuration} \\ \midrule
1 & $4096 \times 4096 * 4096 \times 4096$ square GEMM \\
2 & $2048 \times 8192 * 8192 \times 4096$ GEMM \\
6 & $256 \times 524288 * 524288 \times 256$ large K GEMM \\
9 & $32768 \times 32 * 32 \times 32768$ small K GEMM \\
10 & $16 \times 1024 \times 2048 * 2048 \times 768$ 3DMM \\
50 & Conv2D weight $11 \times 11$, input $256 \times 3 \times 224 \times 224$, output $256 \times 96 \times 55 \times 55$ (stride 4, padding 2) \\
54 & Conv3D weight $3 \times 3 \times 3$, input $16 \times 3 \times 64 \times 64 \times 64$, output $16 \times 64 \times 62 \times 62 \times 26$ (stride 1, padding 0) \\
56 & Conv2D asymmetric weight $5 \times 7$, asymmetric input $8 \times 64 \times 512 \times 256$, output $8 \times 128 \times 508 \times 250$ (stride 1, padding 0) \\
\bottomrule
\end{tabular}
\caption{Descriptions of KernelBench Level 1 Operators Used}
\label{tab:kb-details}
\end{table}

\cref{tab:kb-details} lists the details of the KernelBench benchmarks selected for our GPU evaluation. We use shapes from KernelBench v0.1.
The exact PyTorch operators used can be found in the KernelBench repository\footnote{\url{https://github.com/ScalingIntelligence/KernelBench/tree/6b1cea28ddea79af6b295fb9a3e3413f393efe3d/KernelBench/level1}}.
\section{Ablation Studies}\label{sec:ablations}

\begin{table*}[!h]
    \centering
    \small
    \begin{tabular}{l l l}
        \toprule
        \textbf{Experiment} & \makecell[l]{\textbf{\texttt{12544x64x256} GEMM Speedup}} & \makecell[l]{\textbf{\texttt{4x3x14x256x256} Conv Speedup}} \\
        \midrule
        Baseline (Exo Unopt) & 1.67$\times$ & 0.91$\times$ \\
        No Accelerator ISA & 3.11$\times$ & 2.51$\times$ \\
        No Optimization Menu & 2.34$\times$ & 0.97$\times$ \\
        No Optimization Menu Dropout & 4.72$\times$ & 2.30$\times$ \\
        No LLM Ensemble (o3-mini only) & 4.67$\times$ & 2.08$\times$ \\
        % No Memory Util Feedback & 3.93$\times$ & 2.49$\times$ \\
        No Hardware Perf Feedback & 4.91$\times$ & 2.61$\times$ \\
        LLM Selection (DeepSeek-R1) & 4.89$\times$ &  2.25$\times$\\
        LLM Selection (Gemini 2.5 Flash) & 3.63$\times$ & 2.42$\times$ \\ 
        LLM Selection (Llama 4 Maverick) & 4.14$\times$ & 1.01$\times$ \\
        \textbf{\sys{}} & \textbf{5.53$\times$} & \textbf{2.72$\times$} \\
        Variance Checking & 5.2$\pm$0.33$\times$ & 2.61$\pm$0.11$\times$\\
        \bottomrule
    \end{tabular}
    \vspace{1em}\caption{Speedup relative to Gemmini's software library for each of the studies in this section. We include two representative benchmarks, one GEMM and one convolution, from our initial evaluation.}
    \label{tab:ablations}
\end{table*}

% \subsection{Optimization Parameters}
% We investigate the minimal number of samples (beam size $B$, plan samples $K$, and code samples $N$) needed to generate optimal code.

In this section, we ablate various features of \sys{} to investigate their effect on optimization performance. We focus on two specific benchmarks from \cref{sec:eval}\textemdash{}our \texttt{12544x64x256} GEMM and our \texttt{4x3x14x256x256} convolution on a 16x16 INT8 Gemmini\textemdash{}to isolate the effects of these ablations while limiting the cost of running this extensive exploration.

\subsection{Accelerator ISA}
We find that for GEMM and convolution code, removing the ISA significantly deteriorates performance. Still, \sys{} is able to improve over the original code by a notable margin even without the ISA, given that all its other features are still enabled (see \cref{tab:ablations}). This is because we inherently provide an example of accelerator ISA code at each step via the current code, so the model is able to infer some properties of the accelerator ISA. In addition, many of the nested loop optimizations for the GEMM and convolution workloads are well-understood transformations that operate on the C-syntax loops, addresses, and indices in the code, which resemble general-purpose programming, rather than using accelerator-specific constructs such as configuration instructions. However, full \sys{} performance is not matched as the proportion of functionally correct responses is lower, and instruction-level optimizations cannot easily be identified. For example, the first-compute handling optimization in Appendix~\ref{sec:code-examples-gemm}'s GEMM example and the negative-scaled bias loading in Appendix~\ref{sec:code-examples-admm}'s fine-grained linear algebra example would not have been identified without knowledge of the ISA. Overall, we find that the accelerator ISA is an important part of \sys{}'s prompts.

\subsection{Optimization Menu}
We ablate the optimization menu by removing the menu in the planning prompt, and instead simply asking the model to select one optimization and generate a plan. From this experiment, we find that adding domain knowledge and optimization diversity via the optimization menu is essential to \sys{}. As shown in \cref{tab:ablations}, optimization performance completely deteriorates without the optimization menu. Qualitatively, without the optimization menu, we find that the models tend to repeat similar optimizations, with significantly less diversity and relevance in the generated optimization plans.

\subsection{Optimization Menu Dropout}
``Dropout'' for optimization menu options is a key contribution of this work that increases the diversity of generated optimization plans. \Cref{tab:ablations} shows that menu dropout has a significant effect on performance. Qualitatively, we find that without dropout, models tend to be biased towards a limited set of menu options, a limitation which can be resolved via menu dropout.

\subsection{Hardware Performance Feedback}
As discussed in \cref{sec:2-phase}, during plan generation, we include the latency, scratchpad utilization, and accumulator utilization of the original code. \Cref{tab:ablations} shows that this component is helpful, but in some cases its effects may be limited. This is because the options listed in the optimization menu already capture some of the metrics measured in our performance feedback, for example the menu options which suggest using larger tile sizes. Hardware feedback such as scratchpad and accumulator utilization only serves to augment elaboration of these menu options by providing exact measurements.
% Of course, in cases where scratchpad and accumulator utilization is not a bottleneck, that specific feedback is not needed.

\subsection{LLM Ensembling}
Splitting requests between LLMs in an ensemble also encourages diversity of generated plans and code. Qualitatively, we find that the responses, especially during the planning phase, generated by different LLMs differ substantially. Our experiments in \cref{tab:ablations} show that using individual models, such o3-mini or DeepSeek-R1, on their own results in significantly lower performance. 

\subsection{LLM Selection}
In \cref{sec:eval}, we use an ensemble of gpt-4o and o3-mini for our search. To demonstrate that \sys{} does not depend on a particular family of models, we run \sys{} with several alternative LLMs on the same benchmarks used for other ablation experiments above. In addition to DeepSeek-R1~\cite{deepseekr1}, we run \sys{} with two smaller models, Gemini 2.5 Flash and Llama 4 Maverick. For these experiments, each model was used for both the planning and code generation phases without ensembling, keeping the search parameters identical to those used for matrix multiplication and convolution in~\cref{sec:eval}. As shown in~\cref{tab:ablations}, \sys{} with DeepSeek-R1 is able to optimize both GEMM and convolution, achieving substantial speed-ups over the unoptimized code. Even with smaller models, \sys{} shows substantial speed-ups. Similarly to the LLM ensembling ablation study above, gains are slightly smaller than when all techniques are applied. Nonetheless, this demonstrates that \sys is efficient and flexible across different LLMs.

\subsection{Variance Checking}
To account for the stochasticity in \sys{}'s search, we ran \sys{} on three different seeds using the same benchmarks as those used for other ablation experiments. In \cref{tab:ablations}, we report the mean and standard deviation of the final speedups. The performance spread was minimal, with the lowest-performing run for GEMM still achieving a speedup within 11.8\% of the peak result, and within 8.5\% for convolution. This demonstrates that \sys{}'s search is robust and consistently finds high-performing solutions.

% \subsection{ICL Optimization Examples}

\clearpage
\section{Prompts}
\label{sec:prompts}

\lstset{
  % language=c,
  basicstyle=\ttfamily\scriptsize,
  breaklines=true,
  numbersep=3pt,
  xleftmargin=0em,
  frame=single,
  framexleftmargin=0em,
  extendedchars=\true,
  inputencoding=utf8/latin1
}
\begin{figure}[h]
\begin{lstlisting}[language=C++, linewidth=\textwidth, basicstyle=\ttfamily\tiny]
#define config_ex(dataflow, act, A_stride, A_transpose, B_transpose)
// configure the state of the accelerator
// dataflow is WEIGHT_STATIONARY or OUTPUT_STATIONARY
// act is the activation function, options are NO_ACTIVATION, RELU, LAYERNORM, IGELU, SOFTMAX
// A_stride is the stride with which rows of A in the scratchpad are loaded into the systolic array, during computes. If this stride is 1, then we feed consecutive rows in the scratchpad, starting from the starting address of A, into the systolic array as the A matrix. If the stride is 2, then we feed every other row into the systolic array instead.
// A_transpose is a boolean value that represents whether the matrix A is transposed
// B_transpose is a boolean value that represents whether the matrix B is transposed

#define config_ld(dram_stride, scale_factor, spad_block_stride, id)
// configure mvin instructions
// dram_stride = stride in bytes, with which to load from DRAM
// scale_factor = factor to multiply loaded values
// spad_block_stride = when more than DIM columns are loaded, the distance in rows between each block of DIM columns
// id = id of mvin instruction; id = 0 for mvin, 1 for mvin2, 2 for mvin3

#define mvin(dram_addr, spad_acc_addr, cols, rows)
// mvin from DRAM to scratchpad or accumulator
// mvin, configured by config_ld(..., 0)
// rows must be less than or equal to DIM. if more than DIM rows, multiple mvin instructions are needed
// cols must be less than or equal to 4 * DIM.
// if dram_addr = 0, then zeroes are moved into scratchpad/accumulator, max size DIM x DIM

#define mvin2(dram_addr, spad_acc_addr, cols, rows)
// behavior identical to mvin, but configured by config_ld(..., 1)

#define mvin3(dram_addr, spad_acc_addr, cols, rows)
// behavior identical to mvin, but configured by config_ld(..., 2)

// A = input matrix, B = weight matrix, C = output matrix
// assume a weight-stationary dataflow
// preload, compute_preloaded, and compute_accumulated are used to compute DIM x DIM matrix multiplications.
// if no bias, C = A * B is computed; if there is a bias, C = A * B + bias is computed

#define preload(B_spad_addr, C_acc_addr, B_cols, B_rows, C_cols, C_rows)
// preload weights, B, onto DIM by DIM systolic array
// B must be preloaded before compute
// B must have been moved in to the scratchpad first
// B_cols must be less than or equal to DIM, B_rows must be less than or equal to DIM, C_cols must be less than or equal to DIM, C_rows must be less than or equal to DIM
// must run to change the output address to C_acc_addr 
// if B_spad_addr unchaged from previous preload instruction, can set B_spad_addr = 0xffffffff; must be specified otherwise

#define compute_preloaded(A_spad_addr, bias_spad_addr, A_cols, A_rows, bias_cols, bias_rows)
// compute on DIM by DIM systolic array, with optional added bias (can be used for matrix addition)
// A must have been moved in to the scratchpad first
// first compute after preload to systolic array
// either overwrites or accumulates C depending on bit 30 of C_acc_addr
// A_cols must be less than or equal to DIM, A_rows must be less than or equal to DIM, bias_cols must be less than or equal to DIM, bias_rows must be less than or equal to DIM
// bias_spad_addr = 0xffffffff if no bias
// if there is a bias, bias_cols and bias_rows are probably equal to C_cols and C_rows from preload instruction

#define compute_accumulated(A_spad_addr, bias_spad_addr, A_cols, A_rows, bias_cols, bias_rows) 
// compute on DIM by DIM systolic array
// A must have been moved in to the scratchpad first
// for weight stationary, use when B_spad_addr has not changed
// either overwrites or accumulates C depending on bit 30 of C_acc_addr
// A_cols must be less than or equal to DIM, A_rows must be less than or equal to DIM, bias_cols must be less than or equal to DIM, bias_rows must be less than or equal to DIM
// bias_spad_addr = 0xffffffff if no bias
// if there is a bias, bias_cols and bias_rows are probably equal to B_cols and B_rows from preload instruction

#define config_st(cols)
// configure mvout instruction
// cols = number of columns of matrix in DRAM

#define mvout(dram_addr, spad_acc_addr, cols, rows)
// mvout from scratchpad or accumulator to DRAM
// cols must be less than or equal to DIM
// rows must be less than or equal to DIM

#define fence() asm volatile("fence") 
// fence
\end{lstlisting}
\caption{Accelerator ISA specification for Gemmini accelerators, referenced in \cref{sec:2-phase}.}
\label{fig:gemmini-isa}
\vspace{-1in}
\end{figure}

\begin{figure*}[h]
\begin{lstlisting}[language=C++, linewidth=\textwidth, basicstyle=\ttfamily\tiny]
'''
Gemmini's private memory is "row-addressed", where each row is DIM elements wide, where DIM is the number of PEs across the width of the systolic array. These elements will be of type inputType in the scratchpad, and of type accType in the accumulator.

Every private Gemmini memory address is 32 bits long. The three most signficant bits are reserved, and have special meanings:

    Bit 31 (the MSB) is 0 if we are addressing the scratchpad, and 1 if we are addressing the accumulator.
    Bit 30 is ignored if we are addressing the scratchpad, or if we are reading from the accumulator. If, instead, we are writing to the accumulator, then bit 30 is 0 if we want to overwrite the data at that address, and 1 if we want to accumulate on top of the data already at that address.
    Bit 29 is ignored if we are addressing the scratchpad, or if we are writing to the accumulator. If, instead, we are reading from the accumulator, then bit 29 is 0 if we want to read scaled-down inputType data from the accumulator, and 1 if we want to read accType data from the accumulator.
        If bit 29 is 1 for an accumulator read address, then we do not apply activation functions or scaling to the output of the accumulator.
'''

'''
Gemmini is a decoupled access/execute architecture, which means that "memory-access" and "execute" instructions happen concurrently, in different regions of the hardware.
It has an ExecuteController (for preload and compute instructions), LoadController (mvin), and StoreController (mvout).
Gemmini includes an ROB which is meant to detect hazards between instructions in different controllers. 
Each controller also handles its own dependencies and hazards internally.
'''
```
\end{lstlisting}
\caption{Gemmini Accelerator ISA specification from \cref{sec:2-phase}, continued.}
\label{fig:gemmini-isa-cont}
\end{figure*}
\begin{figure*}[h]
\begin{lstlisting}[]
<optimizations>:
1. modify loop tiling
2. loop reordering
3. split loops
4. fuse loops
5. simplify arithmetic and propagate constants to simplify expressions
6. reorder computations or blocks of computations
7. loop unrolling
8. double buffering
9. move more data to the scratchpad in a more outer loop to increase data reuse
10. spread data throughout the scratchpad rather than loading to the same location repeatedly
11. load data to the scratchpad across outer loop iterations and use if statements to prevent redundant loads on loops inner to those
12. hoist redundant operations out of loops
13. substitute operations with equivalent operations that are faster
14. pipeline operations to better overlap computation and data movement
15. minimize data movement
16. minimize loop overhead
17. other methods not listed here.
\end{lstlisting}
\caption{The list of optimization menu options available (with some probability of dropout) during the planning phase, as described in \cref{sec:2-phase}, used with 16x16 INT8 Gemmini to optimize matrix multiplication and convolution code.}
\label{fig:menu-opts}
\end{figure*}

\begin{figure*}[h]
\begin{lstlisting}[]
<optimizations>:
1. remove unnecessary code
2. simplify arithmetic and propagate constants to simplify expressions
3. merge instructions
4. merge high-level operations
5. reorder operations or blocks of operations
6. move cpu-based computation to the accelerator
7. add or subtract a matrix using the bias
8. hoist redundant operations out of loops
9. substitute operations with equivalent operations that are faster
10. pipeline operations to better overlap computation and data movement
11. eliminate data dependencies and fence operations
12. minimize data movement
13. minimize loop overhead
14. other methods not listed here
\end{lstlisting}
\caption{The optimization menu for TinyMPC code optimization on 4x4 FP32 Gemmini.}
\label{fig:menu-opts-admm}
\end{figure*}
\begin{figure*}[h]
\begin{lstlisting}[language=C++, linewidth=\textwidth, basicstyle=\ttfamily\tiny]
Here is an example of increasing scratchpad tile size for the Y dimension of a 512x512 (X x Z) matrix A and 512x512 (Z x Y) matrix B multiplication. Original code:
    uint32_t b_offset = 16 * 16 * 4 * 8 * sizeof(int8_t); 
    for (int_fast32_t y = 0; y < 8; y++) {
        uint32_t b_base_y = 64 * y; 
        // Load B matrix slice
        for (int_fast32_t zo = 0; zo < 8; zo++) {
            uint32_t b_zo_offset = 4 * 16 * zo; // Number of columns per zo iteration
            for (int_fast32_t z = 0; z < 4; z++) {
                uint32_t b_index = ((zo * 4 + z) * ((16 * 4) * 16)) / 16; // Divide number of elements by 16 since scratchpad is row-indexed
                mvin3(&B[b_zo_offset + 16 * z][b_base_y], b_offset + b_index, 16 * 4, 16);
            }}
        for (int_fast32_t x = 0; x < 32; x++) {
            uint32_t res = 1 << 31;
            uint32_t a_base_x = 16 * x; 
            // Load A matrix slice
            for (int_fast32_t zo = 0; zo < 8; zo++) {
                uint32_t a_index = (zo * (16 * 4) * 16) / 16;
                mvin2(&A[a_base_x][64 * zo], a_index, 16 * 4, 16);
            }
            // Computation
            for (int_fast32_t zo = 0; zo < 8; zo++) {   
                uint32_t a_index = (zo * (16 * 4) * 16) / 16;
                for (int_fast32_t z = 0; z < 4; z++) {
                    uint32_t preload_flag = (zo == 0 && z == 0) ? 0 : 0x40000000;
                    for (int_fast32_t y_in_o = 0; y_in_o < 4; y_in_o++) {
                        uint32_t preload_index = ((zo * 4 + z) * ((16 * 4) * 16) + y_in_o * (16 * 16)) / 16; // Find correct scratchpad index to load B from
                        preload(b_offset + preload_index, res + (y_in_o * (16 * 16)) / 16 | preload_flag, 16, 16, 16, 16);
                        compute_preloaded(a_index + (z * (16 * 16)) / 16, ~((uint32_t)0), 16, 16, 16, 16);
                    }}}
            // Store C matrix slice
            for (int_fast32_t y_in_o = 0; y_in_o < 4; y_in_o++) {
                mvout(&C[a_base_x][b_base_y + 16 * y_in_o], res + (y_in_o * (16 * 16)) / 16, 16, 16); // Divide number of elements by 16 since accumulator is row-indexed
            }}}
Retiled code
    uint32_t b_offset = 16 * 16 * 4 * 8 * sizeof(int8_t); 
    for (int_fast32_t y = 0; y < 2; y++) { // Reduce number of y dimension outer loop iterations
        uint32_t b_base_y = 256 * y;
        // Load larger B matrix slice
        // Tiling reduces redundant loads of B matrix, reducing data movement and increasing data reuse
        for (int_fast32_t zo = 0; zo < 8; zo++) {
            uint32_t b_zo_offset = 4 * 16 * zo; // Number of columns per zo iteration
            for (int_fast32_t z = 0; z < 4; z++) {
                for (int_fast32_t y_in = 0; y_in < 4; y_in++) {
                    uint32_t b_index = (((zo * 4 + z) * 4 + y_in) * ((16 * 4) * 16)) / 16; // Divide number of elements by 16 since scratchpad is row-indexed
                    mvin3(&B[b_zo_offset + 16 * z][b_base_y + 64 * y_in], b_offset + b_index, 16 * 4, 16);
                }}}
        for (int_fast32_t x = 0; x < 32; x++) {
            uint32_t res = 1 << 31;
            uint32_t a_base_x = 16 * x;
            // Load A matrix slice
            // Tiling reduces redundant loads of A matrix, reducing data movement and increasing data reuse
            for (int_fast32_t zo = 0; zo < 8; zo++) {
                uint32_t a_index = (zo * (16 * 4) * 16) / 16;
                mvin2(&A[a_base_x][64 * zo], a_index, 16 * 4, 16);
            }
            // Computation
            for (int_fast32_t zo = 0; zo < 8; zo++) {   
                uint32_t a_index = (zo * (16 * 4) * 16) / 16;
                for (int_fast32_t z = 0; z < 4; z++) {
                    uint32_t preload_flag = (zo == 0 && z == 0) ? 0 : 0x40000000;
                    for (int_fast32_t y_in_o = 0; y_in_o < 16; y_in_o++) { // Increase number of Y dimension inner loop iterations to increase tile size
                        uint32_t preload_index = (((zo * 4 + z) * 4) * ((16 * 4) * 16) + y_in_o * (16 * 16)) / 16; // Find correct scratchpad index to load B from
                        preload(b_offset + preload_index, res + (y_in_o * (16 * 16)) / 16 | preload_flag, 16, 16, 16, 16);
                        compute_preloaded(a_index + (z * (16 * 16)) / 16, ~((uint32_t)0), 16, 16, 16, 16);
                    }}}
            // Store C matrix slice
            for (int_fast32_t y_in_o = 0; y_in_o < 16; y_in_o++) { // Move out a larger tile in the Y dimension
                mvout(&C[a_base_x][b_base_y + 16 * y_in_o], res + (y_in_o * 16 * 16) / 16, 16, 16); // Divide number of elements by 16 since accumulator is row-indexed
            }}}
\end{lstlisting}
\caption{In-context learning example of tiling for Gemmini, provided during the code generation phase in \cref{sec:2-phase}. Inserted in the prompt only when the string \texttt{"tiling"} is detected in the plan generated in Phase 1.}
\label{fig:tiling-icl-example}
\end{figure*}

\begin{figure*}[h]
\begin{lstlisting}[]
Rules:
1. The rewritten program should be semantically equivalent to the original program
2. Limit the scope of the plan to the selected optimization
3. All code must be inside the test() function
4. Do not use C preprocessing directives (#ifdef, #define, etc.)
5. If modifying loops, modify other related loop bounds and adjust address and index calculations to ensure the code is still correct
6. If increasing loaded tile size, ensure that data is spread throughout the scratchpad across all relevant dimensions
7. If loading across new dimensions, add the loop indices of those dimensions to scratchpad address calculations
8. If increasing loaded tile size, update preload and compute instructions to match the new data layout
9. If increasing loaded tile size, update base scratchpad addresses to fit new tile size
\end{lstlisting}
\caption{The list of rules provided during both the planning and code implementation phases for Gemmini code optimization, as described in \cref{sec:2-phase}.}
\label{fig:rules}
\end{figure*}
\begin{figure*}[h]
\begin{lstlisting}[]
<optimizations>:
1. eliminate loads and stores as much as possible, keeping data in SBUF/PSUM instead
2. minimize data movement
3. improve data layout and access patterns
4. loop reordering and restructuring
5. inline a function so it can be more easily optimized and fused
6. skip computation when it is not needed (e.g. it is completely masked out)
7. fuse loops (reordering if necessary)
8. increase reuse by keeping data in SBUF across outer loop iterations
9. hoist redundant operations out of loops
10. delay softmax division until after all reductions are complete
11. Perform nc_matmul on large contiguous blocks within its own affine_range loop to maximize compute throughput
12. Group nc_matmul calls into larger blocks, organizing inputs ahead of time, to maximize Tensor Engine utilization
13. do operations in lower precision such as nl.bfloat16
14. double buffering
15. fuse multiple instructions into one, for example by doing reduction inside nisa.activation()
16. pipeline operations to better overlap computation and data movement (using sequential_range)
17. keep data in SBUF/PSUM instead of storing to and loading from HBM
18. stronger tiling for contraction / moving-free split
19. reorder operations to improve locality
20. fuse dependent operations
21. fuse operations into a single loop so intermediate data does not need to be stored to and loaded from HBM
22. fuse loops that iterate over the same dimension to improve intermediate data reuse
23. allocate a larger tile in SBUF so we can keep data in it rather than storing to and loading from HBM
24. allocate buffers in lower precision such as nl.bfloat16
25. downcast to lower precision during operations that take dtype as an argument
26. keep data in the same layout to avoid transpose operations
27. eliminate intermediate tensor materialization by using in-place operations (storing the output in the same buffer as the input)
28. use the streaming softmax with running max and scaling trick
29. optimize accumulation patterns in PSUM
30. optimize reduction by fusing tile-wise reductions with transformation passes
31. Load larger blocks of data to increase SBUF data reuse and reduce memory traffic
32. Add additional loop levels so larger blocks of data can be loaded (multi-level tiling)
33. Combine adjacent tiles into contiguous blocks before nl.store() to maximize memory throughput.
34. Scan carry-over to parallelize the scan operation
35. Replace general-purpose code with faster specialized instructions
36. Hoist nl.load() operations for reused data (e.g., LHS tiles) outside inner loops to reduce redundant HBM->SBUF transfers.
37. Other methods not listed here.
\end{lstlisting}
\caption{The optimization menu for Trainium code optimization.}
\label{fig:menu-opts-trainium}
\end{figure*}

\begin{figure*}[h]
\begin{lstlisting}[]
1. The rewritten program should be semantically equivalent to the original program, within a small numerical tolerance.
2. Maintain correct tensor shapes and indexing patterns. Remember not to index with affine_range loop variables. Avoid loop carried dependencies.
3. The following imports have already been run: import neuronxcc.nki as nki; import neuronxcc.nki.isa as nisa; import neuronxcc.nki.language as nl; import neuronxcc.nki.typing as nt; import numpy as np;
4. nisa and nl may have similar functions (for example, nisa.nc_matmul() and nl.matmul()), but they may have different arguments or functionality. Make sure to follow the documentation above.
5. Limit the scope of the plan to the selected optimization.
6. Do not count out any of the <optimizations> unless they are clearly irrelevant to the code.
7. Ensure that loop dependencies are not violated inside affine_range loops.
\end{lstlisting}
\caption{The rules for Trainium code optimization, during the Plan stage of prompting.}
\label{fig:rules-trainium-plan}
\end{figure*}

\begin{figure*}[h]
\begin{lstlisting}[]
1. The rewritten program should be semantically equivalent to the original program, within a small numerical tolerance.
2. Maintain correct tensor shapes and indexing patterns. Remember not to index with affine_range loop variables. Avoid loop carried dependencies.
3. The following imports have already been run: import neuronxcc.nki as nki; import neuronxcc.nki.isa as nisa; import neuronxcc.nki.language as nl; import neuronxcc.nki.typing as nt; import numpy as np;
4. nisa and nl may have similar functions (for example, nisa.nc_matmul() and nl.matmul()), but they may have different arguments or functionality. Make sure to follow the documentation above.
5. Optimize the test() function and do not change its name.
6. Wrap the generated code with ```python at the beginning and ``` at the end.
7. Ensure that loop dependencies are not violated inside affine_range loops.
\end{lstlisting}
\caption{The rules for Trainium code optimization, during the Implement stage of prompting.}
\label{fig:rules-trainium-implement}
\end{figure*}

\begin{figure*}[h]
\begin{lstlisting}[]
<optimizations>:
1. Convert a PyTorch operation to inline CUDA C++ code
\end{lstlisting}
\caption{The optimization menu used during the first two iterations of GPU code optimization.}
\label{fig:menu-opts-gpu-translate}
\end{figure*}

\begin{figure*}[h]
\begin{lstlisting}[]
<optimizations>:
1. Convert an operation to optimized CUDA C++ code
2. Convert an operation to CUDA C++ code
3. Convert an operation to optimized Triton code
4. Reduce PyTorch launch overhead
5. Use compilation flags like -O3 and --use_fast_math when compiling CUDA code
6. Minimize global memory accesses
7. Use shared memory to reduce global memory bandwidth usage
8. Cache redundantly computed data in shared memory
9. Use pointers to global memory rather than copying to shared memory
10. Coalesce global memory accesses
11. Avoid bank conflicts in shared memory
12. Use registers efficiently; avoid register spilling
13. Minimize divergent branches within warps
14. Use CUDA warp-level primitives for synchronization
15. Fuse kernels when possible to reduce kernel launch overhead
16. Minimize number of synchronization points
17. Store more data and reduce at the end rather than using atomic operations
18. Use grid-stride loops
19. Tile operations for optimal cache utilization
20. Use L2 persisting cache window to keep frequently reused tensors resident in L2
21. Use multiple CUDA streams to overlap computation and data movement
22. overlap host-to-device transfers with the CUDA-Graph replay
23. Maximize occupancy without excessive register usage
24. Choose optimal block sizes (typically multiples of 32 threads)
25. Use __restrict__ to help compiler with pointer aliasing
26. Loop unrolling (#pragma unroll)
27. Use cuBLASLt for Tensor Core GEMM operations
28. Use cuBLASLt, cuBLAS, or cuDNN for GEMM and convolution operations instead of custom kernels
29. Use Tensor Cores (e.g. wmma APIs) for mixed precision acceleration (FP16, TF32, INT8)
30. Use PyTorch's tensor core APIs (torch.backends.cuda.matmul.allow_tf32, torch.backends.cudnn.allow_tf32, torch.amp) to enable Tensor Cores
31. Use lower precision (e.g. bfloat16, float16, float8_e4m3fn) for computations
32. Quantize weights or activations where accuracy permits (e.g. bfloat16)
33. Leverage fused operations in cuDNN (e.g. convolution + bias + ReLU)
34. Overlap computation and data transfer using CUDA streams and asynchronous copies
35. Use pinned (page-locked) host memory for faster host-device transfers
36. Minimize host-device transfer frequency
37. Choose optimal convolution algorithms (FFT, Winograd, implicit GEMM) based on kernel size
38. Prune unneeded weights for sparse computation
39. Batch inputs to maximize GPU utilization
40. Reuse intermediate results where possible (e.g. shared activations)
41. Vectorize operations by using wider data types
42. Use Tensor core GEMMs for GEMM-like operations
43. Convert convolution operations to Tensor core GEMMs
44. Skip computation when data-dependent execution encounters zero values or a branch that will never be taken
45. Ensure data is stored in contiguous memory blocks
46. Arrange data access patterns to maximize memory bandwidth and minimize latency through techniques like shared memory usage, coalesced global memory access, and memory padding
47. Memory Coalescing: optimize CUDA kernel performance by ensuring threads in the same warp access contiguous memory locations
48. Pre-allocate input and output tensors during graph initialization and reuse them
49. Merge low-level operations
50. Merge high-level operations
51. Reorder operations or blocks of operations
52. Hoist redundant operations out of loops
53. Substitute operations with equivalent operations that are faster
54. Double buffering
55. Pipeline operations to better overlap computation and data movement
56. Minimize data movement
57. Use built-in CUDA primitive functions
58. Call torch:: functions from C++ rather than from Python
59. Use ATen at:: functions rather than PyTorch functions
60. Use CUDA graph capture
61. Use dedicated CUDA streams
62. Profile the code and capture CUDA graphs in the __init__ function
63. Simplify operations where possible
64. Classical compiler optimizations
65. Any other optimizations that you think are relevant
\end{lstlisting}
\caption{The optimization menu used during iteration $t>2$ of GPU code optimization.}
\label{fig:menu-opts-gpu}
\end{figure*}

\begin{figure*}[h]
\begin{lstlisting}[]
1. You will be running the code on an NVIDIA L40S GPU with PyTorch 2.5.0 and CUDA 12.4
2. The rewritten program should be semantically equivalent to the original program, within a small numerical tolerance.
3. All generated code should be contained in a single Python file (inline CUDA code is allowed).
4. Only class ModelNew will be imported during evaluation. Feel free to define other variables, functions, or classes, but make sure they are used by ModelNew.
5. When using torch.utils.cpp_extension load() or load_inline(), make sure to place C++ code in cpp_sources and CUDA code in cuda_sources.
6. Do not use the `function` argument of load_inline(), make a PYBIND11 binding instead.
7. Do not add fallback paths that revert to the original code.
8. Limit the scope of the plan to the selected optimization.
\end{lstlisting}
\caption{The rules for GPU code optimization, during the Plan stage of prompting.}
\label{fig:rules-gpu-plan}
\end{figure*}

\begin{figure*}[h]
\begin{lstlisting}[]
1. You will be running the code on an NVIDIA L40S GPU with PyTorch 2.5.0 and CUDA 12.4
2. The rewritten program should be semantically equivalent to the original program, within a small numerical tolerance.
3. All generated code should be contained in a single Python file (inline CUDA code is allowed).
4. Only class ModelNew will be imported during evaluation. Feel free to define other variables, functions, or classes, but make sure they are used by ModelNew.
5. When using torch.utils.cpp_extension load() or load_inline(), make sure to place C++ code in cpp_sources and CUDA code in cuda_sources.
6. Do not use the `function` argument of load_inline(), make a PYBIND11 binding instead.
7. Do not add fallback paths that revert to the original code.
8. Wrap the generated code with ```python at the beginning and ``` at the end.
\end{lstlisting}
\caption{The rules for GPU code optimization, during the Implement stage of prompting.}
\label{fig:rules-gpu-implement}
\end{figure*}

\begin{figure*}[h]
\begin{lstlisting}[]
Example of using cuBLASLt for Tensor Core GEMM operations

#include <cublasLt.h>
#include <cuda_runtime.h>
#include <cstdint>

#include "sample_cublasLt_LtIgemmTensor.h"
#include "helpers.h"

int roundoff(int v, int d) {
    return (v + d - 1) / d * d;
}

/// Use cublasLtMatmul to perform tensor-op Igemm with memory order transforms on all buffers
///
/// For better performance data order transforms should be offline as much as possible.
///
/// transa, transb assumed N; alpha, beta are host pointers, tensor ops allowed, alpha assumed 1, beta assumed 0,
/// stream assumed 0
void LtIgemmTensor(cublasLtHandle_t ltHandle,
                   cublasOperation_t transa,
                   cublasOperation_t transb,
                   int m,
                   int n,
                   int k,
                   const int8_t *A,
                   int lda,
                   const int8_t *B,
                   int ldb,
                   int32_t *C,
                   int ldc) {
    cublasLtMatmulDesc_t matmulDesc = NULL;
    cublasLtMatrixLayout_t Adesc = NULL, Bdesc = NULL, Cdesc = NULL;
    int32_t alpha = 1, beta = 0;
    cublasOperation_t opTranspose = CUBLAS_OP_T;

    // tensor op igemm kernels require specialized memory order of data
    cublasLtMatrixTransformDesc_t transformDesc = NULL;
    int8_t *Atransform = NULL, *Btransform = NULL;
    int32_t *Ctransform                   = NULL;
    cublasLtMatrixLayout_t AtransformDesc = NULL, BtransformDesc = NULL, CtransformDesc = NULL;
    float transformAlpha = 1.0f, transformBeta = 0.0f;
    cublasLtOrder_t order_COL32       = CUBLASLT_ORDER_COL32;
    cublasLtOrder_t order_COL4_4R2_8C = CUBLASLT_ORDER_COL4_4R2_8C;

    int ldatransform = 32 * m;
    int ldbtransform = 32 * roundoff(n, 8);
    int ldctransform = 32 * m;

    checkCudaStatus(cudaMalloc(reinterpret_cast<void**>(&Atransform), sizeof(int8_t) * roundoff(k, 32) / 32 * ldatransform));
    checkCudaStatus(cudaMalloc(reinterpret_cast<void**>(&Btransform), sizeof(int8_t) * roundoff(k, 32) / 32 * ldbtransform));
    checkCudaStatus(cudaMalloc(reinterpret_cast<void**>(&Ctransform), sizeof(int32_t) * roundoff(n, 32) / 32 * ldctransform));

    checkCublasStatus(cublasLtMatrixTransformDescCreate(&transformDesc, CUDA_R_32F));

    checkCublasStatus(cublasLtMatmulDescCreate(&matmulDesc, CUBLAS_COMPUTE_32I, CUDA_R_32I));
    // tensor op igemm kernels only support NT gemm
    checkCublasStatus(cublasLtMatmulDescSetAttribute(matmulDesc, CUBLASLT_MATMUL_DESC_TRANSA, &transa, sizeof(transa)));
    checkCublasStatus(cublasLtMatmulDescSetAttribute(matmulDesc, CUBLASLT_MATMUL_DESC_TRANSB, &transb, sizeof(transb)));

    // ---------------------------------------------------------------------------------------------
    // create descriptors for original matrices

    checkCublasStatus(cublasLtMatrixLayoutCreate(&Adesc, CUDA_R_8I, m, k, lda));
    checkCublasStatus(cublasLtMatrixLayoutCreate(&Bdesc, CUDA_R_8I, k, n, ldb));
    checkCublasStatus(cublasLtMatrixLayoutCreate(&Cdesc, CUDA_R_32I, m, n, ldc));
    
    ...
\end{lstlisting}
\caption{Excerpt from one of the ICL examples provided during GPU code optimization, when the string \texttt{``tensor core''} is detected in a plan.}
\label{fig:icl-example-gpu}
\end{figure*}

\clearpage
\section{Code Examples}
\label{sec:code-examples}
In this section, we discuss in greater depth what optimizations \sys{} applies in our evaluations and how \sys{} is able to achieve significantly better performance than hand-optimized code.

\subsection{16x16 INT8 Gemmini---\texttt{12544x64x256} GEMM}\label{sec:code-examples-gemm}

\cref{fig:exo-2-unopt} contains the unoptimized Exo-generated code, used as the starting point for search. \cref{fig:exo-2-exo-opt} contains the code generated by Exo after hand-optimization by \citet{exo}.
\cref{fig:exo-2-opt,fig:exo-2-opt-cont,fig:exo-2-opt-cont-cont} contain the result of \sys{} optimization on Exo Unoptimized code. While the code is substantially transformed from the original code, some aspects remain the same. For example, in this case the configuration instructions and loop ordering remain largely the same. 

Of course, many optimizations have been applied to the code. We briefly summarize the optimization menu options selected and plans generated during the optimization process for this code. We also include the speedup after each respective optimization.

\begin{enumerate}[leftmargin=*]
\item 1.67$\times$: initial speedup of Exo Unoptimized code over Gemmini's software library before any optimization.
\item 1.93$\times$ after ``hoist redundant operations out of loops''. This plan hoists constants like \texttt{tile\_dim = 16} and loop-invariant expressions like \texttt{ko * 64} and \texttt{k * 16} out of inner loops. These precomputed values are reused inside Gemmini ops in each iterations, so we should reduce the number of times they must be calculated.
\item 1.95$\times$ after ``double buffering''. This plan defines two buffer regions for matrices A and B in the scratchpad. A \texttt{buffer\_toggle} flag is introduced to alternate between these buffers each iteration. All \texttt{mvin}, \texttt{preload}, and \texttt{compute} instructions are updated to use the active buffer based on the toggle. Data loading for the next iteration is scheduled earlier to overlap with current computation. Address calculations are adjusted to include buffer offsets accordingly.

\item 2.15$\times$ after ``pipeline operations to better overlap computation and data movement''. Moves \texttt{mvin2} (A tile load) to immediately after \texttt{compute\_preloaded} in the ko loop to overlap A prefetch with current compute. Moves \texttt{mvin3} (B tile load) earlier in the k loop, before the next compute, to overlap B prefetch with current compute.

\item 3.13$\times$ after ``load data to the scratchpad across outer loop iterations and use if statements to prevent redundant loads on loops inner to those''. Adds a one-time load of the entire B matrix into a new scratchpad region (specifically, \texttt{new\_B\_base} = 8192) before the main \texttt{i} loop.
Replaces repeated \texttt{mvin3} calls in inner loops with offset calculations into this preloaded B region.

\item 3.54$\times$ after ``move more data to the scratchpad in a more outer loop to increase data reuse''. Before the reduction (\texttt{k}) loop, load the entire 16x256 A-tile into scratchpad using four mvin2 calls to a new \texttt{A\_tile\_base} region.
Remove the double-buffering (\texttt{a\_toggle}) for A since the full tile is now resident.
Replace nested \texttt{ko} and \texttt{k} loops with a single loop over 16 segments, computing each 16x16 A sub-tile address from \texttt{A\_tile\_base}.
Adjust B tile \texttt{preload} and \texttt{compute\_preloaded} calls accordingly, using a simpler tile index.

\item 4.87$\times$ after ``double buffering''. Restores double buffering by reserving two scratchpad regions: \texttt{A\_tile\_base0} and \texttt{A\_tile\_base1}, each holding one full 16x256 A-tile.
In the outer i loop, alternate buffers using \texttt{i \% 2} to select \texttt{current\_buffer} and {next\_buffer}.
In each iteration (except last), while computing with \texttt{current\_buffer}, issue \texttt{mvin2} to load the next A-tile into \texttt{next\_buffer}.
In the inner \texttt{compute} loops, use \texttt{current\_buffer} for A-tile addresses.
After \texttt{compute} and \texttt{mvout}, the next iteration uses the preloaded data.

\item 5.21$\times$ after ``double buffering''. Double-buffers accumulator by allocating two accumulator regions: \texttt{acc\_base0} and \texttt{acc\_base1}.
In each \texttt{i} iteration, compute into \texttt{cur\_acc\_base} and \texttt{mvout} from \texttt{prev\_acc\_base} (except on first iteration), and swap \texttt{cur\_acc\_base} and \texttt{prev\_acc\_base} at the end of the loop.

\item 5.23$\times$ after ``loop unrolling''. Unrolls the innermost loop (\texttt{j\_in\_o}) by a factor of 4.

\item 5.53$\times$ after ``fuse loops''. ``Fuses'' loops by eliminating the loop over \texttt{j\_in\_o} where we \texttt{mvin} 0s to the accumulator. Instead, use the ability of \texttt{preload} to overwrite the values in the accumulator rather than accumulating, when beginning a new partial sum.
\end{enumerate}

From this example, we observe that a diverse set of optimizations is selected, and that speedups are distributed throughout the optimization process rather than concentrated in just one or two steps, showing the importance of a well-designed iterative search process. From here, we summarize the differences between \sys{}-generated code and the previous best code (Exo Opt):

\begin{itemize}
\item \textbf{Tiling.} The Exo Opt code loads 128$\times$256 tiles of A, whereas the \sys{}-generated code loads 32$\times$256 tiles (divided into two 16$\times$256 tiles) of A. While this means there is less reuse for the \sys{}-generated code, there is also less overhead needed for compute instructions to wait for each A tile to be loaded to the scratchpad. In combination with the rest of the optimizations applied by \sys{}, this leads to improved performance.

\item \textbf{Double-buffering.} In the \sys{}-generated code, we see that both the scratchpad and accumulator are explicitly double-buffered. In the schedule, we can see that double buffering is applied 3 times. Initially (in step 3), both the A and B matrices (where matrix multiplication is represented as A$\times$B$=$C), are double buffered in the scratchpad. However, after steps 5 and 6, B and A (respectively) are no longer double-buffered as larger tiles are loaded before beginning computation. The accumulator is double-buffered in step 8, resulting in the code below. The Exo Opt code relies on the accelerator's out-of-order execution to handle executing \texttt{mvin} and {mvout} instructions without dependencies, ahead of the order in which they are issued.

\item \textbf{Software pipelining.} The \sys{}-generated code explicitly issues A \texttt{mvin} instructions before they are needed for computation, whereas as above the Exo Opt code relies on hardware to handle overlapping of data movement and compute. Also, the \sys{}-generated code explictly issues all B \texttt{mvin} instructions at the beginning of the program, whereas the Exo Opt code interleaves these instructions with computation (but still loads the entire B matrix to the scratchpad, once overall). This does not have a significant impact on performance, but LLM-generated code is naturally biased towards such an implementation due to its simplicity.

\item \textbf{First-compute handling.} The \sys{}-generated code utilizes the ability of compute instructions to overwrite the accumulator, whereas Exo Opt code explicitly issues \texttt{mvin} instructions to zero out the accumulator before beginning computation on a tile.

\item \textbf{Arithmetic simplification.} Arithmetic on constants is fully simplified and handled inside shared variables wherever possible in the \sys{}-generated code, reducing the overhead of non-accelerator instructions.
\end{itemize}

Overall, we find that compared to the Exo Opt code, \sys{}-generated code applies more techniques to minimize the amount of CPU overhead during execution. The smaller tiles it uses, in combination with its more explicit application of double-buffering and software pipelining, results in highly tuned, fine-grained overlapping of data movement and computation and a very high level of performance.

\clearpage

% \subsubsection{Unoptimized Code Example}
\begin{figure*}[t]
\begin{lstlisting}[language=C++, linewidth=\textwidth, basicstyle=\ttfamily\tiny]
void test(int8_t A[12544][256], int8_t B[256][64], int8_t C[12544][64]) {
  config_st((64));
  config_ex(WEIGHT_STATIONARY, NO_ACTIVATION, 1, false, false);
  config_ld((64), 1.0f, 16, 2);
  config_ld((256), 1.0f, 16, 1);
  config_ld(0, 1.0f, 0, 0);

  for (int_fast32_t i = 0; i < 784; i++) {
    for (int_fast32_t j = 0; j < 1; j++) {
      uint32_t res = 1 << 31;
      for (int_fast32_t j_in_o = 0; j_in_o < 4; j_in_o++) {
        mvin( 0, res + ((j_in_o) * (256))/16,(16 + 0), (16 + 0) );
      }
      uint32_t a = 0;
      uint32_t b = 16 * 16 * 4 * 4 / 16;
      for (int_fast32_t ko = 0; ko < 4; ko++) {
        mvin2( &A[(16 * i)][64 * ko], a + ((ko) * (1024))/16, 16*(4 + 0), (16 + 0) );
        for (int_fast32_t k = 0; k < 4; k++) {
          mvin3( &B[(64 * ko + 16 * k)][64 * j], b + ((ko) * (4096) + (k) * (1024))/16, 16*(4 + 0), (16 + 0) );
        }
        for (int_fast32_t k = 0; k < 4; k++) {
          for (int_fast32_t j_in_o = 0; j_in_o < 4; j_in_o++) {
            preload(b + ((ko) * (4096) + (k) * (1024) + (j_in_o) * (256))/16, res + ((j_in_o) * (256))/16 | 0x40000000, (16 + 0), (16 + 0), (16 + 0), (16 + 0));
            compute_preloaded(a + ((ko) * (1024) + (k) * (256))/16, ~((uint32_t)0), (16 + 0), (16 + 0), 16, 16);
          }
        }
      }
      for (int_fast32_t j_in_o = 0; j_in_o < 4; j_in_o++) {
        mvout( &C[(16 * i)][16 * j_in_o + 64 * j], res + ((j_in_o) * (256))/16, (16 + 0), (16 + 0) );
      }
    }
  }
  fence();
}
\end{lstlisting}
\caption{Example of Exo-generated unoptimized matrix multiplication code, from the experiments in \cref{sec:gemm}. Achieves 28\% utilization.}
\label{fig:exo-2-unopt}
\end{figure*}

\clearpage

% \subsubsection{Exo Optimized Code Example}
\begin{figure*}[t]
\begin{lstlisting}[language=C++, linewidth=\textwidth, basicstyle=\ttfamily\tiny]
void test(int8_t A[12544][256], int8_t B[256][64], int8_t C[12544][64]) {
  config_st((64));
  config_ex(WEIGHT_STATIONARY, NO_ACTIVATION, 1, false, false);
  config_ld((64), 1.0f, 16, 2);
  config_ld((256), 1.0f, 16, 1);
  config_ld(0, 1.0f, 0, 0);

  uint32_t res = 1 << 31;
  uint32_t a = 0;
  uint32_t b = 16 * 16 * 4 * 4 * 8 * sizeof(int8_t) / 16;
  for (int_fast32_t io = 0; io < 98; io++) {
    for (int_fast32_t i = 0; i < 8; i++) {
      mvin( 0, res + ((i) * (1024))/16, (16), (16) );
      mvin( 0, res + ((i) * (1024) + 256)/16, (16), (16) );
      mvin( 0, res + ((i) * (1024) + (2) * (256))/16, (16), (16) );
      mvin( 0, res + ((i) * (1024) + (3) * (256))/16, (16), (16) );
      for (int_fast32_t ko = 0; ko < 4; ko++) {
        mvin2( &A[(16 * i + 128 * io)][64 * ko], a + ((i) * (4096) + (ko) * (1024))/16, 16*(4), (16) );
        if (io == 0) {
          if (i == 0) {
            mvin3( &B[(64 * ko)][0], b + ((ko) * (4096))/16, 16*(4), (16) );
          }
        }
        if (io == 0) {
          if (i == 0) {
            mvin3( &B[(16 + 64 * ko)][0], b + ((ko) * (4096) + 1024)/16, 16*(4), (16) );
          }
        }
        if (io == 0) {
          if (i == 0) {
            mvin3( &B[(32 + 64 * ko)][0], b + ((ko) * (4096) + (2) * (1024))/16, 16*(4), (16) );
          }
        }
        if (io == 0) {
          if (i == 0) {
            mvin3( &B[(48 + 64 * ko)][0], b + ((ko) * (4096) + (3) * (1024))/16, 16*(4), (16) );
          }
        }
        preload(b + ((ko) * (4096))/16, res + ((i) * (1024))/16 | 0x40000000, (16), (16), (16), (16));
        compute_preloaded(a + ((i) * (4096) + (ko) * (1024))/16, ~((uint32_t)0), (16), (16), 16, 16);
        preload(b + ((ko) * (4096) + 256)/16, res + ((i) * (1024) + 256)/16 | 0x40000000, (16), (16), (16), (16));
        compute_preloaded(a + ((i) * (4096) + (ko) * (1024))/16, ~((uint32_t)0), (16), (16), 16, 16);
        preload(b + ((ko) * (4096) + (2) * (256))/16, res + ((i) * (1024) + (2) * (256))/16 | 0x40000000, (16), (16), (16), (16));
        compute_preloaded(a + ((i) * (4096) + (ko) * (1024))/16, ~((uint32_t)0), (16), (16), 16, 16);
        preload(b + ((ko) * (4096) + (3) * (256))/16, res + ((i) * (1024) + (3) * (256))/16 | 0x40000000, (16), (16), (16), (16));
        compute_preloaded(a + ((i) * (4096) + (ko) * (1024))/16, ~((uint32_t)0), (16), (16), 16, 16);
        preload(b + ((ko) * (4096) + 1024)/16, res + ((i) * (1024))/16 | 0x40000000, (16), (16), (16), (16));
        compute_preloaded(a + ((i) * (4096) + (ko) * (1024) + 256)/16, ~((uint32_t)0), (16), (16), 16, 16);
        preload(b + ((ko) * (4096) + 1024 + 256)/16, res + ((i) * (1024) + 256)/16 | 0x40000000, (16), (16), (16), (16));
        compute_preloaded(a + ((i) * (4096) + (ko) * (1024) + 256)/16, ~((uint32_t)0), (16), (16), 16, 16);
        preload(b + ((ko) * (4096) + 1024 + (2) * (256))/16, res + ((i) * (1024) + (2) * (256))/16 | 0x40000000, (16), (16), (16), (16));
        compute_preloaded(a + ((i) * (4096) + (ko) * (1024) + 256)/16, ~((uint32_t)0), (16), (16), 16, 16);
        preload(b + ((ko) * (4096) + 1024 + (3) * (256))/16, res + ((i) * (1024) + (3) * (256))/16 | 0x40000000, (16), (16), (16), (16));
        compute_preloaded(a + ((i) * (4096) + (ko) * (1024) + 256)/16, ~((uint32_t)0), (16), (16), 16, 16);
        ... 
        // Unrolling continues
      }
      mvout( &C[(16 * i + 128 * io)][0], res + ((i) * (1024))/16, (16), (16) );
      mvout( &C[(16 * i + 128 * io)][16], res + ((i) * (1024) + 256)/16, (16), (16) );
      mvout( &C[(16 * i + 128 * io)][32], res + ((i) * (1024) + (2) * (256))/16, (16), (16) );
      mvout( &C[(16 * i + 128 * io)][48], res + ((i) * (1024) + (3) * (256))/16, (16), (16) );
    }
  }
  fence();
}
\end{lstlisting}
\caption{Example of hand-optimized matrix multiplication code from \citet{exo}, used as a baseline in \cref{sec:gemm}. Achieves 43\% utilization. Note that unrolled \texttt{preload}/\texttt{compute\_preloaded} instructions have been truncated due to length.}
\label{fig:exo-2-exo-opt}
\end{figure*}

\clearpage
% \subsubsection{\sys{}-Optimized Code Example}

\begin{figure*}[t]
\begin{lstlisting}[language=C++, linewidth=1\textwidth, basicstyle=\ttfamily\tiny]
void test(int8_t A[12544][256], int8_t B[256][64], int8_t C[12544][64]) {
  //--------------------------------------------------------------------------
  // Tile and matrix constants
  //--------------------------------------------------------------------------

  const uint32_t tile_dim    = 16;   // tile dimension
  const uint32_t tile_offset = tile_dim; // each row in a tile is tile_dim elements

  // For double buffering, we use two accumulator base addresses separated by 4 tile rows.
  const uint32_t acc_base0 = 1U << 31;
  const uint32_t acc_base1 = (1U << 31) + (4 * tile_dim);

  const uint32_t new_B_base   = 8192;
  const uint32_t A_tile_base0 = 2048;
  const uint32_t A_tile_base1 = 6144;

  //--------------------------------------------------------------------------
  // Gemmini configuration
  //--------------------------------------------------------------------------

  config_st(64);
  config_ex(WEIGHT_STATIONARY, NO_ACTIVATION, 1, false, false);
  config_ld(64, 1.0f, tile_dim, 2);
  config_ld(256, 1.0f, tile_dim, 1);
  config_ld(0, 1.0f, 0, 0);

  //--------------------------------------------------------------------------
  // Load the entire weight matrix B into the scratchpad once.
  //--------------------------------------------------------------------------

  for (int br = 0; br < 256; br += tile_dim) {
    mvin3(&B[br][0],
          new_B_base + (br / tile_dim) * (tile_dim * 4),
          tile_dim * 4,
          tile_dim);
  }

  //--------------------------------------------------------------------------
  // Begin double-buffered accumulator computation.
  //
  // cur_acc_base: where the current tile is computed.
  // prev_acc_base: holds the result of the previously computed tile.
  //--------------------------------------------------------------------------

  uint32_t cur_acc_base  = acc_base0;
  uint32_t prev_acc_base = 0; // Will be set after the first iteration.

  // Outer loop over tiles (784 tiles along A's first dimension)
  for (int i = 0; i < 784; i++) {

    // Determine which scratchpad region holds A for this tile.
    // Alternating between two buffers allows overlapping the load of the next tile.
    uint32_t current_A_buffer = (i % 2 == 0) ? A_tile_base0 : A_tile_base1;
    uint32_t next_A_buffer    = (i % 2 == 0) ? A_tile_base1 : A_tile_base0;

    //-------------------------------------------------------------------------
    // For the first iteration load the A tile into scratchpad.
    // For subsequent iterations, write the previous tile from the accumulator.
    //-------------------------------------------------------------------------
    if (i == 0) {
      for (int b = 0; b < 4; b++) {
        mvin2(&A[tile_dim * i][64 * b],
              current_A_buffer + b * (tile_dim * 4),
              tile_dim * 4,
              tile_dim);
      }
    } else {
      for (int j_in_o = 0; j_in_o < 4; j_in_o++) {
        uint32_t j_off = j_in_o * tile_dim;
        mvout(&C[tile_dim * (i - 1)][tile_dim * j_in_o],
              prev_acc_base + j_off,
              tile_dim,
              tile_dim);
      }
    }
\end{lstlisting}
\caption{Example of an optimized version of the same GEMM from \cref{fig:exo-2-unopt}, generated using \sys{}. Achieves 93\% compute utilization. Continued in \cref{fig:exo-2-opt-cont,fig:exo-2-opt-cont-cont}.}
\label{fig:exo-2-opt}
\end{figure*}
\clearpage
\begin{figure*}[!t]
\begin{lstlisting}
    //-------------------------------------------------------------------------
    // Instead of explicitly zeroing-out the accumulator tile via extended mvin,
    // we fuse the accumulator zeroing into the compute stream.
    //
    // For each accumulator sub-tile (indexed by j_in_o), the very first compute
    // call is issued in overwrite mode (i.e. the accumulator address is used as-is)
    // and then all subsequent compute calls for that sub-tile are issued in accumulate mode.
    //
    // We use an array (first_compute) to track whether a given sub-tile has been updated.
    //-------------------------------------------------------------------------
    int first_compute[4] = { 1, 1, 1, 1 };

    //-------------------------------------------------------------------------
    // Compute the current tile.
    // The complete computation is divided into 16 segments.
    // Loop unrolling by a factor of 4 is applied to the segments loop.
    //-------------------------------------------------------------------------
    for (int seg = 0; seg < 16; seg += 4) {
      for (int j_in_o = 0; j_in_o < 4; j_in_o++) {
        uint32_t j_off = j_in_o * tile_dim;
        uint32_t acc_addr; // will hold the computed accumulator address for preload

        //--- Unrolled segment "seg" -------------------------------------------------
        {
          uint32_t block_id   = seg / 4;
          uint32_t sub_offset = (seg % 4) * tile_dim;
          uint32_t A_seg_addr = current_A_buffer + block_id * (tile_dim * 4) + sub_offset;
          // If this is the very first compute for this sub-tile, use overwrite mode
          if (first_compute[j_in_o]) {
            acc_addr = cur_acc_base + j_off;
            first_compute[j_in_o] = 0;
          } else {
            acc_addr = (cur_acc_base + j_off) | 0x40000000;
          }
          preload(new_B_base + seg * (tile_dim * 4) + j_off,
                  acc_addr,
                  tile_dim, tile_dim,
                  tile_dim, tile_dim);
          compute_preloaded(A_seg_addr,
                            ~(uint32_t)0,
                            tile_dim, tile_dim,
                            tile_dim, tile_dim);
        }

        //--- Unrolled segment "seg+1" -------------------------------------------------
        {
          uint32_t block_id   = (seg + 1) / 4;
          uint32_t sub_offset = ((seg + 1) % 4) * tile_dim;
          uint32_t A_seg_addr = current_A_buffer + block_id * (tile_dim * 4) + sub_offset;
          preload(new_B_base + (seg + 1) * (tile_dim * 4) + j_off,
                  (cur_acc_base + j_off) | 0x40000000,
                  tile_dim, tile_dim,
                  tile_dim, tile_dim);
          compute_preloaded(A_seg_addr,
                            ~(uint32_t)0,
                            tile_dim, tile_dim,
                            tile_dim, tile_dim);
        }

        //--- Unrolled segment "seg+2" -------------------------------------------------
        {
          uint32_t block_id   = (seg + 2) / 4;
          uint32_t sub_offset = ((seg + 2) % 4) * tile_dim;
          uint32_t A_seg_addr = current_A_buffer + block_id * (tile_dim * 4) + sub_offset;
          preload(new_B_base + (seg + 2) * (tile_dim * 4) + j_off,
                  (cur_acc_base + j_off) | 0x40000000,
                  tile_dim, tile_dim,
                  tile_dim, tile_dim);
          compute_preloaded(A_seg_addr,
                            ~(uint32_t)0,
                            tile_dim, tile_dim,
                            tile_dim, tile_dim);
        }
\end{lstlisting}
\caption{Example from \cref{fig:exo-2-opt}, continued.}
\label{fig:exo-2-opt-cont}
\end{figure*}

\begin{figure*}[!t]
\begin{lstlisting}
        //--- Unrolled segment "seg+3" -------------------------------------------------
        {
          uint32_t block_id   = (seg + 3) / 4;
          uint32_t sub_offset = ((seg + 3) % 4) * tile_dim;
          uint32_t A_seg_addr = current_A_buffer + block_id * (tile_dim * 4) + sub_offset;
          preload(new_B_base + (seg + 3) * (tile_dim * 4) + j_off,
                  (cur_acc_base + j_off) | 0x40000000,
                  tile_dim, tile_dim,
                  tile_dim, tile_dim);
          compute_preloaded(A_seg_addr,
                            ~(uint32_t)0,
                            tile_dim, tile_dim,
                            tile_dim, tile_dim);
        }
      } // end inner loop over j_in_o

      //-------------------------------------------------------------------------
      // For seg==0 (i.e. the first unrolled iteration), launch prefetching of the next A tile.
      // This overlaps memory-access with computation.
      //-------------------------------------------------------------------------
      if (seg == 0 && i < 783) {
        for (int b = 0; b < 4; b++) {
          mvin2(&A[tile_dim * (i + 1)][64 * b],
                next_A_buffer + b * (tile_dim * 4),
                tile_dim * 4,
                tile_dim);
        }
      }
    } // end segments loop

    //-------------------------------------------------------------------------
    // Swap accumulator buffers.
    // The tile computed in this iteration (in cur_acc_base) becomes the previous tile,
    // so it must be written back in the next iteration.
    //-------------------------------------------------------------------------
    prev_acc_base = cur_acc_base;
    cur_acc_base = (cur_acc_base == acc_base0) ? acc_base1 : acc_base0;
  } // end outer tile loop

  //--------------------------------------------------------------------------
  // Write back the final computed tile (tile index 783) from the accumulator.
  //--------------------------------------------------------------------------
  for (int j_in_o = 0; j_in_o < 4; j_in_o++) {
    uint32_t j_off = j_in_o * tile_dim;
    mvout(&C[tile_dim * (784 - 1)][tile_dim * j_in_o],
          prev_acc_base + j_off,
          tile_dim,
          tile_dim);
  }

  fence();
}
\end{lstlisting}
\caption{Example from \cref{fig:exo-2-opt,fig:exo-2-opt-cont}, continued.}
\label{fig:exo-2-opt-cont-cont}
\end{figure*}

\clearpage
\subsection{4x4 FP32 Gemmini---TinyMPC Primal Update Forward Pass (Fine-Grained Linear Algebra)}\label{sec:code-examples-admm}

The forward pass of the TinyMPC primal update step computes the following operations:
$$u[i] = - K_{inf} * x[i] - d[i]$$
$$x[i+1] = (A_{dyn} * x[i]) + (B_{dyn} * u[i])$$

Where $A_{dyn}$ is a \texttt{12x12} matrix, $B_{dyn}$ is a \texttt{12x4} matrix, $K_{inf}$ is a \texttt{4x12} matrix, $x$ is an \texttt{NHORIZONx12} matrix (where individual columns are accessed here via indexing), $d$ is an \texttt{NHORIZONx4} matrix, and $u$ is an \texttt{NHORIZONx4} matrix. $A_{dyn}$, $B_{dyn}$, $K_{inf}$, $d$ and the 0th column of $x$ are inputs, and $u$ is the output. The 1st to (\texttt{NHORIZON}-1)th column of $x$ are intermediate values computed over the course of the benchmark. 

This process is repeated until a time horizon, defined as \texttt{NHORIZON} in our code and set to 5 for our evaluations. Note that $x$ is allocated as an \texttt{(NHORIZON+1)x12} matrix in our code since the unoptimized code accesses up to the (\texttt{NHORIZON})th column.

\sys{} generates the code in \cref{fig:admm-1-opt}, optimized over several steps from the starting code in \cref{fig:exo-2-unopt}. The following optimizations are applied, with the following speedups after each optimization:

\begin{enumerate}
    \item 1$\times$: in this case we treat the unoptimized software as the baseline for speedup, so by definition its speedup is 1$\times$.
    
    \item 1.07$\times$ after ``hoist redundant operations out of loops''. Hoists the \texttt{mvin} calls for the constant matrices \texttt{Kinf}, \texttt{Adyn}, and \texttt{Bdyn} above the \texttt{NHORIZON} loop and executes them once rather than in every iteration. Any associated \texttt{config\_ex} and \texttt{config\_ld} calls are also moved outside the loop if needed. The compute calls use the same scratchpad addresses and may set \texttt{B\_spad\_addr = 0xffffffff} to indicate that the weights are already loaded.
    
    \item 1.13$\times$ after ``loop reordering''. This plan claims to move the configuration and \texttt{mvin} instructions for \texttt{Kinf}, \texttt{Adyn}, and \texttt{Bdyn} before the \texttt{NHORIZON} loop, but this has already been handled in the previous step. In reality, only some configuration instructions that were unnecessarily left behind by the previous step are hoisted.
    
    \item 2.00$\times$ after ``move CPU-based computation to the accelerator''. Replaces CPU-based element-wise negation and addition with equivalent Gemmini compute instructions. Specifically, when \texttt{x\_i} and \texttt{d\_i} are loaded, a scaling factor of -1 is used to negate it in the scratchpad. The product of \texttt{Kinf} and \texttt{x\_i} is kept in the accumulator and the negated \texttt{d\_i} is multiplied by a dummy 0 vector in order accumulate it on top of this product, enabling data to be kept resident in the accumulator rather than moving it back and forth between the CPU and accelerator memory.

    \item 2.12$\times$ after ``hoisting redundant operations out of loops''. Identifies a few \texttt{config\_ld} instructions that are still redundant and removes them from the loop.

    \item 2.66$\times$ after ``loop unrolling''. Changes the outer loop to increment by 2 instead of 1 and duplicate the loop body so each iteration computes two time steps: first u[i] and x[i+1], then u[i+1] and x[i+2]. Also merges \texttt{fence} calls at the end of the unrolled loop body if possible. The implementation very aggressively removes \texttt{fence} instructions, which is actually correct as reuse of the same accelerator memory instructions by subsequent \texttt{mvin} and \texttt{mvout} instructions means that dependencies can be handled internally to the accelerator, rather than via synchronization of both the CPU and accelerator.
    
    \item 2.95$\times$ after ``loop fusion''. Undoes the loop unrolling from the previous step, but keeps the reduced \texttt{fence} instructions. Additionally, this eliminates the unnecessary calculation of \texttt{x[NHORIZON]} during execution, which saves cycles. This optimization makes sense since it is likely the reduced number of \texttt{fence} instructions that improved performance in step 6, rather than the loop unrolling (which usually provides limited benefit, if any).
\end{enumerate}

We further discuss the similarities and differences between the hand-optimized hardware FSM-based implementation in \cref{fig:admm-1-hw-opt} and the \sys{}-optimized code:

\begin{itemize}
    \item \textbf{Data orchestration.} Due to the coarse-grained nature of the hardware FSM operations and the fact that data movement between the accelerator and main memory is handled within hardware, we are not able to hoist shared data loads of the \texttt{Kinf}, \texttt{Adyn}, and \texttt{Bdyn} matrices out of the loop. This is the main advantage the \sys{}-generated software ISA-based implementation has over the hardware FSM-based implementation.

    \item \textbf{Operator fusion.} Both implementations are able to handle all computation on the accelerator, but notably the hardware FSM-based implementation fuses the addition of \texttt{d[i]} into the bias when multiplying \texttt{Kinf} and \texttt{x[i]} (then the whole result is negated while storing to main memory), whereas the addition of \texttt{d[i]} is handled as a separate compute instruction in the \sys{}-generated code. So, the hardware FSM-based implementation actually has an advantage in this regard and the \sys{}-generated code has further room for improvement. Both implementations make use of negative scaling of loaded data (via \texttt{config\_ld} instructions) in order to handle subtraction.

    \item \textbf{Fence instructions.} The \sys{}-generated code is able to remove CPU-based fence instructions and instead handle dependencies within the accelerator, whereas the hardware FSM-based code is forced to place a fence after each matrix multiplication, as accumulated results must be moved back to main memory and loaded to the scratchpad for the next operation.

    \item \textbf{Configuration overhead.} While they do not have as much overhead as fence instructions, configuration instructions can also be blocking. The hand-optimized code hoists configuration instructions out of the loop where possible, but since different matrix sizes must be loaded inside the loop, configuration instructions cannot be completely eliminated, giving \sys{}'s code the advantage in this aspect.

    \item \textbf{Dead code elimination.} Both implementations eliminate the extra computation of \texttt{x[NHORIZON]} that is present in the unoptimized code.
\end{itemize}

Overall, we find that \sys{} identifies all major optimization opportunities available in the code, with the exception of handling subtraction of \texttt{d[i]} slightly suboptimally. Qualitatively, optimizing this code by hand can be difficult due to a lack of readability and the difficulty of debugging low-level accelerator code. The \sys{}-generated code is well-commented and the sequence of optimizations applied can be easily understood. This will be helpful for further optimization of this benchmark or future optimization of other benchmarks, via methods like the schedule reuse demonstrated in \cref{sec:reuse-eval}.

\clearpage
% \subsubsection{Unoptimized Software ISA-Based Code Example}

\begin{figure*}[!h]
\begin{lstlisting}[language=C++, linewidth=1\textwidth, basicstyle=\ttfamily\tiny]
void test(float Adyn[12][12], float Bdyn[12][4] float Kinf[4][12], float x[NHORIZON + 1][12][1], float d[NHORIZON][4][1], float u[NHORIZON][4][1]) {
    static elem_t Kinf_x[4][1];
    static elem_t A_x[12][1];
    static elem_t B_u[12][1];

    for (int i = 0; i < NHORIZON; i++) {
        // define spad addresses for cached matrices
        // spad is row addressed and each row is 4 elements wide
        static uint32_t A_sp_addr = 0; // 144 elements, 0 to 35
        static uint32_t B_sp_addr = 36; // 48 elements, 36 to 47
        static uint32_t Kinf_sp_addr = 48; // 48 elements, 48 to 59
        static uint32_t C1_sp_addr = 60; // 16 elements, 60 to 63
        static uint32_t C2_sp_addr = 64; // 144 elements, 64 to 99
        static uint32_t x_sp_addr = 100; // 12 elements (at a time), 100 to 111
        static uint32_t u_sp_addr = 112; // 12 elements (at a time), 112 to 123
        static uint32_t acc_start_addr = 1 << 31;

        // tiled_matmul_spad_dram(Kinf, x[i], Kinf_x, NINPUTS, false, false);
        config_ex(WEIGHT_STATIONARY, NO_ACTIVATION, 1, false, false);
        config_st(4, 1.0);
        config_ld(48, 1.000000, 4, 0);
        config_ld(4, 1.000000, 4, 1);
        config_ld(4, 1.000000, 4, 2);
        mvin(Kinf, Kinf_sp_addr, 12, 4);
        mvin2(x[i][0], x_sp_addr, 1, 4);
        preload(x_sp_addr, acc_start_addr, 1, 4, 1, 4);
        compute_preloaded(Kinf_sp_addr, 0xffffffff, 4, 4, 4, 4);
        mvin2(x[i][4], x_sp_addr + 4, 1, 4);
        preload(x_sp_addr + 4, acc_start_addr | (1 << 30), 1, 4, 1, 4);
        compute_preloaded(Kinf_sp_addr + 4, 0xffffffff, 4, 4, 4, 4);
        mvin2(x[i][8], x_sp_addr + 8, 1, 4);
        preload(x_sp_addr + 8, acc_start_addr | (1 << 30), 1, 4, 1, 4);
        compute_preloaded(Kinf_sp_addr + 8, 0xffffffff, 4, 4, 4, 4);
        mvout(Kinf_x[0], acc_start_addr | (1 << 30), 1, 4);
        fence();

        static acc_t Kinf_x_negated[4][1] row_align_acc(1);
        static acc_t d_i_negated[4][1] row_align_acc(1);
        negate_matrix(Kinf_x, Kinf_x_negated, 4, 1);
        negate_matrix(d[i], d_i_negated, 4, 1);
        add_matrix(Kinf_x_negated, d_i_negated, u[i], 4, 1);
\end{lstlisting}
\caption{Unoptimized software ISA-based starting code for the TinyMPC primal update forward pass from \cref{sec:admm}. Achieves 5.7\% of theoretical maximum utilization. Continued in \cref{fig:admm-1-unopt-cont}.}
\label{fig:admm-1-unopt}
\end{figure*}

\begin{figure*}
\begin{lstlisting}[language=C++, linewidth=1\textwidth, basicstyle=\ttfamily\tiny]
        // tiled_matmul_spad_dram(Adyn, x[i], A_x, NSTATES, false, false);
        config_ex(WEIGHT_STATIONARY, NO_ACTIVATION, 1, false, false);
        config_st(4, 1.0);
        config_ld(48, 1.000000, 4, 0);
        config_ld(4, 1.000000, 4, 1);
        config_ld(4, 1.000000, 4, 2);
        for (int chunk = 0; chunk < 3; chunk++) {
            mvin(Adyn[chunk*4], A_sp_addr + chunk*12, 12, 4);
        }
        mvin2(x[i][0], x_sp_addr, 1, 4);
        mvin2(x[i][4], x_sp_addr + 4, 1, 4);
        mvin2(x[i][8], x_sp_addr + 8, 1, 4);

        preload(x_sp_addr, acc_start_addr, 1, 4, 1, 4);
        compute_preloaded(A_sp_addr, 0xffffffff, 4, 4, 4, 4);
        preload(0xffffffff, acc_start_addr + 4, 1, 4, 1, 4);
        compute_accumulated(A_sp_addr + 12, 0xffffffff, 4, 4, 4, 4);
        preload(0xffffffff, acc_start_addr + 8, 1, 4, 1, 4);
        compute_accumulated(A_sp_addr + 24, 0xffffffff, 4, 4, 4, 4);

        preload(x_sp_addr + 4, acc_start_addr | (1 << 30), 1, 4, 1, 4);
        compute_preloaded(A_sp_addr + 4, 0xffffffff, 4, 4, 4, 4);
        preload(0xffffffff, (acc_start_addr + 4) | (1 << 30), 1, 4, 1, 4);
        compute_accumulated(A_sp_addr + 4 + 12, 0xffffffff, 4, 4, 4, 4);
        preload(0xffffffff, (acc_start_addr + 8) | (1 << 30), 1, 4, 1, 4);
        compute_accumulated(A_sp_addr + 4 + 24, 0xffffffff, 4, 4, 4, 4);

        preload(x_sp_addr + 8, acc_start_addr | (1 << 30), 1, 4, 1, 4);
        compute_preloaded(A_sp_addr + 8, 0xffffffff, 4, 4, 4, 4);
        preload(0xffffffff, (acc_start_addr + 4) | (1 << 30), 1, 4, 1, 4);
        compute_accumulated(A_sp_addr + 8 + 12, 0xffffffff, 4, 4, 4, 4);
        preload(0xffffffff, (acc_start_addr + 8) | (1 << 30), 1, 4, 1, 4);
        compute_accumulated(A_sp_addr + 8 + 24, 0xffffffff, 4, 4, 4, 4);

        mvout(A_x[0], acc_start_addr, 1, 4);
        mvout(A_x[4], acc_start_addr + 4, 1, 4);
        mvout(A_x[8], acc_start_addr + 8, 1, 4);
        fence();

        // tiled_matmul_spad_dram(Bdyn, u[i], B_u, NSTATES, false, false);
        config_ex(WEIGHT_STATIONARY, NO_ACTIVATION, 1, false, false);
        config_st(4, 1.0);
        config_ld(16, 1.000000, 4, 0);
        config_ld(4, 1.000000, 4, 1);
        config_ld(4, 1.000000, 4, 2);
        for (int chunk = 0; chunk < 3; chunk++) {
            mvin(Bdyn[chunk*4], B_sp_addr + chunk*4, 4, 4);
        }
        mvin2(u[i][0], x_sp_addr, 1, 4);
        preload(x_sp_addr, acc_start_addr, 1, 4, 1, 4);
        compute_preloaded(B_sp_addr, 0xffffffff, 4, 4, 4, 4);
        preload(0xffffffff, acc_start_addr + 4, 1, 4, 1, 4);
        compute_accumulated(B_sp_addr + 4, 0xffffffff, 4, 4, 4, 4);
        preload(0xffffffff, acc_start_addr + 8, 1, 4, 1, 4);
        compute_accumulated(B_sp_addr + 8, 0xffffffff, 4, 4, 4, 4);
        mvout(B_u[0], acc_start_addr, 1, 4);
        mvout(B_u[4], acc_start_addr + 4, 1, 4);
        mvout(B_u[8], acc_start_addr + 8, 1, 4);
        fence();

        add_matrix(A_x, B_u, x[i+1], 12, 1);
    }
}
\end{lstlisting}
\caption{Unoptimized software ISA-based starting code for the TinyMPC primal update forward pass, continued from \cref{fig:admm-1-unopt}.}
\label{fig:admm-1-unopt-cont}
\end{figure*}

\clearpage
% \subsubsection{Optimized Hardware FSM-Based Code Example}

\begin{figure*}[h]
\begin{lstlisting}[language=C++, linewidth=1\textwidth, basicstyle=\ttfamily\tiny]
void test(float Adyn[12][12], float Bdyn[12][4] float Kinf[4][12], float x[NHORIZON + 1][12][1], float d[NHORIZON][4][1], float u[NHORIZON][4][1]) {
    static elem_t B_u[12][1];

    gemmini_extended_config_ex(1, 0, 0, 1, false, false);
    gemmini_extended3_config_ld(4, 1.0, false, 1);
    gemmini_extended3_config_ld(4, 1.0, false, 2);
    for (int i = 0; i < NHORIZON; i++)
    {
        gemmini_extended_config_st(4, 0, -1.0);
        gemmini_extended3_config_ld(48, 1.0, false, 0);
        gemmini_loop_ws(1, 1, 3, 0, 3, 0, Kinf, x[i], d[i], u[i], 12, 1, 1, 1, false, false, false, false, true, 0, 1, 1, false);
        gemmini_fence();

        if (i < NHORIZON - 1) {
            gemmini_extended_config_st(4, 0, 1.0);
            gemmini_extended3_config_ld(16, 1.0, false, 0);
            gemmini_loop_ws(3, 1, 1, 0, 3, 0, Bdyn, u[i], NULL, B_u, 4, 1, 1, 1, false, false, false, false, false, 0, 1, 1, false);
            gemmini_fence();

            gemmini_extended3_config_ld(48, 1.0, false, 0);
            gemmini_loop_ws(3, 1, 3, 0, 3, 0, Adyn, x[i], B_u, x[i+1], 12, 1, 1, 1, false, false, false, false, true, 0, 1, 1, false);
            gemmini_fence();
        }
    }
}
\end{lstlisting}
\caption{Example of the TinyMPC primal update forward pass from \cref{fig:admm-1-unopt}, hand-optimized and using the hardware FSM, from the experiments. Achieves 10.9\% of theoretical maximum utilization.}
\label{fig:admm-1-hw-opt}
\end{figure*}

% \subsubsection{\sys{}-Generated Code Example}
\begin{figure*}[h]
\begin{lstlisting}[language=C++, linewidth=1\textwidth, basicstyle=\ttfamily\tiny]
void test(float Adyn[12][12],
          float Bdyn[12][4],
          float Kinf[4][12],
          float x[NHORIZON + 1][12][1],
          float d[NHORIZON][4][1],
          float u[NHORIZON][4][1]) {

  // Define scratchpad addresses for all matrices.
  uint32_t A_sp_addr      = 0;
  uint32_t B_sp_addr      = 36;
  uint32_t Kinf_sp_addr   = 48;
  uint32_t x_sp_addr      = 100;
  uint32_t acc_start_addr = 1 << 31;  // MSB = 1 indicates accumulator address

  //------------------------------------------------------------------
  // Hoisted Invariant Matrix Loads: These matrices do not change over
  // the horizon. We load them once into the Gemmini scratchpad.
  //------------------------------------------------------------------
  config_ex(WEIGHT_STATIONARY, NO_ACTIVATION, 1, false, false);
  config_st(4, 1.0);

  // Load constant weight matrix Kinf.
  config_ld(48, 1.000000, 4, 0);
  mvin(Kinf, Kinf_sp_addr, 12, 4);

  // Load constant system dynamics matrix Adyn.
  config_ld(48, 1.000000, 4, 0);
  for (int chunk = 0; chunk < 3; chunk++) {
    mvin(Adyn[chunk * 4], A_sp_addr + chunk * 12, 12, 4);
  }

  // Load constant control dynamics matrix Bdyn.
  config_ld(16, 1.000000, 4, 0);
  for (int chunk = 0; chunk < 3; chunk++) {
    mvin(Bdyn[chunk * 4], B_sp_addr + chunk * 4, 4, 4);
  }
\end{lstlisting}
\caption{Example of an \sys{}-optimized version of the TinyMPC primal update forward pass from \cref{fig:admm-1-unopt}. Achieves 15.7\% of theoretical maximum compute utilization. Continued in \cref{fig:admm-1-opt-cont}.}
\label{fig:admm-1-opt}
\end{figure*}

\begin{figure*}
\begin{lstlisting}[language=C++, linewidth=1\textwidth, basicstyle=\ttfamily\tiny]
  //------------------------------------------------------------------
  // Outer horizon loop with fusion of iterations and operations.
  //------------------------------------------------------------------
  for (int i = 0; i < NHORIZON; i++) {
    // Compute u[i] = -(Kinf*x[i]) - d[i]
    config_ld(4, -1.000000, 4, 1);
    mvin2(x[i][0], x_sp_addr, 1, 4);
    preload(x_sp_addr, acc_start_addr, 1, 4, 1, 4);
    compute_preloaded(Kinf_sp_addr, 0xffffffff, 4, 4, 4, 4);

    mvin2(x[i][4], x_sp_addr + 4, 1, 4);
    preload(x_sp_addr + 4, acc_start_addr | (1 << 30), 1, 4, 1, 4);
    compute_preloaded(Kinf_sp_addr + 4, 0xffffffff, 4, 4, 4, 4);

    mvin2(x[i][8], x_sp_addr + 8, 1, 4);
    preload(x_sp_addr + 8, acc_start_addr | (1 << 30), 1, 4, 1, 4);
    compute_preloaded(Kinf_sp_addr + 8, 0xffffffff, 4, 4, 4, 4);

    mvin2(d[i], x_sp_addr, 1, 4);
    config_ld(4, 1.000000, 4, 1);
    mvin2(0, x_sp_addr + 4, 1, 4);
    preload(x_sp_addr + 4, acc_start_addr | (1 << 30), 1, 4, 1, 4);
    compute_accumulated(x_sp_addr + 4, x_sp_addr, 1, 4, 1, 4);

    mvout(u[i], acc_start_addr, 1, 4);

    if (i < NHORIZON - 1) {
      // Compute A_x = Adyn * x[i]
      mvin2(x[i][0], x_sp_addr, 1, 4);
      mvin2(x[i][4], x_sp_addr + 4, 1, 4);
      mvin2(x[i][8], x_sp_addr + 8, 1, 4);

      preload(x_sp_addr, acc_start_addr, 1, 4, 1, 4);
      compute_preloaded(A_sp_addr, 0xffffffff, 4, 4, 4, 4);
      preload(0xffffffff, acc_start_addr + 4, 1, 4, 1, 4);
      compute_accumulated(A_sp_addr + 12, 0xffffffff, 4, 4, 4, 4);
      preload(0xffffffff, acc_start_addr + 8, 1, 4, 1, 4);
      compute_accumulated(A_sp_addr + 24, 0xffffffff, 4, 4, 4, 4);

      preload(x_sp_addr + 4, acc_start_addr | (1 << 30), 1, 4, 1, 4);
      compute_preloaded(A_sp_addr + 4, 0xffffffff, 4, 4, 4, 4);
      preload(0xffffffff, (acc_start_addr + 4) | (1 << 30), 1, 4, 1, 4);
      compute_accumulated(A_sp_addr + 4 + 12, 0xffffffff, 4, 4, 4, 4);
      preload(0xffffffff, (acc_start_addr + 8) | (1 << 30), 1, 4, 1, 4);
      compute_accumulated(A_sp_addr + 4 + 24, 0xffffffff, 4, 4, 4, 4);

      preload(x_sp_addr + 8, acc_start_addr | (1 << 30), 1, 4, 1, 4);
      compute_preloaded(A_sp_addr + 8, 0xffffffff, 4, 4, 4, 4);
      preload(0xffffffff, (acc_start_addr + 4) | (1 << 30), 1, 4, 1, 4);
      compute_accumulated(A_sp_addr + 8 + 12, 0xffffffff, 4, 4, 4, 4);
      preload(0xffffffff, (acc_start_addr + 8) | (1 << 30), 1, 4, 1, 4);
      compute_accumulated(A_sp_addr + 8 + 24, 0xffffffff, 4, 4, 4, 4);

      // Compute B_u = Bdyn * u[i] and accumulate onto A_x
      mvin2(u[i][0], x_sp_addr, 1, 4);
      preload(x_sp_addr, acc_start_addr | (1 << 30), 1, 4, 1, 4);
      compute_preloaded(B_sp_addr, 0xffffffff, 4, 4, 4, 4);
      preload(0xffffffff, (acc_start_addr + 4) | (1 << 30), 1, 4, 1, 4);
      compute_accumulated(B_sp_addr + 4, 0xffffffff, 4, 4, 4, 4);
      preload(0xffffffff, (acc_start_addr + 8) | (1 << 30), 1, 4, 1, 4);
      compute_accumulated(B_sp_addr + 8, 0xffffffff, 4, 4, 4, 4);

      mvout(x[i + 1][0], acc_start_addr, 1, 4);
      mvout(x[i + 1][4], acc_start_addr + 4, 1, 4);
      mvout(x[i + 1][8], acc_start_addr + 8, 1, 4);

      fence();
    }
  }
}
\end{lstlisting}
\caption{Example from \cref{fig:admm-1-opt}, continued.}
\label{fig:admm-1-opt-cont}
\end{figure*}

%%%%%%%%%%%%%%%%%%%%%%%%%%%%%%%%%%%%%%%%%%%%%%%%%%%%%%%%%%%%%%%%%%%%%%%%%%%%%%%
%%%%%%%%%%%%%%%%%%%%%%%%%%%%%%%%%%%%%%%%%%%%%%%%%%%%%%%%%%%%%%%%%%%%%%%%%%%%%%%

\end{document}